\documentclass[1p]{elsarticle}
\usepackage{threeparttable}
\usepackage{longtable}
\usepackage{placeins}
\usepackage{rotating}
\usepackage{lineno,hyperref,natbib}
\usepackage[dvipsnames]{xcolor}
\modulolinenumbers[10]

\newcommand\araa{{ARA\&A}}
\newcommand\apj{{ApJ}}
\newcommand\aap{{A\&A}}
\newcommand\icarus{{Icarus}}
\newcommand\mnras{{MNRAS}}
\newcommand\nat{{Nature}}
\newcommand\grl{{Geophys.~Res.~Lett.}}
\newcommand\jgr{{J.~Geophys.~Res.}}

\journal{Journal of \LaTeX\ Templates}

\begin{document}

\begin{frontmatter}

\title{Planet Four: Terrains - Discovery of Araneiforms Outside of the South Polar Layered Deposits}

\author[1,2]{Megan E. Schwamb\corref{mycorrespondingauthor}}
\cortext[mycorrespondingauthor]{Corresponding author}
\ead[]{mschwamb.astro@gmail.com}

\author[3]{Klaus-Michael Aye}
\author[3]{Ganna Portyankina}
\author[4]{Candice J. Hansen}
\author[5]{Campbell Allen}
\author[6]{Sarah Allen}
\author[7]{Fred J. Calef III}
\author[5]{Simone Duca}
\author[5]{ Adam McMaster}
\author[5]{Grant R. M. Miller}

\address[1]{Gemini Observatory,  Northern Operations Center, 670 North A'ohoku Place, Hilo, HI 96720, USA}
\address[2]{Institute for Astronomy and Astrophysics, Academia Sinica; 11F AS/NTU,  National Taiwan University, 1 Roosevelt Rd., Sec. 4, Taipei 10617, Taiwan}
\address[3]{Laboratory for Atmospheric and Space Physics, University of Colorado at Boulder, Boulder, Colorado, 80303, USA}
\address[4]{Planetary Science Institute, 1700 E. Fort Lowell, Suite 106, Tucson, AZ 85719, USA}
\address[5]{Oxford Astrophysics, Denys Wilkinson Building, Keble Road, Oxford OX1 3RH, UK}
\address[6]{Adler Planetarium, 1300 S. Lake Shore Drive, Chicago, IL 60605, USA}
\address[7]{Jet Propulsion Laboratory, California Institute of Technology, Pasadena, CA 91109, USA}

\begin{abstract}
We present the results of a systematic mapping of seasonally sculpted terrains on the South Polar region of Mars with the Planet Four: Terrains (P4T) online citizen science project. P4T enlists members of the general public to visually identify features in the publicly released \emph{Mars Reconnaissance Orbiter} Context Camera (CTX) images. In particular, P4T volunteers are asked to identify:  1) araneiforms (including  features with a central pit and radiating channels known as `spiders'); 2) erosional depressions, troughs, mesas, ridges, and quasi-circular pits characteristic of the South Polar Residual Cap (SPRC) which we collectively refer to as `Swiss cheese terrain', and 3) craters. In this work we present the distributions of our high confidence classic spider araneiforms and Swiss cheese terrain identifications in 90 CTX images covering 11$\%$ of the South polar regions at latitudes $\le$ -75$^{\circ}$ N. We find no locations within our high confidence spider sample that also have confident Swiss cheese terrain identifications. Previously spiders were reported as being confined to the South Polar Layered Deposits (SPLD). Our work has provided the first identification of spiders at locations outside of the SPLD, confirmed with high resolution HiRISE (High Resolution Imaging Science Experiment) imaging. We find araneiforms on the Amazonian and Hesperian polar units and the Early Noachian highland units, with 75$\%$ of the identified araneiform locations in our high confidence sample residing on the SPLD. With our current coverage, we cannot confirm whether these are the only geologic units conducive to araneiform formation on the Martian South Polar region. Our results are consistent with the current CO$_2$ jet formation scenario with the process exploiting weaknesses in the surface below the seasonal CO$_2$ ice sheet to carve araneiform channels into the regolith over many seasons. These new regions serve as additional probes of the conditions required for channel creation in the CO$_2$ jet process. 
\end{abstract}


\end{frontmatter}

\linenumbers

\section{Introduction}
The seasonal processes sculpting the Martian South Polar region are driven by the sublimation and deposition of carbon dioxide (CO$_2$) ice.  A significant portion of the Martian atmosphere, of which CO$_2$ is the predominant species, freezes or snows out on to the winter pole during the fall and winter. During the spring and summer, all or part of the deposited CO$_2$ returns to the atmosphere \citep{1966Sci...153..136L, 1992mars.book..934J, 2015Icar..251..164P,2016Icar..272..228G}. This is observed by the large seasonal variations in atmospheric pressure, with amplitudes up to 25$\%$, first measured by the \emph{Viking} landers \citep{1979JGR....84.2923H,1992Icar...99....1W}. Combined multi-season gravity field observations from \emph{Mars Global Survey} (\emph{MGS}),  \emph{Mars Odyssey}, and \emph{Mars Reconnaissance Orbiter} (\emph{MRO}) point to approximately 12 to 16$\%$ of the mass of the entire Martian atmosphere solidifying out on to the surface of the winter pole \citep{2016Icar..272..228G}. This cycle is directly linked to the current Martian climate. Thus, studying how the Martian South Pole region's inventory of CO$_2$ ice changes and evolves throughout the season and from Mars year to Mars year provides insight into processes driving Mars' climate and atmosphere. 

The Martian South Pole's CO$_2$ inventory can be divided into buried ice deposits and two broad surface ice caps, the temporary seasonal cap and the more permanent South Polar Residual Cap (SPRC). The buried CO$_2$ deposits vary from tens to thousands of meters in thickness and are topped with a 10-60 m  layer of water ice. The mass of these buried CO$_2$  ice reservoirs if sublimated is estimated to double the planet's current atmospheric pressure \citep{2016GeoRL..43.4172B}. These CO$_2$  ice deposits are located below the South Polar Layered Deposits (SPLD) \citep{ 2011Sci...332..838P,2016GeoRL..43.4172B}.  The SPLD is comprised mostly of  bands of  dust and water ice in addition to the buried subsurface  CO$_2$ ice reservoirs \citep{1973JGR....78.4231C,2000Icar..144..210C, 2011Sci...332..838P,2016GeoRL..43.4172B}. The SPLD is thought to have formed through repeated deposition linked to Mars' orbital/obliquity variations produced by the planet's  Milankovitch cycles \citep{2000Icar..144..243H}.  Recent modeling of Mars' climate variations is also able to produce wide-spread deposition of CO$_2$ on the South Polar region,  \citep{2011Sci...332..838P,2016GeoRL..43.4172B}, indicating the formation of the  buried CO$_2$  reservoirs is also linked to Mars' Milankovitch cycles. 

The SPRC is located between -84 to -89 degrees latitude and 220 E and 50 degrees E longitude. It is  primarily made of carbon dioxide ice \citep{1966Sci...153..136L, 2003Sci...299.1051B,2003Sci...299.1048T, 2009Icar..203..352T}. The extent of the SPRC has been observed to expand and to retreat during various Mars years, but the structure as a whole  survives past the spring and summer season \citep{1979JGR....84.8263K, 1992mars.book..934J, 1992mars.book.1180K, 2001JGR...10623635J, 2005Icar..174..513B, 2010Icar..208...82J}. The temporary seasonal ice sheet  on the other hand completely sublimates away by the end of the Southern summer \citep{1979JGR....84.8263K, 2003JGRE..108.5084P, 2006JGRE..111.3S07K,2007JGRE..112.3S13L, 2009JGRE..114.8005P,2013JGRE..118.2520P,2015Icar..251..164P}. The SPRC is thicker and higher albedo than most of the temporary seasonal cap, with thickness ranging between $\sim$0.5 and 10 meters \citep{2000Natur.404..161T,2003Sci...299.1051B, 2009Icar..203..352T,2016Icar..268..118T}. The surface of the SPRC is heavily eroded with smooth edged  quasi-circular flat bottomed pits, mesas, troughs, and other depressions \citep{1992mars.book..934J,2000Natur.404..161T,2001Sci...294.2146M,2001JGR...10623429M, 2005Icar..174..535T, 2009Icar..203..352T,2013Icar..225..923T,2016Icar..268..118T}. \cite{2009Icar..203..352T} and \cite{2016Icar..268..118T} provide a detailed description of morphologies of the SPRC based on orbital imagery.  The SPRC pits, depressions, and troughs have been observed to change in depth and areal coverage indicative of active mass loss \citep{2001Sci...294.2146M,2009Icar..203..352T,2013Icar..225..923T,2017Icar..286...69B}.  Mass balance modeling suggests these errosional features are due to the uneven sublimation and deposition of CO$_2$ ice on the SPRC \citep{2003Sci...299.1051B,2015Icar..251..211B,2016Icar..268..118T}. 

The seasonal cap is a temporary CO$_2$ ice sheet that extends from the pole to latitudes as far north as -50$^{\circ}$, and in cold protected patches to -22$^{\circ}$ \citep{2006Icar..180..321S,2010Icar..208...82J}. Mars Orbiter Laser Altimeter (MOLA) observations place the seasonal ice sheet thickness at $\sim$0.9-2.5 m \citep{2001Sci...294.2141S,2004JGRE..109.5004A}, with compaction decreasing the thickness over the winter \citep{2009Icar..202...90M}. A portion of the seasonal cap covers an area referred to as the cryptic terrain, areas where the albedo is low but has the temperatures of CO$_2$ ice ($\sim$150 K), indicating  the presence of semi-translucent slab ice \citep{2000JGR...105.9653K,2007JGRE..112.8005K}. Every Mars year, the spring  sublimation  of  the seasonal polar cap results in the formation of CO$_2$ jets and dark seasonal fans. In the generally accepted CO$_2$ jet model, sunlight penetrates through the slab of  CO$_2$ ice to the base regolith layer,  heating the ground  \citep{2000mpse.conf...93K, 2003JGRE..108.5084P,2006Natur.442..793K, 2007JGRE..112.8005K,  2008JGRE..113.6005P, 2010Icar..205..296T, 2010Icar..205..311P, 2011Icar..213..131P,  2011GeoRL..38.8203T}. This results in sublimation at the base of the ice sheet, forming a trapped layer of gas between the ice and the regolith. The trapped CO$_2$ gas is thought to exploit any weaknesses in the ice above, breaking through to the top of the ice sheet as a CO$_2$ jet. Dust and dirt from below the ice sheet are carried by the jet and expelled into the atmosphere.  It is thought that the local surface winds carry the particles as they settle onto the top of the ice sheet producing the dark fan-like streaks and blotches observed from orbit during the spring and summer.  When the seasonal cap disappears, the majority of the seasonal fans and blotches fade and blend into the background regolith, further supporting the idea that the fan material is the same as the regolith below the ice sheet \citep{2006Natur.442..793K,2010Icar..205..296T,2011JGRE..116.8007P}. Recent laboratory experiments by \cite{Kaufmann2017118} were able to trigger dust eruptions from a layer of dust inside a CO$_2$ ice slab under Martian conditions, lending further credence to the proposed CO$_2$ jet and fan production model. 

\begin{figure}[!]
\begin{center}
\includegraphics[width=1.0\columnwidth]{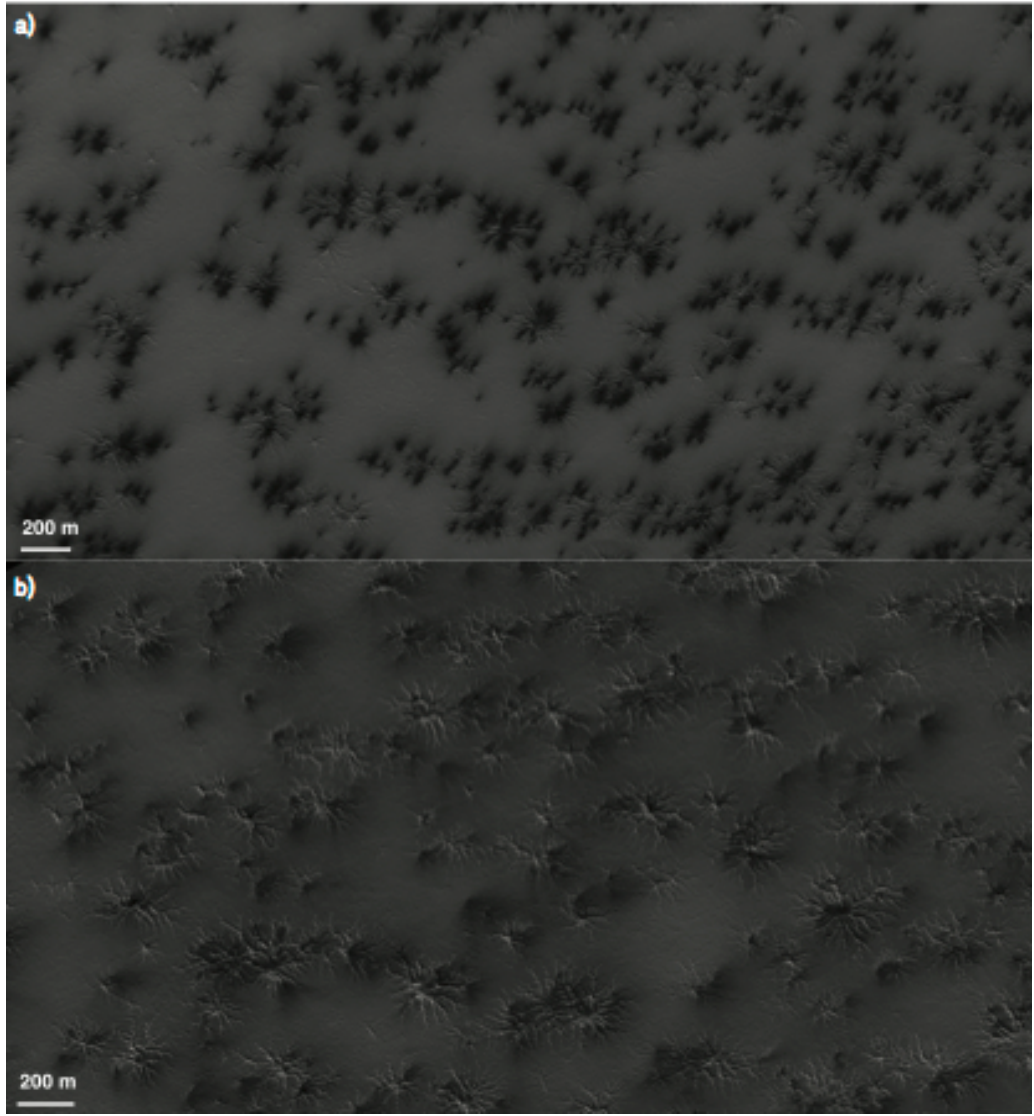}
 \caption{ \label{fig:spider_hirise_known_locations1} Examples of araneiform features imaged by HiRISE. Figures a and b show radially-organized 1-m deep channels emanating from a slightly deeper center. More colloquially these dendritic channel structures have been referred to as 'spiders.' At these times the terrain in the images was covered by a semi-translucent seasonal layer of CO$_2$ ice.  The araneiforms are channels below the ice carved into the underlying regolith. Seasonal fans of fine particles from ruptures in the ice are deposited on top of the seasonal ice layer are present and visible in the images as the dark black  fan-like and more blotch-like features. The location of seasonal fans are not always directly associated with  araneiform channels, but araneiform channels have been observed to have seasonal fans emanating at the surface of the ice sheet above their positions at HiRISE resolution \citep{2010Icar..205..283H}.  From top to bottom:  ESP$\_$020558$\_$0930  (L$_{\rm s}$=198.9$^{\circ}$) and  ESP$\_$011420$\_$0930 (L$_{\rm s}$=184.3$^{\circ}$). }
 \end{center}
 \end{figure}

\begin{figure}
\begin{center}
\includegraphics[width=0.9\columnwidth]{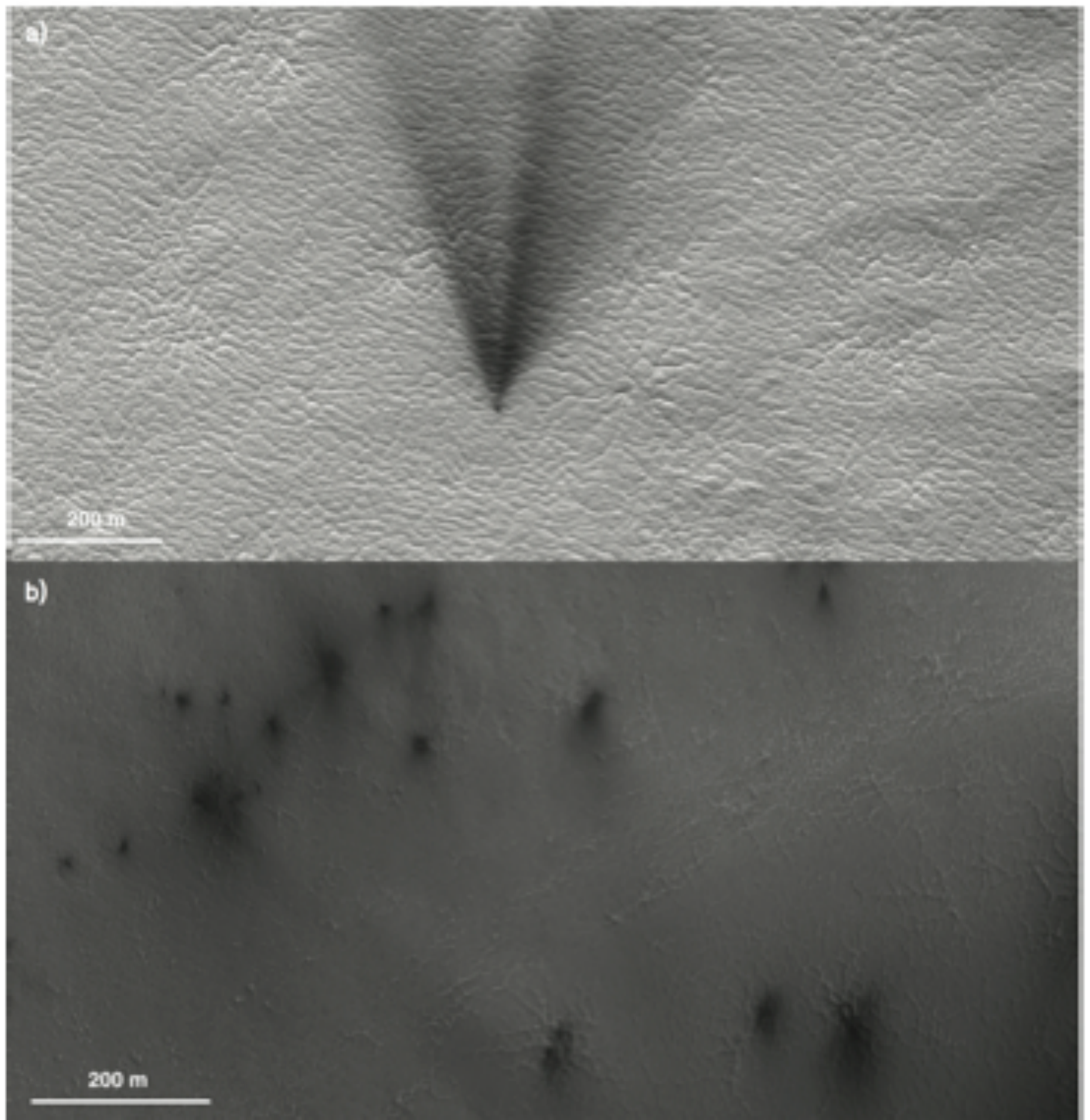}
 \caption{ \label{fig:spider_hirise_known_locations2}  Connected araneiforms, a diffferent end-member of the spectrum of araneiforms  with no centralized channel organization, but bearing the characteristic hallmark of sinuous channels with seasonal fans as imaged by HiRISE. At these times the terrain in the images was covered by a semi-translucent seasonal layer of CO$_2$ ice.  The araneiforms are channels below the ice carved into the underlying regolith. Seasonal fans of fine particles from ruptures in the ice are deposited on top of the seasonal ice layer are present and visible in the images as the dark black  fan-like and more blotch-like features. The location of seasonal fans are not always directly associated with  araneiform channels, but araneiform channels have been observed to have seasonal fans emanating at the surface of the ice sheet above their positions at HiRISE resolution \citep{2010Icar..205..283H}. From top to bottom: PSP$\_$002850$\_$0935  (L$_{\rm s}$=195.4$^{\circ}$) and  ESP$\_$038022$\_$0985  (L$_{\rm s}$=191.2$^{\circ}$).    }
 \end{center}
 \end{figure}

Small pits in the surface with radiating channels a few meters deep,  colloquially known as `spiders,' have also been identified in spacecraft imagery in many of the same areas as where the seasonal fans are present \citep{2000mpse.conf...93K, 2003JGRE..108.5084P,2006Natur.442..793K, 2010Icar..205..283H}. Spiders range in diameter from tens of meters to 1 km\citep{2010Icar..205..283H}. Many seasonal fans appear to originate from the  spider `legs', but not all observed fans do \citep{2003JGRE..108.5084P,2010Icar..205..283H}. With the arrival of \emph{MRO} and the HiRISE \citep[High Resolution Imaging Science Experiment;][]{2007JGRE..112.5S02M} camera, with a pixel scale of $\sim$30 cm/pixel at 300 km altitude, new morphologies of spider-like channels have been found \citep{2010Icar..205..283H}. This includes `lace terrain', where the dendritic-like channels of spiders are connected with no visible central pit.  Spiders and these other spider-like dendritic channels are now collectively referred to as araneiforms \citep{2010Icar..205..283H}, and they are  thought to form via the CO$_2$ jet process \citep{2000mpse.conf...93K, 2003JGRE..108.5084P,2006Natur.442..793K, 2007JGRE..112.8005K,  2008JGRE..113.6005P, 2010Icar..205..296T, 2010Icar..205..311P, 2011Icar..213..131P,  2011GeoRL..38.8203T, 2012GeoRL..3913204D}. A sample of araneiform features from high resolution imaging is shown in Figures \ref{fig:spider_hirise_known_locations1} and  \ref{fig:spider_hirise_known_locations2}.

 Through a survey of over 5,000  Mars Orbiter Camera (MOC) Narrow Angle (NA) \citep{1992JGR....97.7699M,2010IJMSE...5....1M} images, \cite{2003JGRE..108.5084P} linked the presence of CO$_2$ slab ice with araneiform formation. They found that araneiforms are located in regions  where the seasonal CO$_2$ ice cap becomes cryptic for at least some part of the Southern spring and summer, lending further support to the idea that araneiforms are gradually carved into the ground by the trapped  CO$_2$ gas during the formation of CO$_2$  jets. \cite{2003JGRE..108.5084P} also found that spiders are confined to the top of the SPLD. The erosional mechanism forming araneiforms is a slow process; over many spring/summer seasons  the trapped gas underneath the sublimating ice sheet carves these features into the top of the SPLD. Estimates from modeling by \cite{2008JGRE..113.6005P} and recent HiRISE observations of araneiform channel formation by \cite{Portyankina201793} place araneiform ages at $10^3-10^4$ years. 

\cite{2003JGRE..108.5084P} argue that with their areal coverage they would have seen spiders in the MOC NA images they visually inspected outside of the SPLD. Why araneiforms appear to only be constrained to the top of the SPLD is an open question. \cite{2003JGRE..108.5084P} postulate that araneiforms may be restricted to the SPLD because the SPLD is composed of more loosely consolidated material than other geologic units on the South Polar region  \citep{2000JGR...105.6961V}; perhaps making it easier to erode by the CO$_2$ gas than in other areas. In the past decade with the arrival of the Context Camera \citep[CTX;][]{2007JGRE..112.5S04M} aboard \emph{MRO}, many areas of the  South Polar region have been imaged multiple times each Mars Year with 6-8 m per pixel scale, better resolution than some of the MOC NA observations searched by \cite{2003JGRE..108.5084P} and also in areas that were not covered in the original \cite{2003JGRE..108.5084P} search. 

More widely  distributed  coverage will allow us to expand upon the previous maps of araneiform locations. With newly discovered  araneiform locales, we can further explore what conditions (weather, types of terrain, erodibility  of  the  ground,  latitude,  and other surface and climate properties) are key for araneiform development through comparison to previous regions monitored by HiRISE for 5 Mars Years such as the informally named `Manhattan' and  `Inca City' \citep[e.g.][]{2010Icar..205..283H,2012JGRE..117.2006P,Portyankina201793}.  We can also compare the distribution of araneiforms to other features on the South Polar region produced by the sublimation of  CO$_2$ ice,  such as the Swiss cheese terrain.  In addition, finding new areas with past CO$_2$  jets and seasonal fans activity is an important resource for future mission and target planning. 

We created Planet Four: Terrains\footnote{\url{http://terrains.planetfour.org} or \url{https://www.zooniverse.org/projects/mschwamb/planet-four-terrains}}  (P4T), an online citizen science project which enlists the general public to map the locations of  1) araneiforms (including  features with a central pit and radiating channels known as `spiders'); 2) erosional depressions, troughs, mesas, ridges, and quasi-circular pits characteristic of the South Polar Residual Cap (SPRC) which we collectively refer to as `Swiss cheese terrain', and 3) craters in  publicly available CTX observations. The human brain is ideally suited for this task and with very little training is easily capable of identifying these features in orbital images of Mars. Previous studies have visually identified features like araneiforms, Swiss cheese terrain, or recurring slope lineae  using a single person or groups of researchers reviewing observations taken from orbit \cite[e.g.][]{2003JGRE..108.5084P,2005Icar..174..535T, 2009Icar..203..352T,2010Icar..205..283H, 2014Icar..231..365O,2011Sci...333..740M}.  With the Internet, tens of thousands of people across the globe can be enlisted in such tasks to create a larger sample and  in particular review images typically in more detail than in the time a  single researcher or group of researchers can. This citizen science or crowd-sourcing approach, where independent assessments from multiple non-expert classifiers are combined, has been applied to nearly all areas in astronomy and planetary science \citep{2015ARA&A..53..247M} (see references therein) including galaxy morphology \citep{2008MNRAS.389.1179L,2013MNRAS.435.2835W}, exoplanet searches \citep{2012MNRAS.419.2900F, 2012ApJ...754..129S}, circumstellar disk  identification \citep{2016arXiv160705713K}, and crater counting \citep{2014Icar..234..109R,2016Icar..271...30B}. 

In this Paper we present the first results from P4T, examining the distribution of araneiforms with spider morphology and  comparing to the Swiss cheese terrain on the Martian South Pole.  In Section 2 we provide an overview of the image dataset used in this work. We describe the P4T project and web classification interface in Section 3. In Section 4 we detail the process of combining the multiple volunteer assessments to identify surface features in the Mars image data reviewed. In Section 5 we present our map of spider locations within ~15$^{\circ}$ of the Martian South Pole,  and  in Section 6 we compare  these locations to the distribution of secure Swiss cheese terrain identifications. We report the discovery of araneiforms outside of the SPLD and in Section 7 present higher resolution confirmation imaging from HiRISE. In Section 8 we discuss the implications of this result for the CO$_2$  jet model.  All place names referred to in this Paper are informal and not approved by the International Astronomical Union. All reported latitudes are areographic, and all longitudes are reported in reference to East longitude.  Full machine-readable versions of the catalogs and tables presented in this Paper are also available from \url{https://www.zooniverse.org/projects/mschwamb/planet-four-terrains/about/results}\footnote{Note to the editor: When this manuscript is accepted we will make these online links accessible. All material has been submitted in the supplementary information as well}. 

\section{Dataset}
\label{sec:dataset}

For our analysis, we used publicly available observations from CTX \citep{2007JGRE..112.5S04M} aboard \emph{MRO} obtained from NASA's Planetary Data System (PDS)\footnote{\url{http://pds-imaging.jpl.nasa.gov/}}. CTX has widespread coverage of the Martian south polar region at a variety of solar longitudes (L$_{ \rm S}$). The camera provides the best balance between resolution  and areal coverage with a single observation typically spanning a $\sim$30$\times$60 km swath at 6 m/pixel spatial scale. \cite{2003JGRE..108.5084P} found araneiforms were constrained to the SPLD; we thus restricted our study to CTX observations with latitudes southward of -75$^{\circ}$ latitude in order to encompass the majority of the SPLD defined by \cite{Tanaka:2014wd}\footnote{\url{http://pubs.usgs.gov/sim/3292/}}. We selected 90 CTX images  for review on P4T for this work. The area covered by our search images is shown in Figure \ref{fig:coverage}. We have surveyed CTX observations covering 303,192 km$^2$ within -70$^\circ$ latitude and 11$\%$ of the South Polar region within -75$^\circ$ N.  The areal coverage of the surveyed CTX images as a function of latitude is shown in Figure \ref{fig:fraccoverage}. The observing circumstances and planetographic coordinates of the selected CTX observations are summarized in Table \ref{tab:CTXobs}.  

\subsection{CTX Image Selection}

We provide a brief overview of the process used to select the 90 CTX images used in this analysis. Since araneiform formation takes thousands of Mars years \citep{2008JGRE..113.6005P,Portyankina201793}, we do not restrict ourselves to a single Mars year. We chose from publicly available CTX observations taken between October 2007 and October 2013, Mars years (MY) 28-32 according to the convention defined by \cite{2000JGR...105.9553C} and \cite{2015Icar..251..332P}. The CO$_2$ jet process on the seasonal ice cap produces fans which appear as dark streaks and blotches in CTX images throughout the Southern spring and summer. We attempted to select CTX observations of locations where the  CO$_2$ seasonal cap had already sublimated in order to reduce the number of obscuring fans. The extent of the seasonal CO$_2$ cap in latitude and longitude varies as as a function of L$_{ \rm S}$.  We used previous thermal measurements of the South polar region \citep{2015Icar..251..164P} and a brief visual inspection of the CTX images to determine latitude/L$_{ \rm S}$ ranges that were ice free. 

\begin{figure}
\begin{center}
\includegraphics[width=1.0\columnwidth]{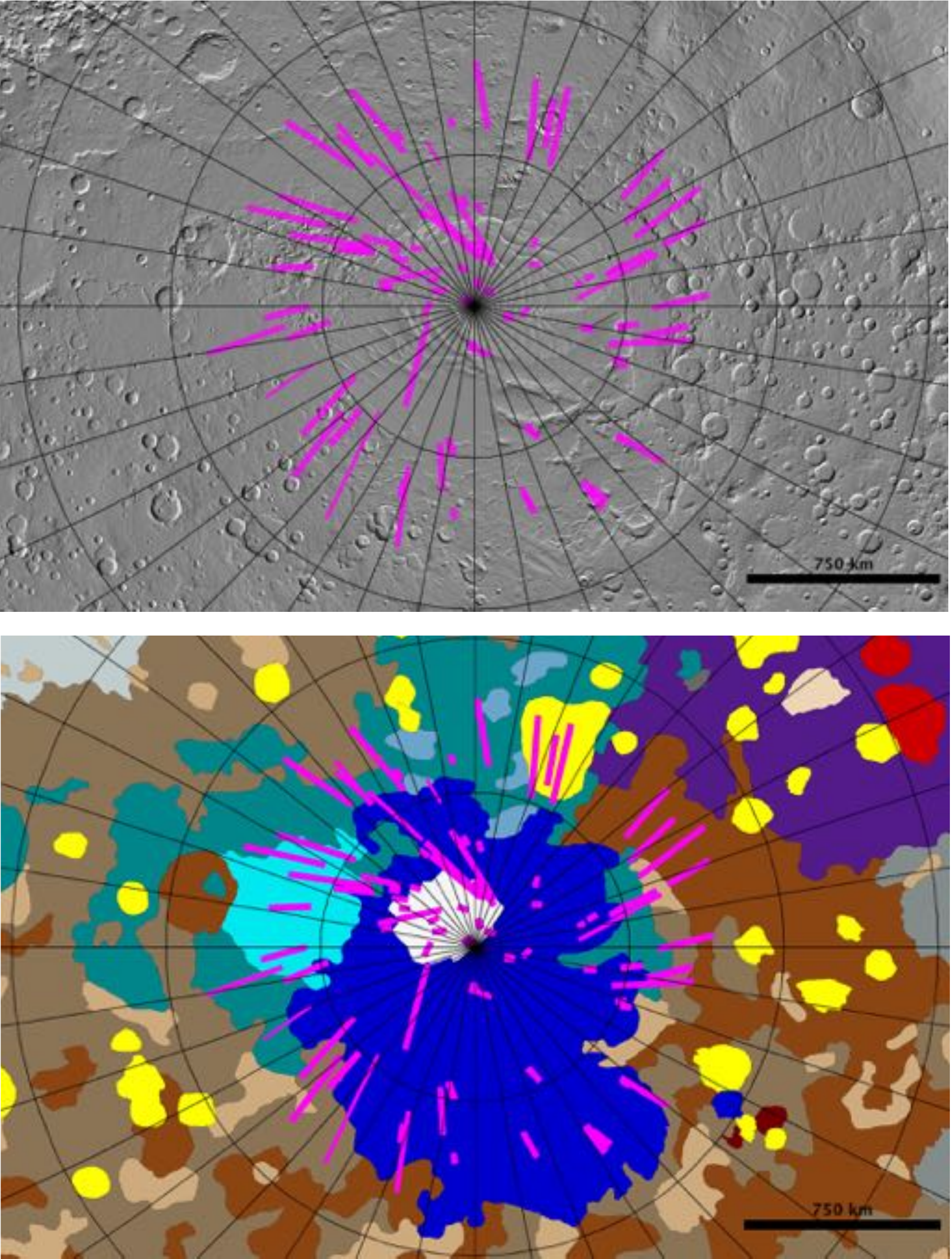}
 \caption{  \label{fig:coverage} Top: MOLA shaded relief map \citep{1992JGR....97.7781Z,2001JGR...10623689S}   with the CTX footprints in magenta. Bottom: The footprints of the surveyed CTX images in magenta overlaid on top of the geologic map from \cite{Tanaka:2014wd}. For both plots, latitude and longitude lines are plotted every 10  degrees. The zero meridian is pointing straight up. A legend for the geologic map is provided in \ref{ref:appendix}. } 
 \end{center}
 \end{figure}

\begin{figure}
\begin{center}
\includegraphics[width=1.0\columnwidth]{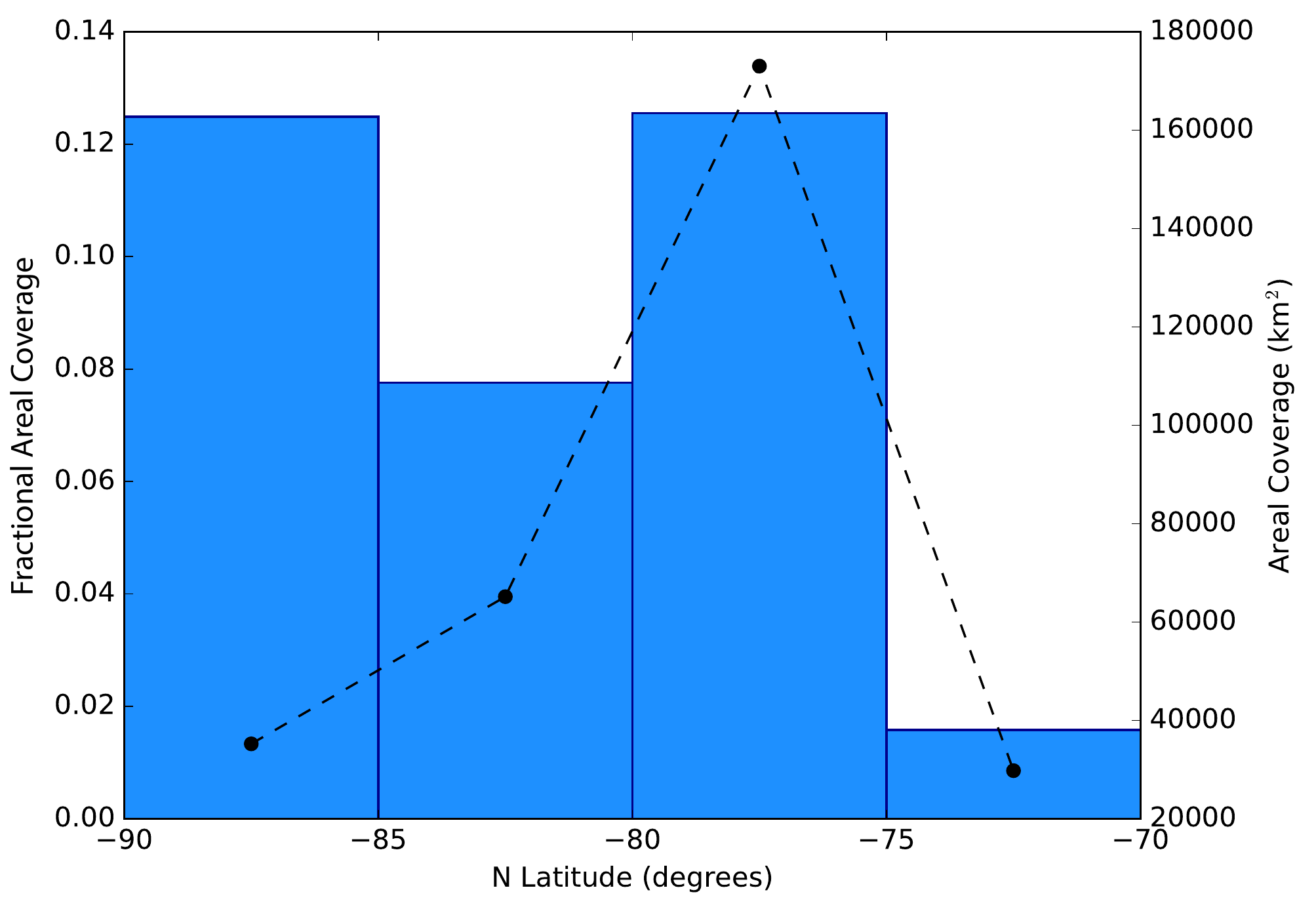}
 \caption{  \label{fig:fraccoverage} The fractional areal coverage of the Martian South Polar region (blue in the online version) histogram of the P4T CTX images searched  in 5-degree latitude bins. The dashed line shows the areal coverage in km$^2$. } 
 \end{center}
 \end{figure}

\begin{figure}
\begin{center}
\includegraphics[width=1.0\columnwidth]{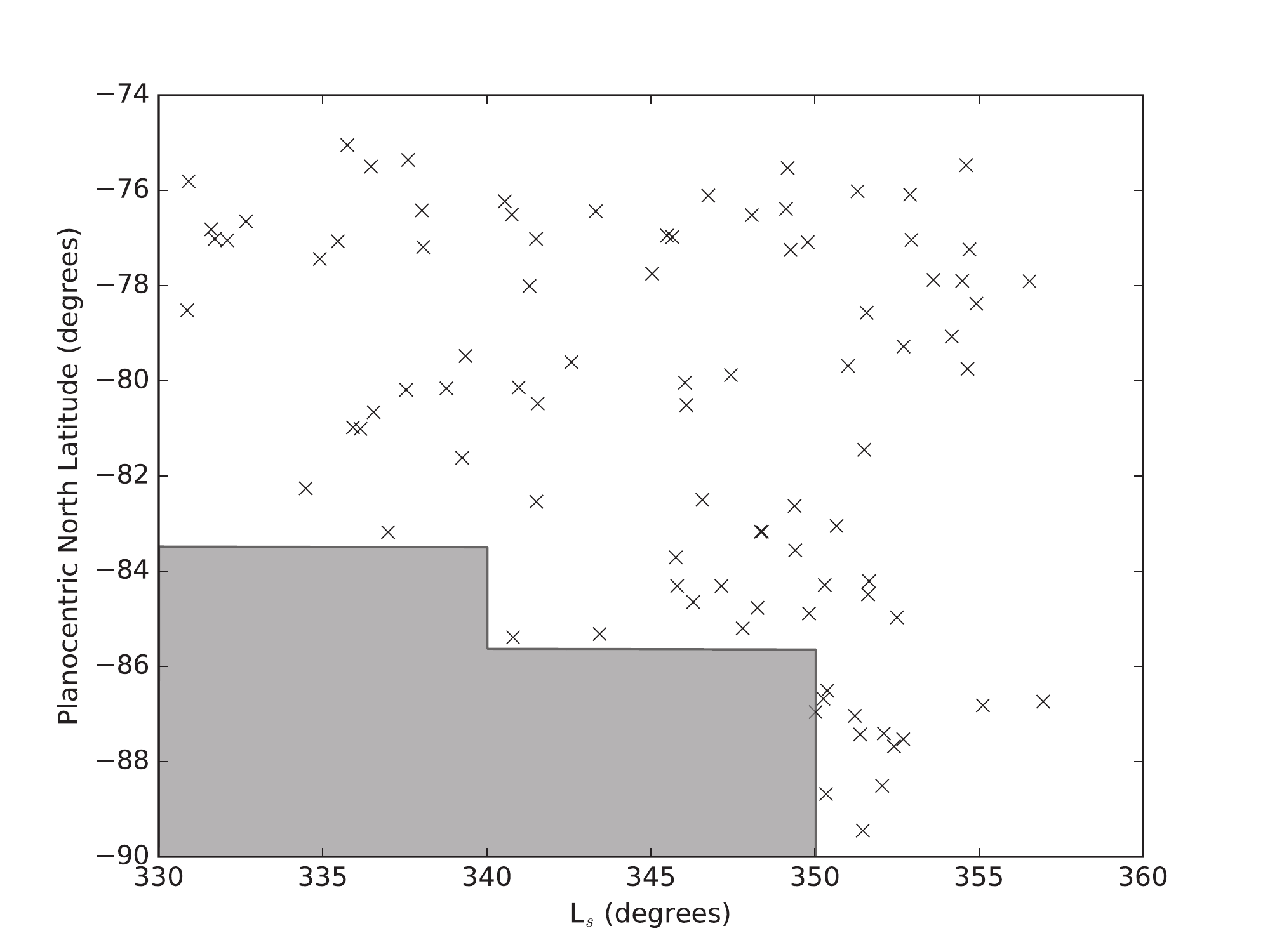}
 \caption{ \label{fig:ctx_ls} The distribution of center latitudes for the 90 CTX images selected as a function of  L$_S$. The greyed region represents the excluded latitudes chosen to avoid as much as possible the presence of the dark seasonal fans on the seasonal cap. }
 \end{center}
 \end{figure}

With the remaining CTX images that pass the selection cuts, we attempt to achieve as widespread coverage as possible distributed across the South Polar region. We divide the South polar region southward of -75 $^{\circ}$ latitude  into  30$^{\circ}$ longitude and 5$^{\circ}$  latitude bins. The 90 CTX images used in this study were randomly selected from each bin with slightly more images picked with field centers between -85$^{\circ}$ and -75$^{\circ}$ latitude. The final area covered by our search images is shown in Figure \ref{fig:coverage}. Figure \ref{fig:ctx_ls} shows the latitude distribution of the selected CTX images for this study as a function of L$_{ \rm S}$.

A visual inspection of the final 90 CTX images selected finds that majority of the images are free from clouds or obscuring seasonal fans. We do not have criteria for assessing the fraction of cloud cover, atmospheric opacity, or presence of seasonal fans in  each of the selected CTX images used in our search. Thus the lack of a positive identification of araneiforms, Swiss Cheese Terrain, or craters  by P4T does not necessarily mean the feature is not present in a given CTX image. Our analysis instead identifies locations where araneiforms, Swiss Cheese Terrain, and craters are confidently identified by P4T but does not provide a complete sample. 

\subsection{CTX Image Processing and P4T Subject Creation}

The selected full frame CTX images searched by P4T  are subdivided into smaller subimages that are subsequently presented to volunteers on the P4T website. The raw CTX image products or Experiment Data Records (EDRs) were processed with Python using the United States Geological Survey's (USGS) Integrated Software for Imagers and Spectrometers (ISIS) \citep[ISIS-3][]{2004LPI....35.2039A,2007LPI....38.1779B}\footnote{\url{http://isis.astrogeology.usgs.gov/}} and the ISIS-3 python wrapper Pysis\footnote{\url{https://github.com/wtolson/Pysis}}. After radiometric calibration and noise removal, the CTX frames were divided into smaller non-overlapping 800$\times$600 pixel ($\sim$4.8$\times3.6$ km) PNG (Portable Network Graphics) subimages to be presented on the P4T website. We refer to these subimages as `subjects.' The PNG exporter  from the ISIS toolset (isis2std) was configured to output at 8-bit output resolution, meaning that the dynamic ratios of the more dynamic CTX data were visually reduced compared to scientific display systems. By default, isis2std cuts off the lowest and highest 0.5 percent of image values to exclude cold and hot pixels. After visual inspection by the science team of varying dynamic ratios, we settled on a full zero to hundred percent stretch for each subject generated to get the increased dynamic ratio which helps in identifying patterns in very dark or bright areas. The fact that we did not identify any problems with leaving the image stretch at 100 percent of the original values in the CTX subimage ISIS cube is a testament to the quality of the CTX camera and its calibration. We also experimented with over-stretching the generated subjects but did not find any improvement in image quality.

\begin{figure}
\begin{center}
\includegraphics[width=1.0\columnwidth]{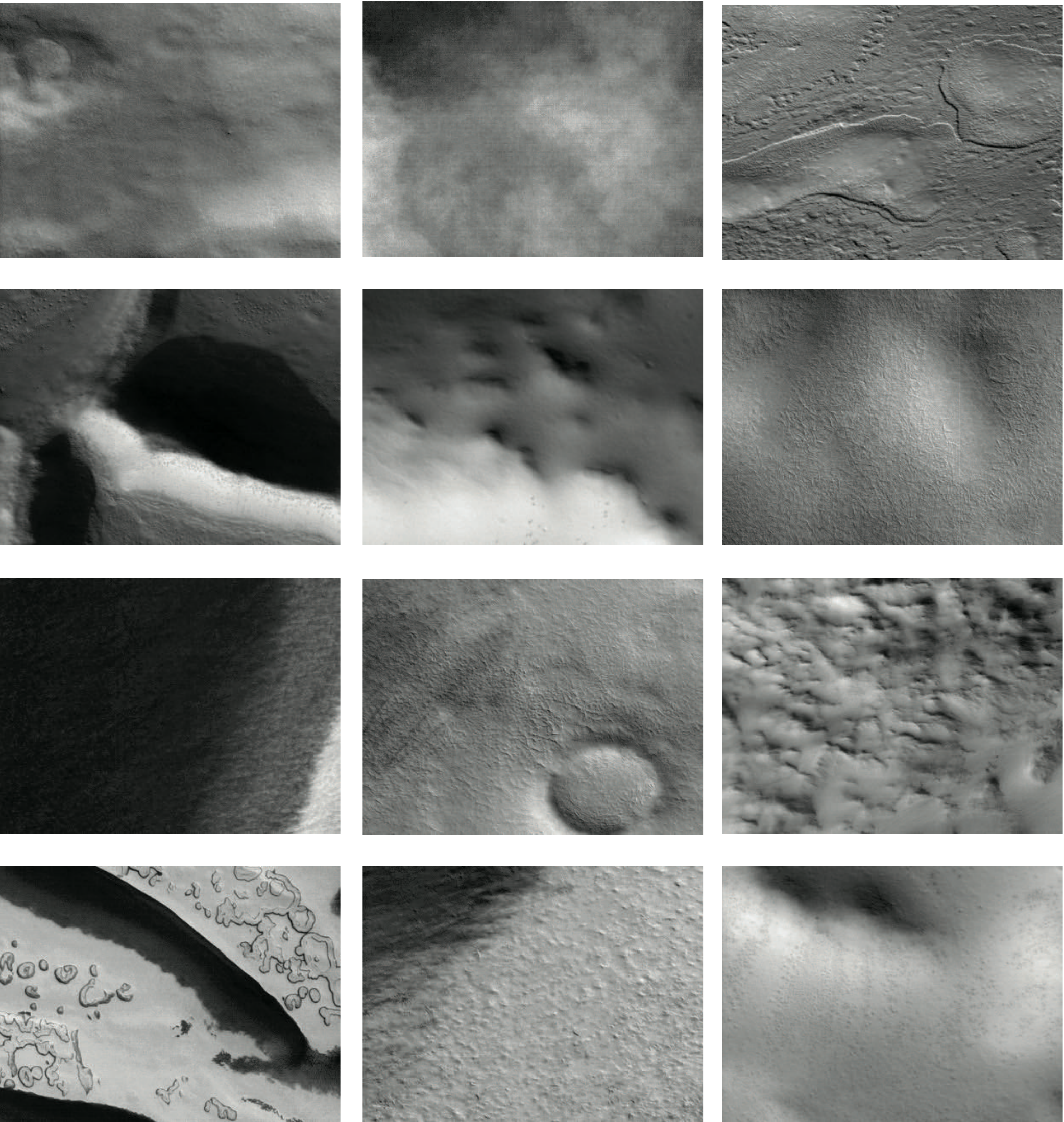}
 \caption{ \label{fig:subject_examples} Selection of subjects derived from full frame CTX images shown on the P4T website. Each subject is an 800$\times$600 pixel ($\sim$4.8$\times3.6$ km) subimage. Starting at top left, Subject IDs (starting with top row, left to right): 489465, 1041974, 489329, 488393, 484326, 1319951, 489919, 1058815, 1040473, 1491825, 1514939,  and 1515625. CTX images (starting with top row, left to right): P13$\_$006119$\_$1035, G14$\_$023676$\_$1023,G14$\_$023538$\_$1006, P13$\_$006204$\_$0986, 
 G14$\_$023634$\_$1036,P13$\_$006240$\_$1030, P12$\_$005747$\_$1035, P13$\_$006239$\_$1040,P13$\_$006123$\_$0953, G14$\_$023835$\_$0906, P13$\_$006234$\_$1008,P13$\_$006282$\_$1046}
 \end{center}
 \end{figure}

A characteristic sample of P4T subjects is presented in Figure \ref{fig:subject_examples}. In total 20,122 subjects were generated, and Table \ref{tab:CTXobs} provides a list of the 90 CTX observations and the number of subjects associated with each full frame CTX image. A CTX observation contained a mean of  224 P4T subjects with a minimum of 24 and maximum of 522 subjects. Due to the variable length and width of CTX observations, there are typically small regions on the right and bottom edges of the CTX full frame image that did not make it into a subject image, and thus not searched by P4T.  Supplemental Table 1 summarizes the P4T subjects used in this work, including the center latitude, center longitude, and location within the full frame CTX observation.

\begin{small}
\begin{longtable}{llllll}
\caption[short]{CTX Observations Examined in This Study \label{tab:CTXobs} } \\ 

\hline
\hline \\[-2ex]
CTX image  & Latitude & Longitude & $L_s$ & Observation & $\#$ of \\
 & (degrees) & (degrees) & (degrees) & Time & P4T Subjects \\
  \hline 
\endfirsthead

\multicolumn{3}{c}{{\tablename} \thetable{} -- Continued} \\[0.5ex]
  \hline \hline \\[-2ex]
CTX image  & Latitude & Longitude & $L_s$ & Observation & $\#$ of \\
 & (degrees) & (degrees) & (degrees) & Time & P4T Subjects \\
   \hline 
 \endhead

 \hline
  \multicolumn{3}{l}{{Continued on Next Page\ldots}} \\
\endfoot

  \\[-1.8ex] \hline \hline
  \caption[]{The center coordinates for all CTX images searched used in the analysis presented in this paper. The table includes the latitude and longitude, UTC time and date of observation, and number of P4T subjects generated from the observation.  We include the full CTX filename here; the first 15 characters are the unique CTX observation identifier.  }  \\
\endlastfoot

D13$\_$032173$\_$1031$\_$XN$\_$76S227W & -77.02 & 132.51 & 331.71 & 2013-06-07T10:18:40.139 & 66\\
D13$\_$032182$\_$1030$\_$XN$\_$77S112W & -77.05 & 247.78 & 332.09 & 2013-06-08T03:08:08.698 & 183\\
D13$\_$032278$\_$0991$\_$XN$\_$80S204W & -81.01 & 155.56 & 336.15 & 2013-06-15T14:40:39.298 & 108\\
D13$\_$032298$\_$0969$\_$XN$\_$83S028W & -83.18 & 331.6 & 336.99 & 2013-06-17T04:04:24.753 & 66\\
D13$\_$032311$\_$0999$\_$XN$\_$80S031W & -80.19 & 329.04 & 337.54 & 2013-06-18T04:23:39.968 & 304\\
D13$\_$032352$\_$0985$\_$XN$\_$81S063W & -81.62 & 296.36 & 339.25 & 2013-06-21T09:04:37.135 & 162\\
D14$\_$032510$\_$0963$\_$XN$\_$83S054W & -83.71 & 305.21 & 345.76 & 2013-07-03T16:34:14.369 & 66\\
D14$\_$032511$\_$0959$\_$XI$\_$84S078W & -84.31 & 282.04 & 345.8 & 2013-07-03T18:26:12.728 & 48\\
D14$\_$032517$\_$1000$\_$XN$\_$80S249W & -80.04 & 110.55 & 346.04 & 2013-07-04T05:40:49.097 & 66\\
D14$\_$032518$\_$0995$\_$XN$\_$80S281W & -80.51 & 78.42 & 346.08 & 2013-07-04T07:32:06.155 & 392\\
D14$\_$032523$\_$0954$\_$XN$\_$84S045W & -84.65 & 314.82 & 346.29 & 2013-07-04T16:52:42.650 & 66\\
D14$\_$032530$\_$0975$\_$XN$\_$82S259W & -82.5 & 100.75 & 346.57 & 2013-07-05T05:59:04.972 & 78\\
D14$\_$032574$\_$0969$\_$XN$\_$83S005W & -83.17 & 354.99 & 348.35 & 2013-07-08T16:16:02.560 & 66\\
D14$\_$032575$\_$0969$\_$XN$\_$83S028W & -83.17 & 331.6 & 348.39 & 2013-07-08T18:08:13.419 & 66\\
D14$\_$032593$\_$1037$\_$XN$\_$76S174W & -76.39 & 185.51 & 349.12 & 2013-07-10T03:50:10.687 & 66\\
D14$\_$032600$\_$0965$\_$XN$\_$83S357W & -83.56 & 2.88 & 349.4 & 2013-07-10T16:53:27.794 & 66\\
D14$\_$032640$\_$1003$\_$XN$\_$79S014W & -79.69 & 345.3 & 351.01 & 2013-07-13T19:42:56.864 & 87\\
D14$\_$032656$\_$0959$\_$XI$\_$84S078W & -84.21 & 281.5 & 351.65 & 2013-07-15T01:36:53.707 & 66\\
D14$\_$032666$\_$0916$\_$XN$\_$88S350W & -88.51 & 9.05 & 352.05 & 2013-07-15T20:17:50.497 & 66\\
D14$\_$032675$\_$0924$\_$XN$\_$87S253W & -87.68 & 106.58 & 352.41 & 2013-07-16T13:08:05.533 & 66\\
D14$\_$032682$\_$0925$\_$XN$\_$87S066W & -87.53 & 293.1 & 352.69 & 2013-07-17T02:13:30.246 & 90\\
D14$\_$032733$\_$1028$\_$XN$\_$77S036W & -77.24 & 323.18 & 354.71 & 2013-07-21T01:38:54.660 & 204\\
D14$\_$032790$\_$0933$\_$XN$\_$86S110W & -86.74 & 248.82 & 356.96 & 2013-07-25T12:11:45.417 & 24\\
G13$\_$023338$\_$1043$\_$XI$\_$75S229W & -75.81 & 131.06 & 330.91 & 2011-07-20T00:01:53.559 & 336\\
G13$\_$023354$\_$1032$\_$XN$\_$76S301W & -76.82 & 58.54 & 331.6 & 2011-07-21T05:56:38.165 & 396\\
G13$\_$023432$\_$1026$\_$XI$\_$77S262W & -77.44 & 97.75 & 334.91 & 2011-07-27T07:49:28.555 & 294\\
G13$\_$023452$\_$1049$\_$XN$\_$75S098W & -75.05 & 261.45 & 335.75 & 2011-07-28T21:14:17.312 & 435\\
G13$\_$023456$\_$0990$\_$XN$\_$81S204W & -80.98 & 155.23 & 335.92 & 2011-07-29T04:42:00.456 & 108\\
G13$\_$023469$\_$1046$\_$XI$\_$75S211W & -75.5 & 148.2 & 336.47 & 2011-07-30T05:02:26.465 & 210\\
G14$\_$023496$\_$1047$\_$XI$\_$75S213W & -75.36 & 147.08 & 337.6 & 2011-08-01T07:32:15.443 & 138\\
G14$\_$023506$\_$1036$\_$XN$\_$76S132W & -76.42 & 227.99 & 338.02 & 2011-08-02T02:13:28.855 & 522\\
G14$\_$023507$\_$1029$\_$XN$\_$77S158W & -77.19 & 201.15 & 338.06 & 2011-08-02T04:06:06.023 & 120\\
G14$\_$023524$\_$0999$\_$XN$\_$80S262W & -80.16 & 97.56 & 338.77 & 2011-08-03T11:52:54.656 & 66\\
G14$\_$023538$\_$1006$\_$XN$\_$79S285W & -79.48 & 74.31 & 339.35 & 2011-08-04T14:03:39.719 & 336\\
G14$\_$023567$\_$1039$\_$XN$\_$76S358W & -76.23 & 2.04 & 340.55 & 2011-08-06T20:18:46.307 & 438\\
G14$\_$023577$\_$0999$\_$XN$\_$80S262W & -80.14 & 97.62 & 340.97 & 2011-08-07T15:00:08.617 & 66\\
G14$\_$023590$\_$0975$\_$XN$\_$82S259W & -82.54 & 100.99 & 341.51 & 2011-08-08T15:18:11.704 & 66\\
G14$\_$023591$\_$0996$\_$XN$\_$80S284W & -80.48 & 75.43 & 341.55 & 2011-08-08T17:10:49.067 & 132\\
G14$\_$023616$\_$1004$\_$XN$\_$79S248W & -79.61 & 111.85 & 342.58 & 2011-08-10T15:56:35.480 & 66\\
G14$\_$023634$\_$1036$\_$XN$\_$76S027W & -76.44 & 333.0 & 343.32 & 2011-08-12T01:36:48.168 & 522\\
G14$\_$023676$\_$1023$\_$XN$\_$77S091W & -77.75 & 268.09 & 345.04 & 2011-08-15T08:09:24.912 & 294\\
G14$\_$023687$\_$1031$\_$XN$\_$76S036W & -76.95 & 323.86 & 345.49 & 2011-08-16T04:44:01.680 & 232\\
G14$\_$023691$\_$1031$\_$XN$\_$76S143W & -76.98 & 216.96 & 345.65 & 2011-08-16T12:12:04.519 & 381\\
G14$\_$023718$\_$1039$\_$XN$\_$76S160W & -76.11 & 199.16 & 346.75 & 2011-08-18T14:42:24.470 & 348\\
G14$\_$023728$\_$0958$\_$XI$\_$84S057W & -84.31 & 302.44 & 347.15 & 2011-08-19T09:22:32.460 & 102\\
G14$\_$023735$\_$1002$\_$XN$\_$79S262W & -79.88 & 97.18 & 347.44 & 2011-08-19T22:29:22.283 & 138\\
G14$\_$023794$\_$0952$\_$XN$\_$84S061W & -84.89 & 298.47 & 349.82 & 2011-08-24T12:48:10.629 & 66\\
G14$\_$023807$\_$0886$\_$XN$\_$88S313W & -88.68 & 47.11 & 350.34 & 2011-08-25T13:05:07.603 & 66\\
G14$\_$023815$\_$0970$\_$XN$\_$83S282W & -83.05 & 78.09 & 350.66 & 2011-08-26T04:05:02.101 & 48\\
G14$\_$023833$\_$0926$\_$XN$\_$87S017W & -87.43 & 342.35 & 351.38 & 2011-08-27T13:43:17.984 & 66\\
G14$\_$023835$\_$0906$\_$XN$\_$89S050W & -89.45 & 310.24 & 351.46 & 2011-08-27T17:27:10.913 & 66\\
G14$\_$023851$\_$0926$\_$XN$\_$87S180W & -87.41 & 179.52 & 352.1 & 2011-08-28T23:23:15.422 & 66\\
G15$\_$023911$\_$1021$\_$XN$\_$77S023W & -77.9 & 336.45 & 354.49 & 2011-09-02T15:38:24.799 & 66\\
G15$\_$023927$\_$0932$\_$XI$\_$86S057W & -86.82 & 302.59 & 355.12 & 2011-09-03T21:30:31.425 & 66\\
G15$\_$023963$\_$1021$\_$XN$\_$77S006W & -77.91 & 353.12 & 356.54 & 2011-09-06T16:52:57.448 & 60\\
P12$\_$005705$\_$1016$\_$XI$\_$78S133W & -78.52 & 226.28 & 330.87 & 2007-10-14T23:21:44.255 & 294\\
P12$\_$005747$\_$1035$\_$XI$\_$76S195W & -76.65 & 164.52 & 332.66 & 2007-10-18T05:55:17.203 & 138\\
P12$\_$005790$\_$0978$\_$XI$\_$82S284W & -82.26 & 75.67 & 334.48 & 2007-10-21T14:18:20.245 & 66\\
P12$\_$005813$\_$1030$\_$XI$\_$77S195W & -77.07 & 164.97 & 335.46 & 2007-10-23T09:20:46.192 & 66\\
P12$\_$005839$\_$0994$\_$XI$\_$80S170W & -80.66 & 189.37 & 336.55 & 2007-10-25T09:56:39.216 & 120\\
P13$\_$005940$\_$1035$\_$XN$\_$76S064W & -76.51 & 295.91 & 340.76 & 2007-11-02T06:50:28.896 & 522\\
P13$\_$005941$\_$0947$\_$XI$\_$85S065W & -85.39 & 295.0 & 340.8 & 2007-11-02T08:39:53.181 & 408\\
P13$\_$005953$\_$1020$\_$XN$\_$78S057W & -78.01 & 302.45 & 341.3 & 2007-11-03T07:08:57.500 & 522\\
P13$\_$005958$\_$1030$\_$XI$\_$77S195W & -77.02 & 164.87 & 341.5 & 2007-11-03T16:30:56.168 & 150\\
P13$\_$006005$\_$0947$\_$XN$\_$85S015W & -85.32 & 344.59 & 343.44 & 2007-11-07T08:22:07.760 & 408\\
P13$\_$006112$\_$0852$\_$XN$\_$85S304W & -85.2 & 55.76 & 347.8 & 2007-11-15T16:27:46.904 & 66\\
P13$\_$006119$\_$1035$\_$XN$\_$76S271W & -76.52 & 88.75 & 348.08 & 2007-11-16T05:38:06.661 & 366\\
P13$\_$006123$\_$0953$\_$XN$\_$84S001W & -84.77 & 358.59 & 348.25 & 2007-11-16T13:04:04.676 & 522\\
P13$\_$006146$\_$1045$\_$XN$\_$75S289W & -75.53 & 70.6 & 349.17 & 2007-11-18T08:08:17.767 & 306\\
P13$\_$006148$\_$1028$\_$XN$\_$77S342W & -77.25 & 17.18 & 349.26 & 2007-11-18T11:51:51.830 & 522\\
P13$\_$006151$\_$0974$\_$XN$\_$82S055W & -82.63 & 305.13 & 349.38 & 2007-11-18T17:27:16.677 & 252\\
P13$\_$006161$\_$1030$\_$XN$\_$77S338W & -77.09 & 22.03 & 349.78 & 2007-11-19T12:11:05.625 & 324\\
P13$\_$006167$\_$0931$\_$XN$\_$86S092W & -86.96 & 267.2 & 350.02 & 2007-11-19T23:21:18.515 & 102\\
P13$\_$006173$\_$0934$\_$XN$\_$86S263W & -86.68 & 96.7 & 350.26 & 2007-11-20T10:34:49.912 & 66\\
P13$\_$006174$\_$0958$\_$XN$\_$84S316W & -84.29 & 43.28 & 350.3 & 2007-11-20T12:28:01.397 & 66\\
P13$\_$006176$\_$0935$\_$XN$\_$86S350W & -86.51 & 10.06 & 350.38 & 2007-11-20T16:11:29.064 & 204\\
P13$\_$006197$\_$0930$\_$XN$\_$87S187W & -87.04 & 173.3 & 351.22 & 2007-11-22T07:27:46.120 & 150\\
P13$\_$006199$\_$1040$\_$XN$\_$76S296W & -76.02 & 63.9 & 351.3 & 2007-11-22T11:15:22.194 & 522\\
P13$\_$006204$\_$0986$\_$XN$\_$81S065W & -81.45 & 295.05 & 351.5 & 2007-11-22T20:35:13.014 & 252\\
P13$\_$006206$\_$1016$\_$XN$\_$78S124W & -78.57 & 235.47 & 351.58 & 2007-11-23T00:20:04.979 & 504\\
P13$\_$006207$\_$0956$\_$XN$\_$84S136W & -84.49 & 223.45 & 351.62 & 2007-11-23T02:10:20.857 & 522\\
P13$\_$006229$\_$0951$\_$XN$\_$84S014W & -84.97 & 345.69 & 352.5 & 2007-11-24T19:19:03.593 & 522\\
P13$\_$006234$\_$1008$\_$XN$\_$79S168W & -79.28 & 191.3 & 352.7 & 2007-11-25T04:42:19.717 & 306\\
P13$\_$006239$\_$1040$\_$XN$\_$76S308W & -76.09 & 51.62 & 352.9 & 2007-11-25T14:04:14.736 & 408\\
P13$\_$006240$\_$1030$\_$XN$\_$77S335W & -77.04 & 25.01 & 352.94 & 2007-11-25T15:56:00.916 & 522\\
P13$\_$006257$\_$1034$\_$XN$\_$76S079W & -77.88 & 282.15 & 353.61 & 2007-11-26T23:43:54.535 & 264\\
P13$\_$006271$\_$1010$\_$XN$\_$79S098W & -79.07 & 261.59 & 354.17 & 2007-11-28T01:54:24.701 & 306\\
P13$\_$006282$\_$1046$\_$XN$\_$75S043W & -75.47 & 316.98 & 354.61 & 2007-11-28T22:29:37.821 & 522\\
P13$\_$006283$\_$1003$\_$XN$\_$79S065W & -79.75 & 294.76 & 354.65 & 2007-11-29T00:20:28.301 & 522\\
P13$\_$006290$\_$1017$\_$XN$\_$78S258W & -78.38 & 101.84 & 354.92 & 2007-11-29T13:26:24.488 & 522\\
\end{longtable}
\end{small}

\section{Planet Four: Terrains (P4T)}
\label{sec:P4T}
\begin{sidewaysfigure}
\includegraphics[width=0.9\columnwidth]{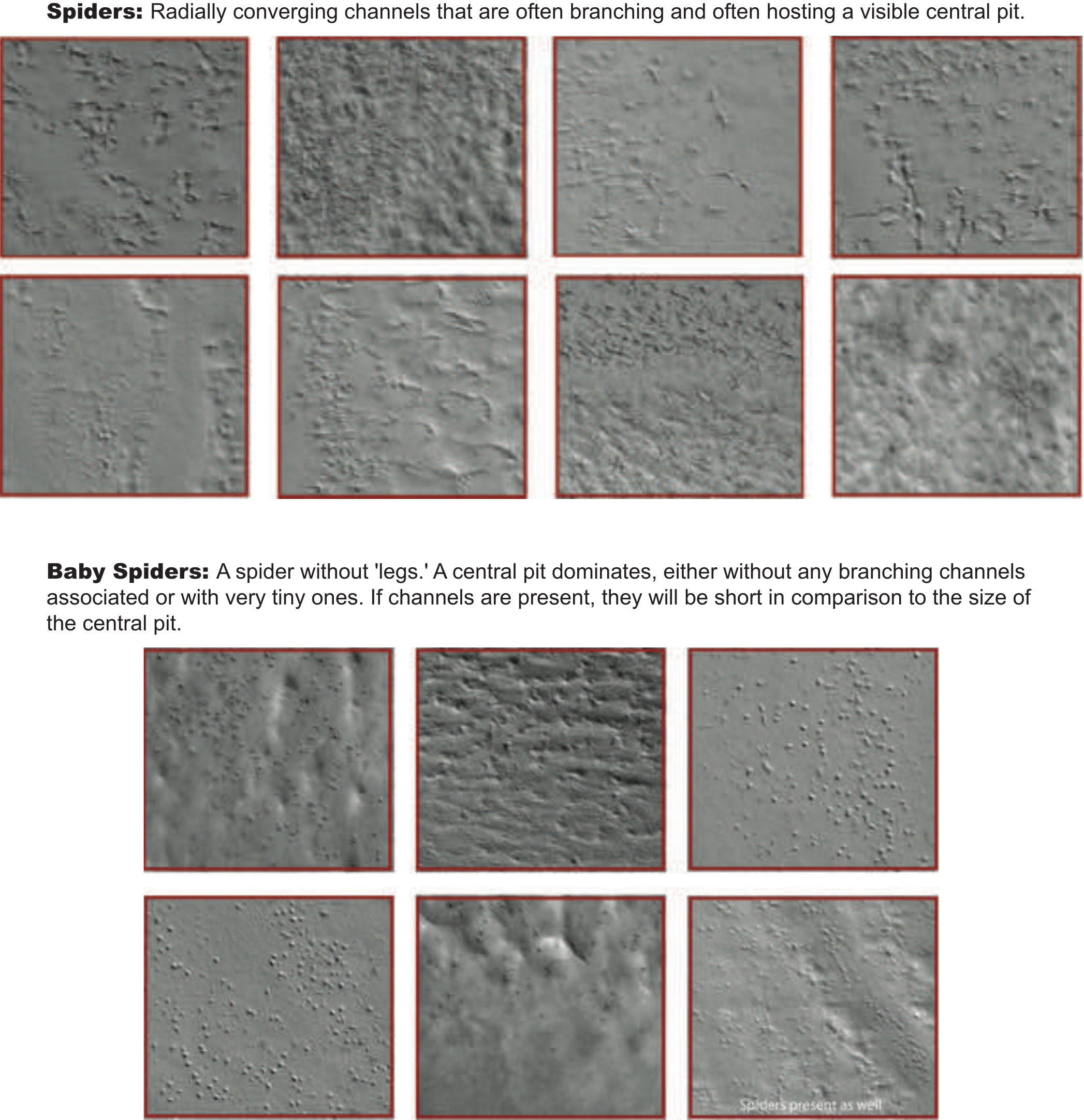}
 \caption{ \label{fig:examples1} Examples of CTX surface morphologies searched for on P4T continued with the help text provided on the P4T project site guide.}
 \end{sidewaysfigure}

\begin{sidewaysfigure}
\includegraphics[width=0.8\columnwidth]{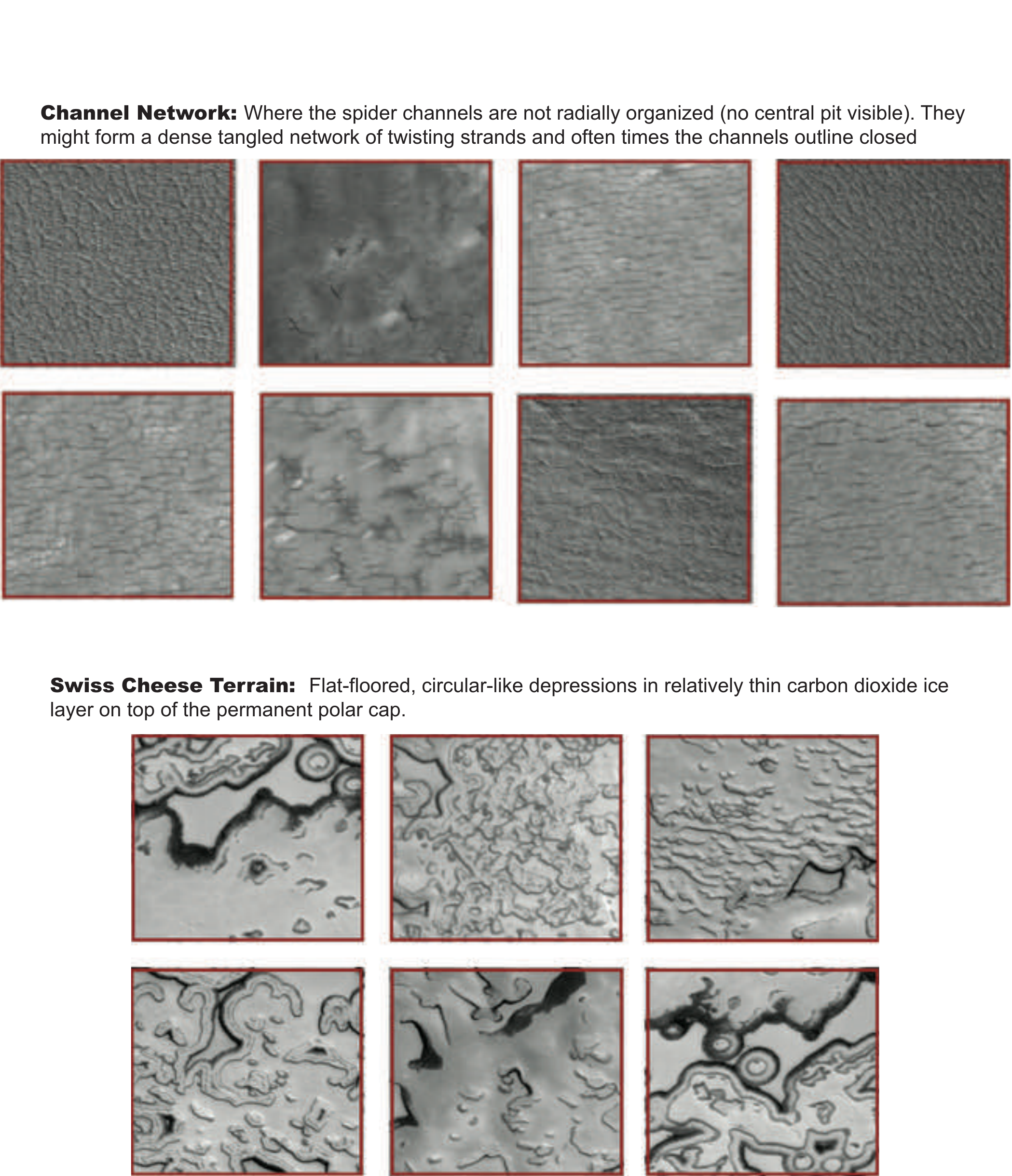}
 \caption{ \label{fig:examples2} Examples of CTX surface morphologies searched for on P4T continued with the help text provided on the P4T project site guide. }
 \end{sidewaysfigure}

\begin{figure}
\begin{center}
\includegraphics[width=1.0\columnwidth]{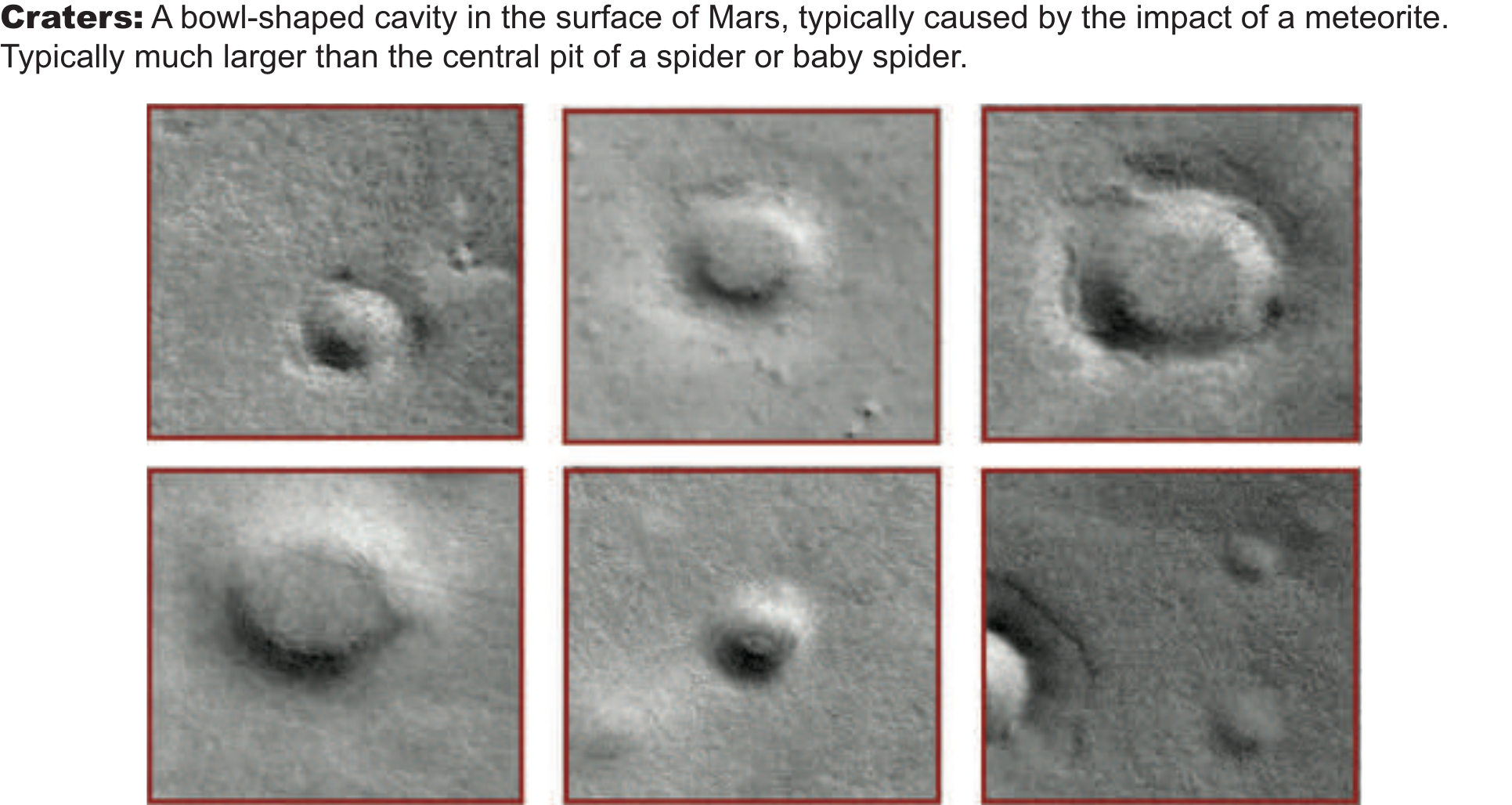}
 \caption{ \label{fig:examples3} Examples of CTX surface morphologies searched for on P4T continued with the help text provided on the P4T project site guide.}
 \end{center}
 \end{figure}

The aim of P4T is to identify features of interest in CTX observations of the South Polar region. For this endeavor we focused on three types of surface features and their distribution on the Martian South Polar region: 1) araneiforms 2) erosional depressions, troughs, mesas, ridges, and quasi-circular pits characteristic of the SPRC which we collectively refer to as `Swiss cheese terrain', and 3) craters. We aim to study the distribution of the araneiforms on the South Polar region and explore their locations compared to the locations of other CO$_2$ ice sublimation features The crater identifications can aide with the surface age dating of the SPLD, similarly to what has been done for the North Polar region \citep{2016GeoRL..43.3060L}.  Examples of each of the  three types of surfaces features (taken from the P4T site guide\footnote{\url{http://terrains-guide.planetfour.org/}}) are shown at the resolution of CTX in Figures \ref{fig:examples1}, \ref{fig:examples2}, and \ref{fig:examples3}. 

Several different types of araneiform structures have been identified on the South Polar region \citep{2001JGR...10623429M,2003JGRE..108.5084P,2007JGRE..112.8005K, 2010Icar..205..283H}. Similar to \cite{2010Icar..205..283H}'s categories identified in HiRISE imaging, we divide araneiform terrain broadly into three araneiform morphologies distinguishable at CTX resolution: `baby spiders', `spiders', and `lace terrain'. Spiders are defined as radially converging channels that are often branching and often hosting a visible central pit (see Figure \ref{fig:examples1}). We note for the reader, that any subsequent reference to `spiders'  in the text uses this definition. Baby spiders are spiders with `short legs', where a central pit dominates with short or no radial channels (see Figure \ref{fig:examples1}). In the help documentation, we recommend to volunteers that if the channels are shorter than the extent of the central pit, then it is a baby spider. As identification with P4T is through visual inspection, we acknowledge that there is not always a clear dividing line between spiders and baby spiders. The criteria separating the two categories is more qualitative than quantitative. With the resolution of CTX, it is difficult to distinguish patterned ground, formed by the repeating freezing and thawing of soil, from lace araneiforms, formed by the CO$_2$ jet process. In HiRISE images one can observe the more sinuous nature of the lace araneiforms differentiating these features from polygonal channels, but this is typically not visible in CTX observations. For P4T, we combine lace araneiforms and pattern ground together as one category, referring to them collectively as a  `channel network' (see Figure \ref{fig:examples2}).  When needed for the channel network regions identified by P4T, we plan to use higher resolution imaging from HiRISE, to distinguish polygonal channels from interconnected araneiforms. 

The SPRC's top surface layers have been categorized into different groups of characteristic smooth-walled features including: troughs, mesas, and quasi-circular pits visible in orbital imagery \citep[e.g][]{1992mars.book..934J, 2009Icar..203..352T,2016Icar..268..118T}. A recent inventory of SPRC surface morphology based on HiRISE and CTX imagery is provided in \cite{2009Icar..203..352T,2016Icar..268..118T}. The main priority of P4T is to identify new aranieform locations, thus we ask the P4T volunteers to sort araneiforms in more detail by distinguishing araneiforms of different morphology from each other. Given the spatial resolution of CTX and the size of each P4T subject image, we chose to not task P4 volunteers with distinguishing  between the different categories established by  \cite{2009Icar..203..352T,2016Icar..268..118T}. For the sublimation features of the SPRC, we combined the morphological categories together into one for P4T which we simply refer to as `Swiss cheese terrain'. In the P4T help content (see Figure \ref{fig:examples2}), we describe the Swiss cheese terrain as flat-floored, circular-like depressions and visual examples for P4T show the majority of the different sublimation morphologies visible on the SPRC with CTX. We note for the reader, that any subsequent reference to `Swiss cheese terrain' refers to the combined SPRC sublimation features previously identified. 

\subsection{Web Interface}
\label{sec:interface}
\begin{figure}[!]
\begin{center}
\includegraphics[width=1.0\columnwidth]{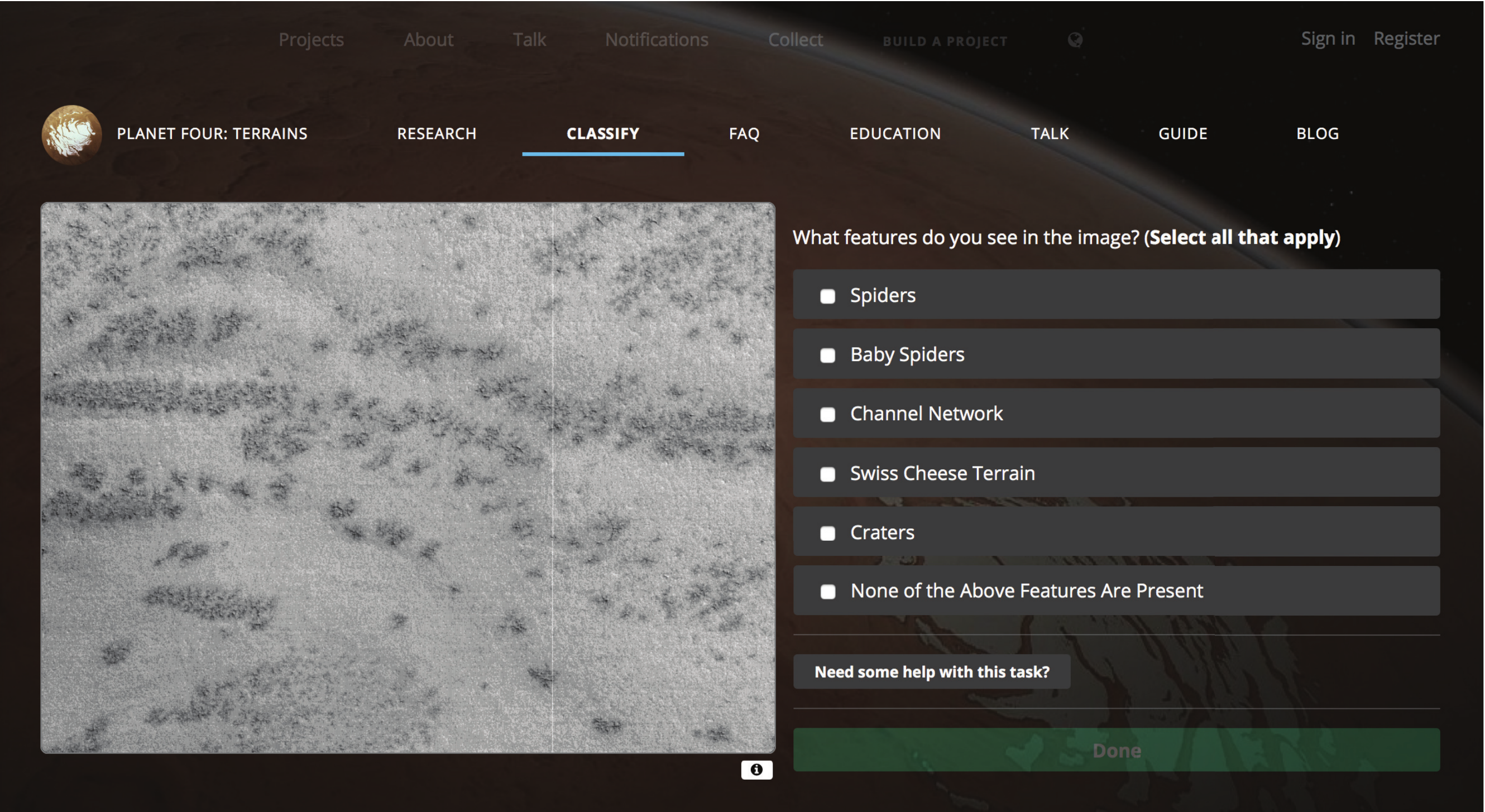}
 \caption{ \label{fig:interface} Main classification interface of the P4T website.}
 \end{center}
 \end{figure}

The Planet Four: Terrains website\footnote{\\url{http://terrains.planetfour.org} or url{https://www.zooniverse.org/projects/mschwamb/planet-four-terrains}} and online classification interface is built upon the Zooniverse\footnote{
\url{http://www.zooniverse.org}} \citep{2011MNRAS.410..166L,2012amld.book..213F} Project Builder platform\footnote{ \url{http://www.zooniverse.org/lab}. The code base for the Zooniverse Project Builder Platform is available under an open-source license at \url{https://github.com/zooniverse/Panoptes} and \url{https://github.com/zooniverse/Panoptes-Front-End}}. The Zooniverse Project Builder platform enables the rapid development of online citizen science projects by providing a set of web-tools for scientists to create and maintain their own citizen science projects. The platform and its Application Program Interface (API) is built upon Amazon Web Services which allows the P4T website to quickly and  efficiently scale to handle the load from varying numbers of visitors on the site at the same time; it is capable of supporting tens of thousands of simultaneous users. When a volunteer arrives at the P4T website, the web interface (see Figure \ref{fig:interface}) displays a selected 800$\times$600 pixel CTX subject. The Zooniverse API pseudo randomly selects a new set of subjects for each classifier upon request, in order to distribute the volunteer effort across the known dataset. In addition, the API algorithm also selects subjects the volunteer has not previously reviewed and that have not been viewed by enough volunteers to mark them as complete. Each subject is typically assessed independently by 20 classifiers before it is retired from review on the P4T website. 

Volunteers are tasked with assessing the image and determining what surface features of interest are present in the subject selecting from a choice of: `spiders', `baby spiders', `channel network', `Swiss cheese terrain', `craters' and `none of the above'. The volunteer is able to select more than one response that best describes the subject image. In this Paper, a `classification' is defined as the total amount of information collected about one subject by a single volunteer answering the question presented in the P4T classification interface. Help content and example images of each feature/answer choice can be accessed by clicking on the `Need some help with this task?' button. Additional examples and help content are provided on a linked site guide\footnote{\url{http://terrains-guide.planetfour.org/}}.  To minimize external information influencing or biasing a volunteer's response, no identifying information about the original parent CTX image including the filename or observing circumstances (such as location coordinates, time of day, or L$_{\rm s}$) are provided to the volunteer before they submit their classification. Thus the classifier cannot assess whether the image is from a location that previously has been identified as having araneiforms (such as the SPLD) or was previously imaged by HiRISE in previous south polar monitoring observations. To keep the multiple volunteer assessments independent for each subject, the classifier is kept blind to previous people's responses for the presented subject, and the subject's internal Zooniverse identifier is hidden from the classification interface. 

Once the volunteer selects the categories that best describe the presented subject and hits the `Done' button, the classification is submitted through the Zooniverse API and stored in the Zooniverse PostgreSQL database. The subject identifier, volunteer's IP (Internet Protocol) address, Zooniverse username if available, timestamp, web browser and operating system information, and user response are recorded. At this point, the volunteer cannot go back and revise their classification.  P4T volunteers can classify in two modes: registered with a Zooniverse account or unregistered. The P4T classification interface is presented the same for both registered and non-registered classifiers. The only difference is that non-registered classifiers are reminded from time-to-time to log-in/register for a Zooniverse account. Registered classifications are easily linked by their associated Zooniverse account. For non-registered classifications, a unique identifier is generated and used to link the classifications completed by a given IP address. We note because of the IP tracking, a non-registered classifications from a single IP address may not necessarily equate to a single individual. Additionally, if a volunteer initially classifies non-registered and then logs-in to a Zooniverse account, the previous classifications are not linked  with their registered account and remain attributed to an unregistered classifier. 
 
\subsection{Talk Discussion Tool}

After submitting a classification on the P4T website, the classification interface presents the volunteer with two options: `Talk' or `Next'. `Next' will load a new subject image in the classification interface. Selecting the `Talk' button instead loads the P4T Talk discussion tool\footnote{\url{https://www.zooniverse.org/projects/mschwamb/planet-four-terrains/talk}}. Talk enables volunteers to further explore the P4T dataset beyond the main tasks and aims of the classification interface. The discussion tool hosts message boards that support interactions with the science team and others in the P4T volunteer community. Each subject has a dedicated page on Talk where a registered volunteer can initiate a new discussion or add commentary to an on-going discussion about the subject. Volunteers can also associate the subject with searchable Twitter-like hashtags and link multiple subjects together into groups. Reading previously posted commentary on Talk might bias a volunteer's assessment in the classification interface. To maintain the independence of the classifications, the Zooniverse identifier for the subject is not presented in the classification interface and the direct link for the Talk subject page is only revealed after a volunteer submits their classification to the Zooniverse database. For this work, we focus primarily on the results from the main classification interface. 
 
\subsection{Site History and Statistics}

\begin{figure}
\begin{center}
\includegraphics[width=0.93\columnwidth]{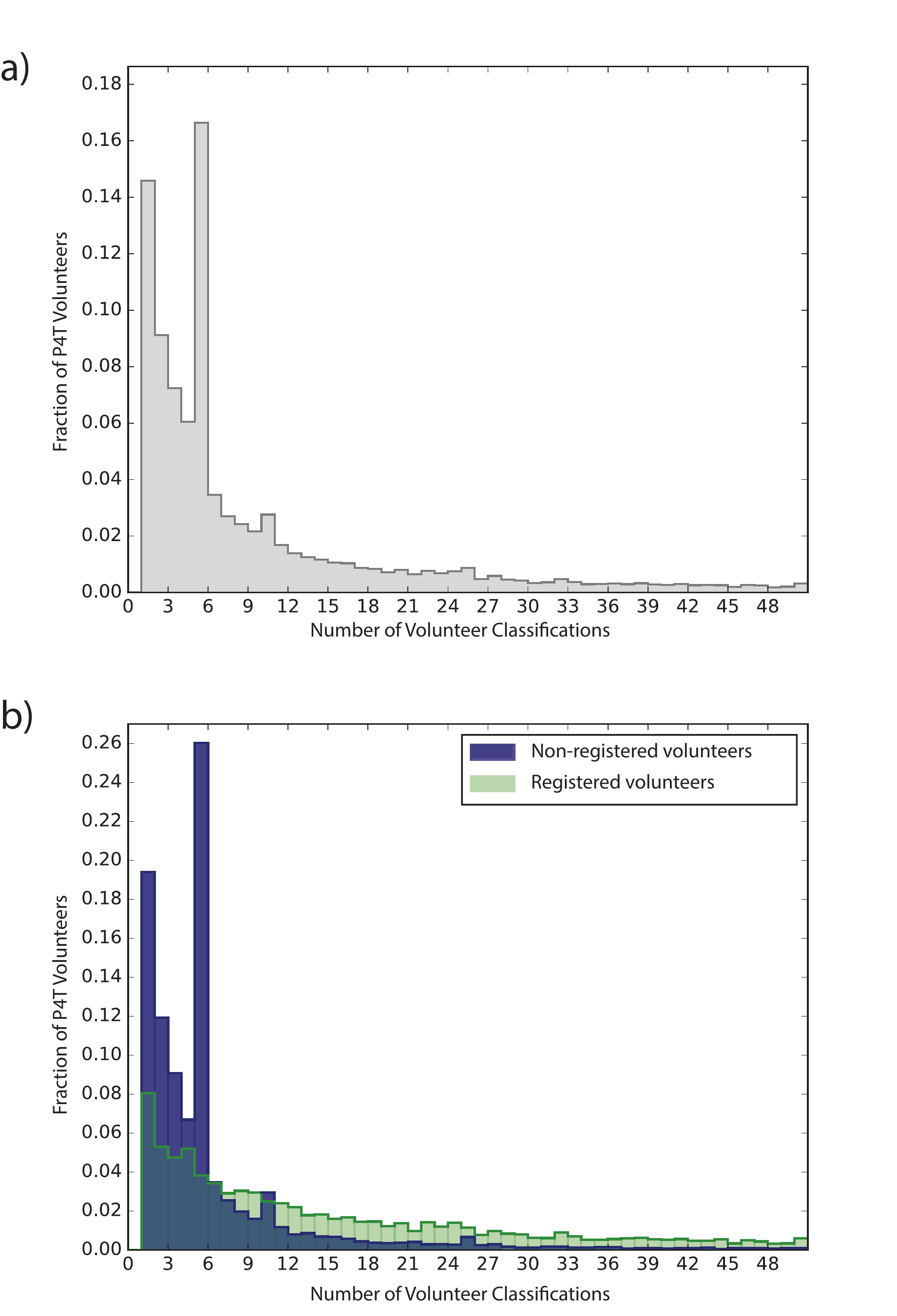}
 \caption{ \label{fig:classification_counts}  Distribution of P4T volunteer classifications used in this work. Figure a shows the combined distribution of both logged-in and non-logged in sessions. Figure b shows the volunteer classification count individually for registered and non-logged volunteers. Both distributions use a bin size of 1, the distributions plotted are cut off beyond 50 classifications.}
 \end{center}
 \end{figure}

The 20,122 subjects derived from the 90 CTX full frame images used in this study were classified by 6,309 registered Zooniverse users and 8,513 non-logged-in sessions (tracked by IP address). The classifications were collected from June 24, 2015 to August 10, 2016. We plot the distribution of registered volunteers and non-logged-in sessions as a function of number of classifications in Figure \ref{fig:classification_counts}. Registered volunteers classified a mean of 61 subjects with a median of 14. Non-logged-in sessions classified an average of 19 and median of 5 subjects. 20$\%$ of registered volunteers classified more than 50 subjects, while only 3$\%$ of non-logged-in sessions classified more than 50 subjects. 

\begin{figure}
\begin{center}
\includegraphics[width=1.0\columnwidth]{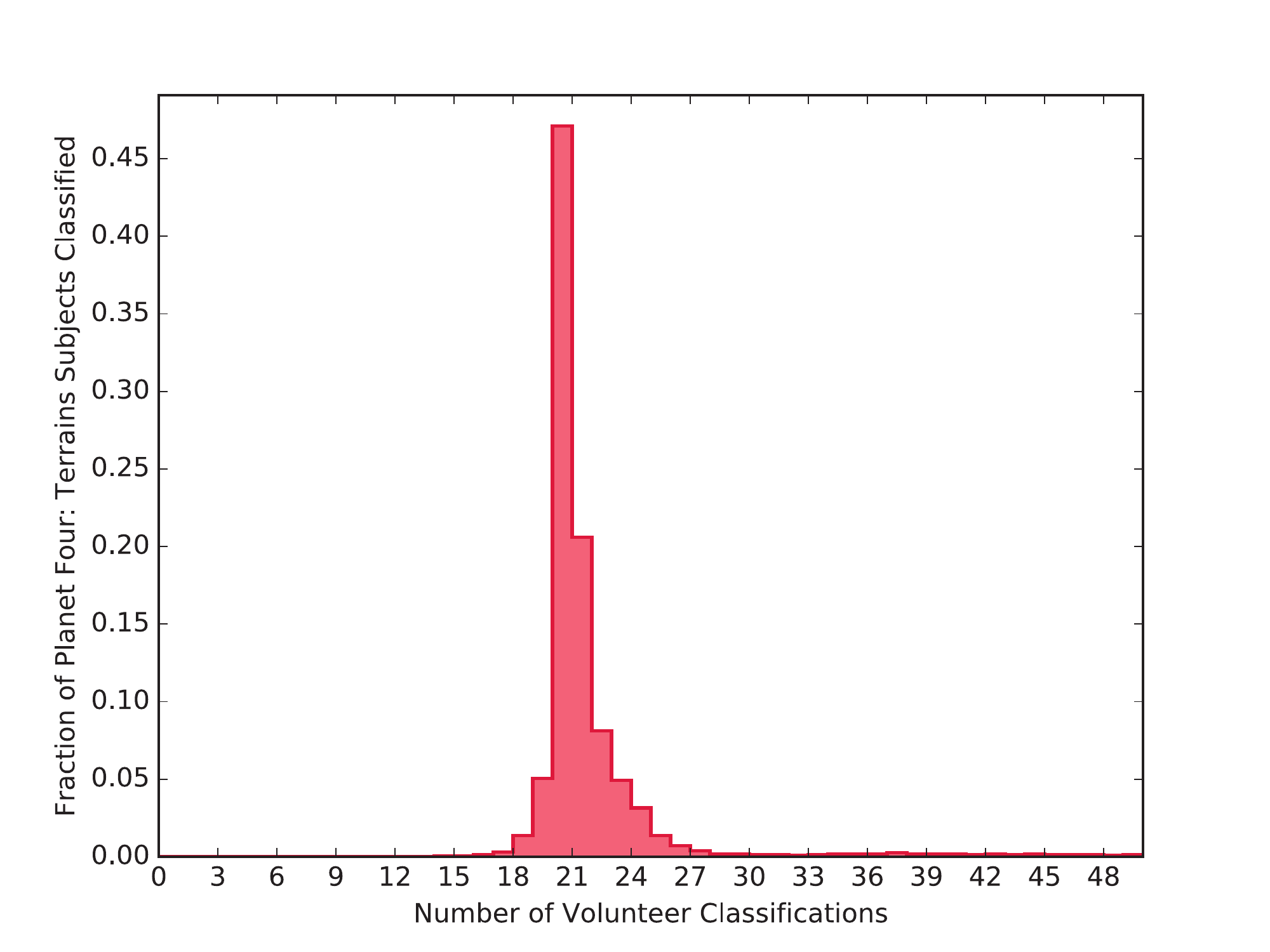}
 \caption{ \label{fig:subject_counts} Histogram of the number of P4T classifications per subject used in this work. The distribution was generated with a bin size of 1 and for clarity the plot is cut off beyond 50 classifications.}
 \end{center}
 \end{figure}

We plot the distribution of classifications per subject in Figure \ref{fig:subject_counts}. The majority of the subjects received 20 classifications or more. Due to a bug in the backend of the Zooniverse platform, some volunteers were shown the same subject to classify twice or more. To ensure the assessments for each subject remain independent, we filter the classification database and remove any duplicate classifications keeping the first response from the registered username or IP address in the case of non-logged classifications. For all the values reported in this Paper, we use the filtered classifications. In some cases, this will leave a subject with less than 20 independent assessments, but the impact is negligible with only 7$\%$ of the subjects in this study having less than 20 unique classifications and 0.07$\%$ with less than 17 unique classifications. We also note that due to a glitch in the backend of the Zooniverse platform a portion of P4T subjects were not retired after 20 classifications. 8$\%$ of the subjects in this study received more than 25 independent assessments, with only 3$\%$ of the subjects receiving more than 50 classifications.

\section{Data Analysis}

For this work, we focus solely on the P4T identification of spiders and Swiss cheese terrain. We refer the reader to Section \ref{sec:P4T} for the specific definitions of spiders and Swiss cheese terrain we use in this analysis. We combine the multiple volunteer classifications of a given subject together, examining the number of volunteers who selected the `spiders' or `Swiss cheese terrain' buttons for each subject in order to identify spiders  and Swiss cheese terrain present  in the P4T data. Some volunteers may be better at spotting these features than others. Rather than treat all volunteer assessments equality, we apply a user weighting scheme that enables us to pay more attention to those volunteers who are better at identifying spiders or Swiss cheese terrain and also reduce the influence of potentially unreliable classifiers, such as those who did not engage in or understand the task.  We then apply these weights when combining the volunteer assessments for spiders or Swiss cheese to determine how likely a P4T image is to have these features of interest present.

\subsection{User Weighting Scheme}
We use a modified version of the iterative user weighting schemes developed by \cite{2008MNRAS.389.1179L} and \cite{2011MNRAS.410..166L} for the visual morphological classifications of galaxies in the Galaxy Zoo project and by \cite{2012ApJ...754..129S} for the visual identification of planet transits in NASA \emph{Kepler} data in the Planet Hunters project. The weighting scheme evaluates the ability of each classifier and assigns a weight based on their tendency to agree with the majority opinion, distinguishing those volunteers who are better at spotting Swiss cheese or spiders in order to pay attention to their responses more than others when identifying which subjects have these features present. 

We define a `user' for our case as either a volunteer with a registered Zooniverse account or the collective behavior of non-logged-in sessions with a unique-IP address. A unique non-logged-in IP address may not necessarily by a single individual (see \ref{sec:interface}), but for the weighting scheme we link the classifications together and determine a single weight. If a user excels at identifying spiders, it does not  necessarily mean they would be as good as identifying Swiss cheese terrain in the P4T data. We therefore treat the identification of spiders and Swiss cheese terrain independently as two separate classes of responses, and determine separate user weights for each class. For the analysis presented here, the volunteer classifications are effectively divided into two responses per class: `found' and `not found'. For spider identification this breaks down into  `spiders found' if the volunteer selected the `spider' button while classifying the subject and `no spiders found' if the volunteer did not select the `spiders' button. For this analysis, if a volunteer clicked on the `baby spiders' or `channel network' buttons without also clicking on the 'spiders' button, this would count as a 'no spiders found' response for the subject image. We do the same thing with the responses for Swiss cheese terrain, with a vote of `Swiss cheese found' if the 'Swiss cheese terrain' button was selected when the subject image was reviewed or `no Swiss cheese found' if volunteer didn't mark the image as having Swiss cheese terrain. 

Each user $j$ is assigned two weights, one for each class: $w_j(\rm spider)$ and $w_j(\rm Swiss)$. Initially, all users start out with each of those weights equal to 1. Then for each P4T subject  \emph{i}  scores for spiders, $s_i(\rm spider)$,  and Swiss cheese pits,  $s_i(\rm Swiss)$, are calculated. We define these scores per subject as follows:
\begin{eqnarray}
s_i(\rm spider) &=& \frac{1}{S_i } \displaystyle \sum_{k} w_k(\rm spiders) \qquad    k \textrm{=users who selected `spider' for subject \emph{i}} \\
s_i(\rm Swiss) &=&  \frac{1}{C_i } \displaystyle \sum_{m} w_m(\rm Swiss)\qquad    m \textrm{=users who selected `Swiss cheese' for subject \emph{i}}\\
\end{eqnarray}
 where $S_i$  and  $C_i$ are the sum of the respective user weights for all the users who classified subject \emph{i}:
\begin{eqnarray}
S_i &=& \displaystyle \sum_{k=j} w_k(\rm spiders) \qquad    \\
C_i &=& \displaystyle \sum_{m=j} w_m(\rm Swiss) \qquad \\   
 j &=& \textrm{all users who classified subject \emph{i}} \nonumber 
\end{eqnarray}
Subject scores vary between the values of 0 and 1, inclusive. A subject score of 1 is assigned if all volunteers who classified the subject agree and identified the same features of interest in the subject image. 

Once the subject scores $s_i(\rm spider)$ and $s_i(\rm Swiss)$ are calculated, we next assign new user weights for each user $j$  by the prescription below:
\begin{eqnarray}
w_j(spiders)  &=& \left\{ \begin{array}{ll}
 \displaystyle  {\frac{A}{N_j}  \sum_{i=p} s_i (\rm spider) + \frac{A}{N_j}  \sum_{i=q} [1- s_i (\rm spider)]} & \textrm{if $N_j >1$}\\
 1 & \textrm{if $N_j =1$} 
  \end{array} \right. \\
     p &=& \textrm{subjects volunteer classified as `spiders found' } \nonumber  \\
      q &=& \textrm{subjects volunteer classified as `spiders not found' } \nonumber \\
\nonumber \\
w_j(Swiss)  &=& \left\{ \begin{array}{ll}
 \displaystyle \frac{B}{N_j}   \sum_{i=t} s_i (\rm Swiss) + \frac{B}{N_j}   \sum_{i=u} [1- s_i (\rm Swiss)]& \textrm{if $N_j >1$}\\
 1 & \textrm{if $N_j =1$} 
  \end{array} \right. \\
    t &=& \textrm{subjects volunteer classified as `Swiss cheese found' } \nonumber  \\
    u &=& \textrm{subjects volunteer classified as `Swiss cheese not found' } \nonumber
  \end{eqnarray}
where $N_j$ is the number of subjects classified by user \emph{j}. The scaling factors $A$ and $B$ are chosen to be such that the median user weight for volunteers who classify more than one subject will be 1. We only adjust the weights of those volunteers who have classified more than one subject, which constitutes 85$\%$ of P4T  users.  With a single classification there is not much information to use to evaluate a  volunteer's ability to discern spiders and Swiss cheese terrain. Thus for those users who have classified a single subject,  we choose to keep the user weights static at a value of 1, where the median of the adjusted weights will lie.  A volunteer who has classified more than one subject is upweighted strongly when they agree with the majority weighted vote and down weighted more harshly when their response is at odds with the majority of the volunteers who reviewed the subject image.

\begin{figure}
\begin{center}
\includegraphics[width=1.0\columnwidth]{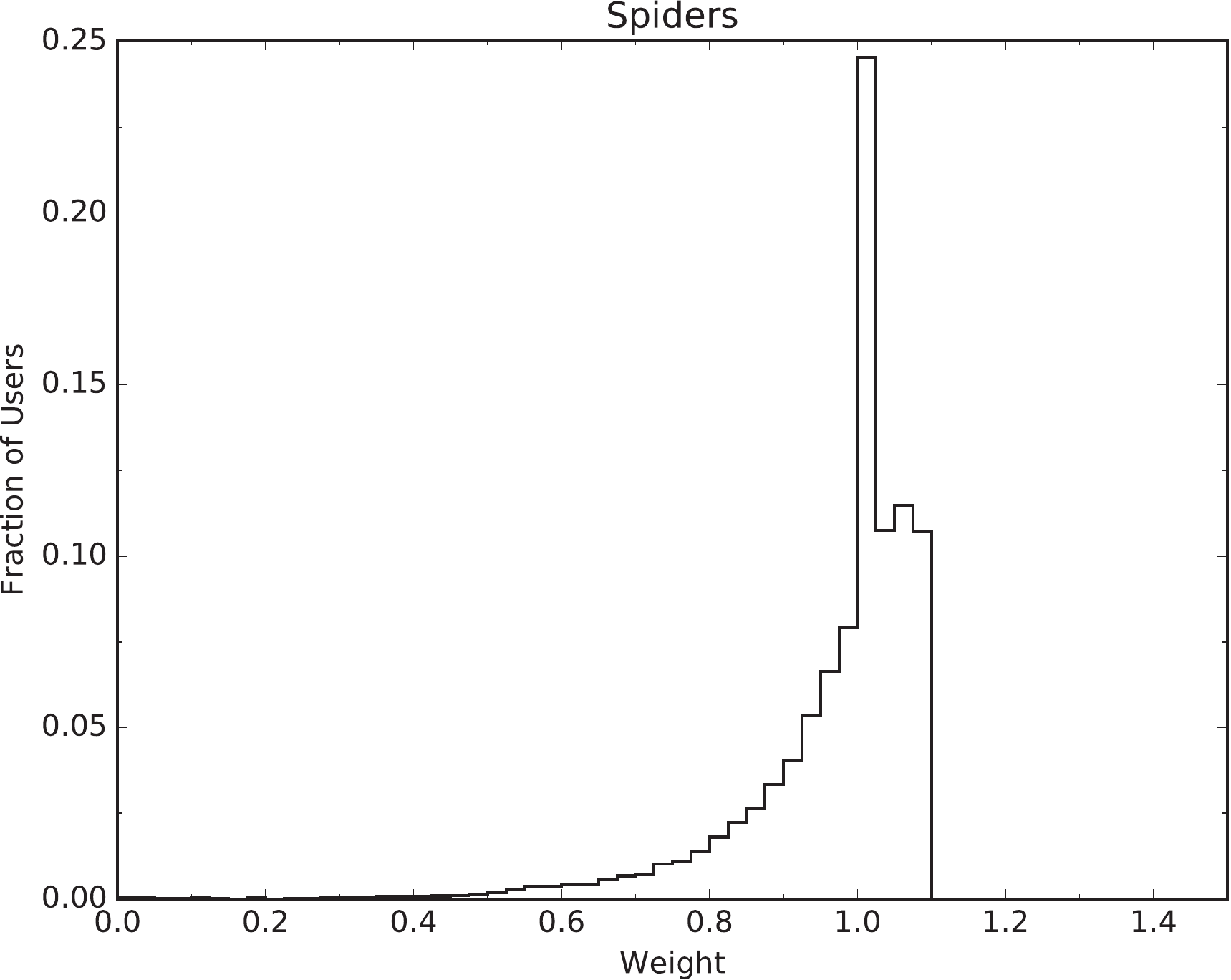}
 \caption{ \label{fig:spider_weights} Histogram of spider user weights, $w(\rm spider)$, plotted with a bin size of 0.025 }
 \end{center}
 \end{figure}

\begin{figure}
\begin{center}
\includegraphics[width=1.0\columnwidth]{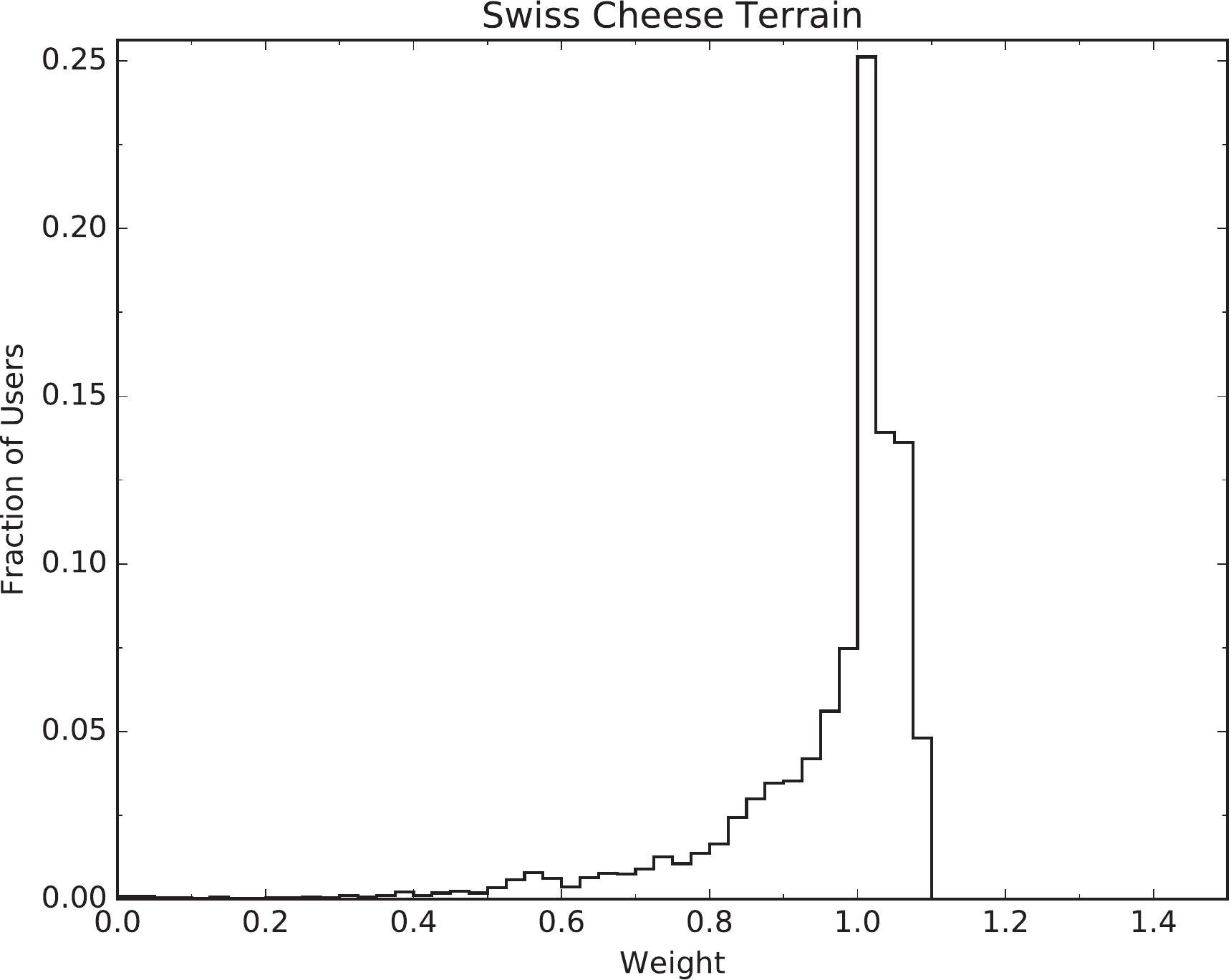}
 \caption{ \label{fig:swiss_weights} Histogram of Swiss Cheese user weights, $w(\rm Swiss)$,  plotted with a bin size of 0.025 }
 \end{center}
 \end{figure}

After the user weights are first adjusted, the subject scores in each class, $s_i(\rm spider)$ and $s_i(\rm Swiss)$, are recalculated using the updated user weights with Equations 1 and 2. Then new user weights are assigned with Equations 6 and 7.  The process is iterated until convergence is achieved, when the median absolute difference between the old and updated user weights is less than or equal to 1x10$^{-4}$, in this case after four iterations for both spiders and Swiss cheese. We plot the distribution of user weights for spiders in Figure \ref{fig:spider_weights} and for Swiss cheese terrain in Figure \ref{fig:swiss_weights}. With this scheme, a user weight can never be zero. For $w_j(\rm spider)$,  91$\%$ of users have weights greater than  0.8 and 43$\%$  of user weights are greater than 1.  For $w_j(\rm Swiss)$, 89$\%$ of users have weights greater than  0.8 and 43$\%$  of user weights are greater than 1.

\subsection{Combining User Classifications and Gold Standard Data}

\begin{figure}
\begin{center}
\includegraphics[width=1.0\columnwidth]{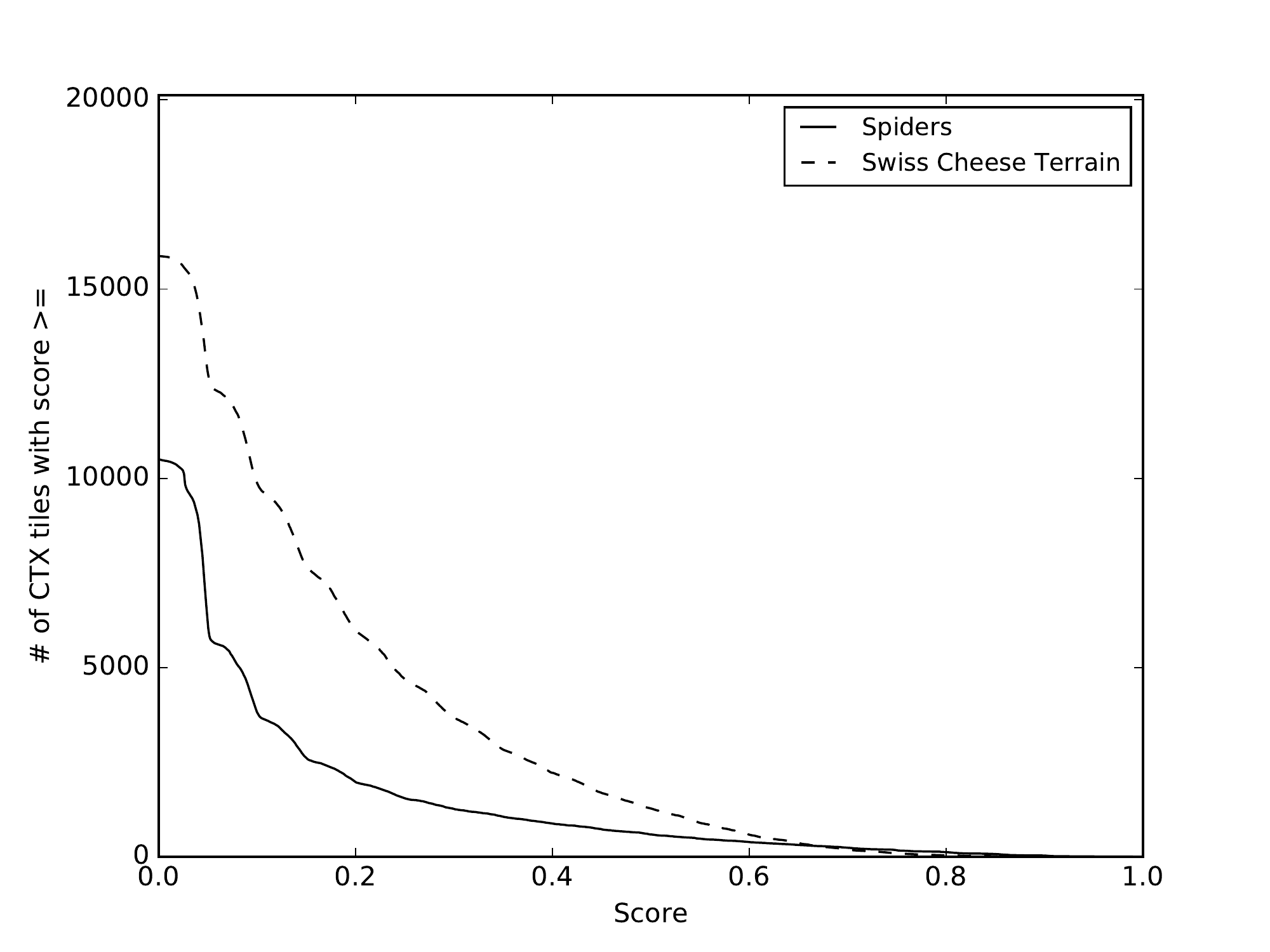}
 \caption{ \label{fig:scores} Cumulative distribution of spider and Swiss cheese terrain scores ($s_i(\rm spider)$ and $s_i(\rm spider)$) for P4T subjects.}
 \end{center}
 \end{figure}

\begin{figure}
\begin{center}
\includegraphics[width=1.0\columnwidth]{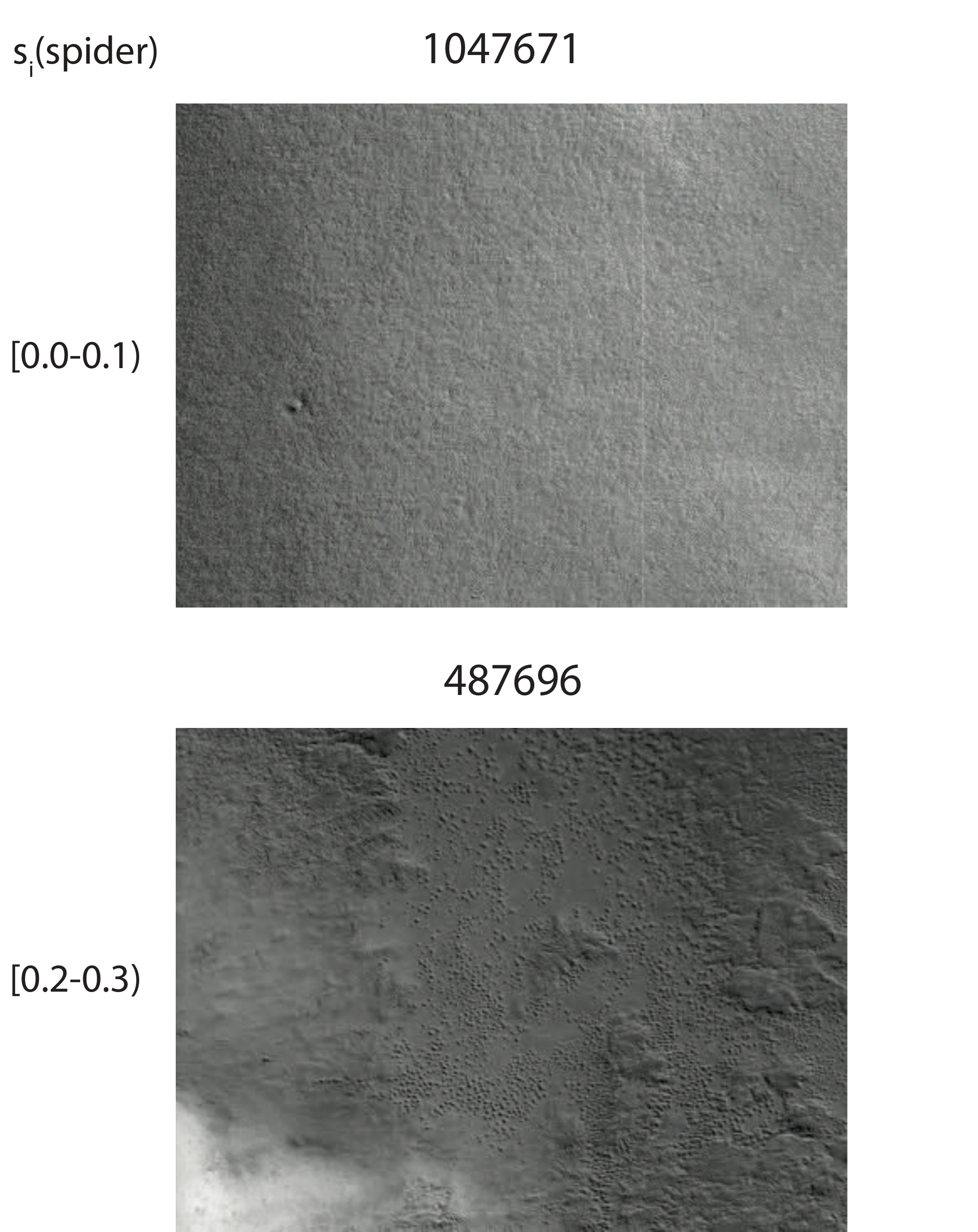}
 \caption{ \label{fig:spider_score_ex1} Randomly selected sample of subjects per spider score  ($s_i(\rm spider)$) binned with bin size of 0.1. CTX images D14$\_$032593$\_$1037 (top)  and P13$\_$005953$\_$1020 (bottom). Each P4T subject image is $\sim$4.8$\times3.6$ km.}
 \end{center}
 \end{figure}

\begin{figure}
\begin{center}
\includegraphics[width=.9\columnwidth]{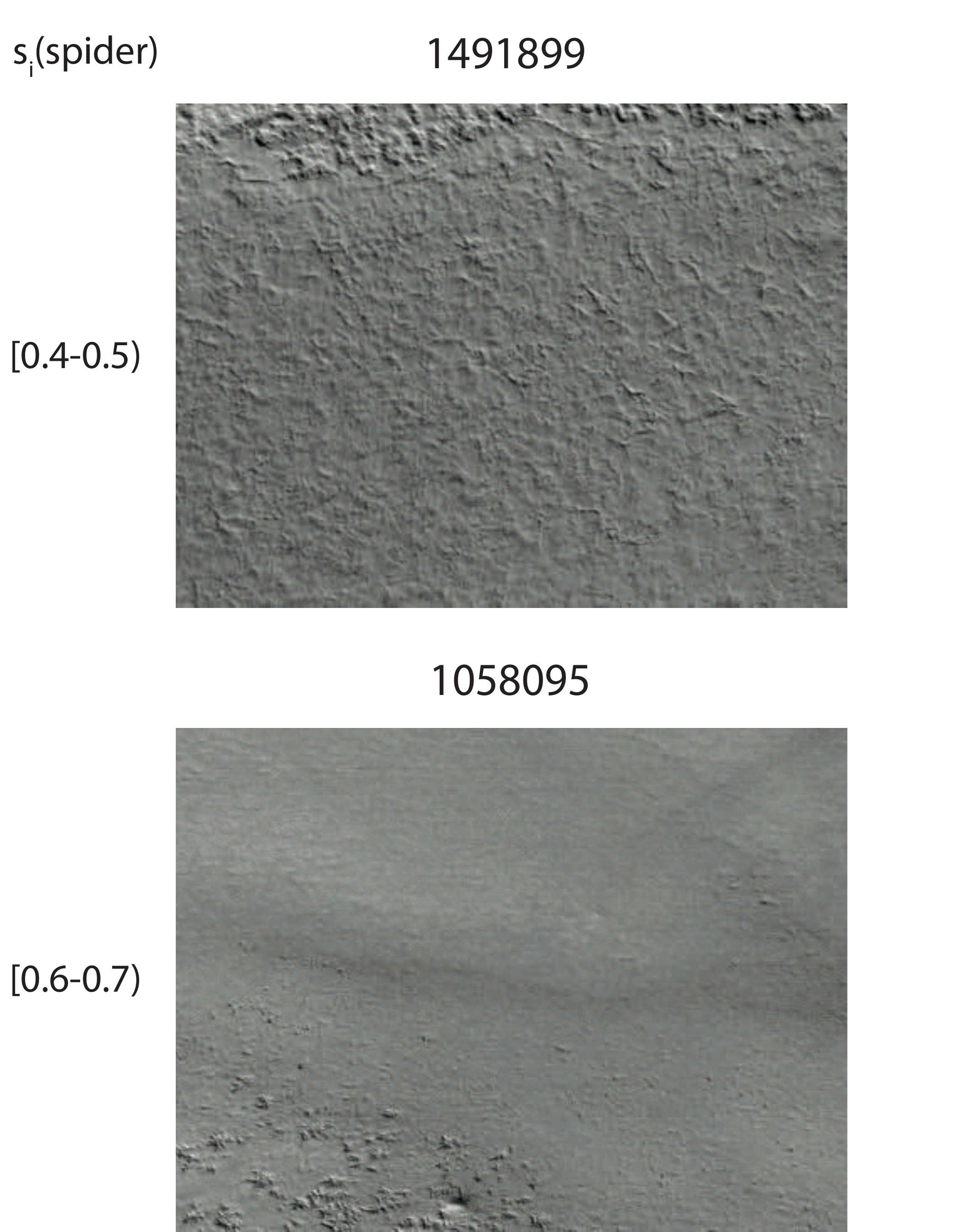}
 \caption{ \label{fig:spider_score_ex2} Randomly selected sample of subjects per spider score  ($s_i(\rm spider)$) binned with bin size of 0.1 continued. CTX images: G14$\_$023524$\_$0999 (top) and P13$\_$006151$\_$0974(bottom). Each P4T subject image is $\sim$4.8$\times3.6$ km.}
 \end{center}
 \end{figure}

\begin{figure}
\begin{center}
\includegraphics[width=1.0\columnwidth]{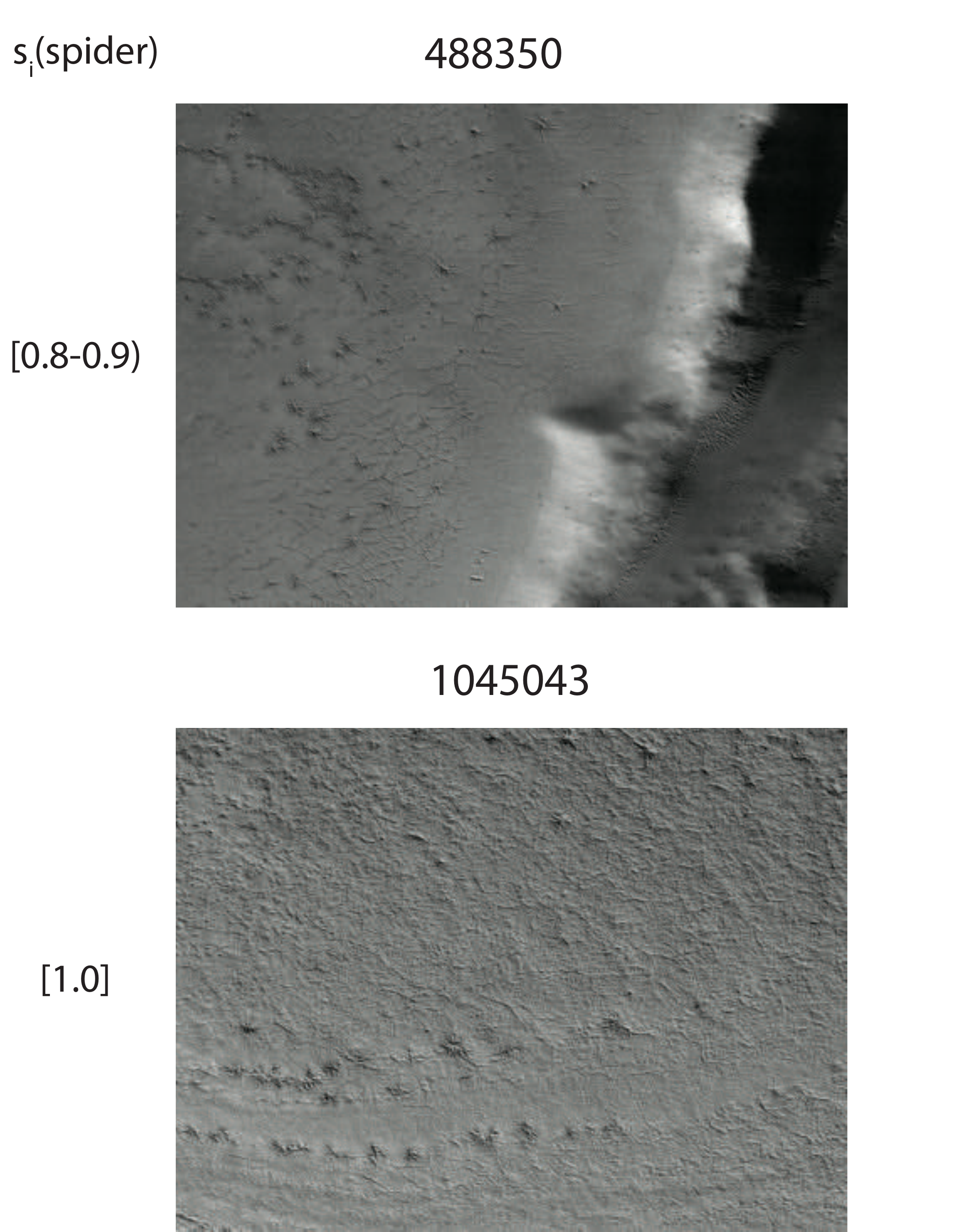}
 \caption{ \label{fig:spider_score_ex3} Randomly selected sample of subjects per spider score  ($s_i(\rm spider)$) binned with bin size of 0.1 continued.  CTX images: P13$\_$006204$\_$0986 (top) and P13$\_$006151$\_$0974 (bottom). Each P4T subject image is $\sim$4.8$\times3.6$ km.}
 \end{center}
\end{figure}

\begin{figure}
\begin{center}
\includegraphics[width=1.0\columnwidth]{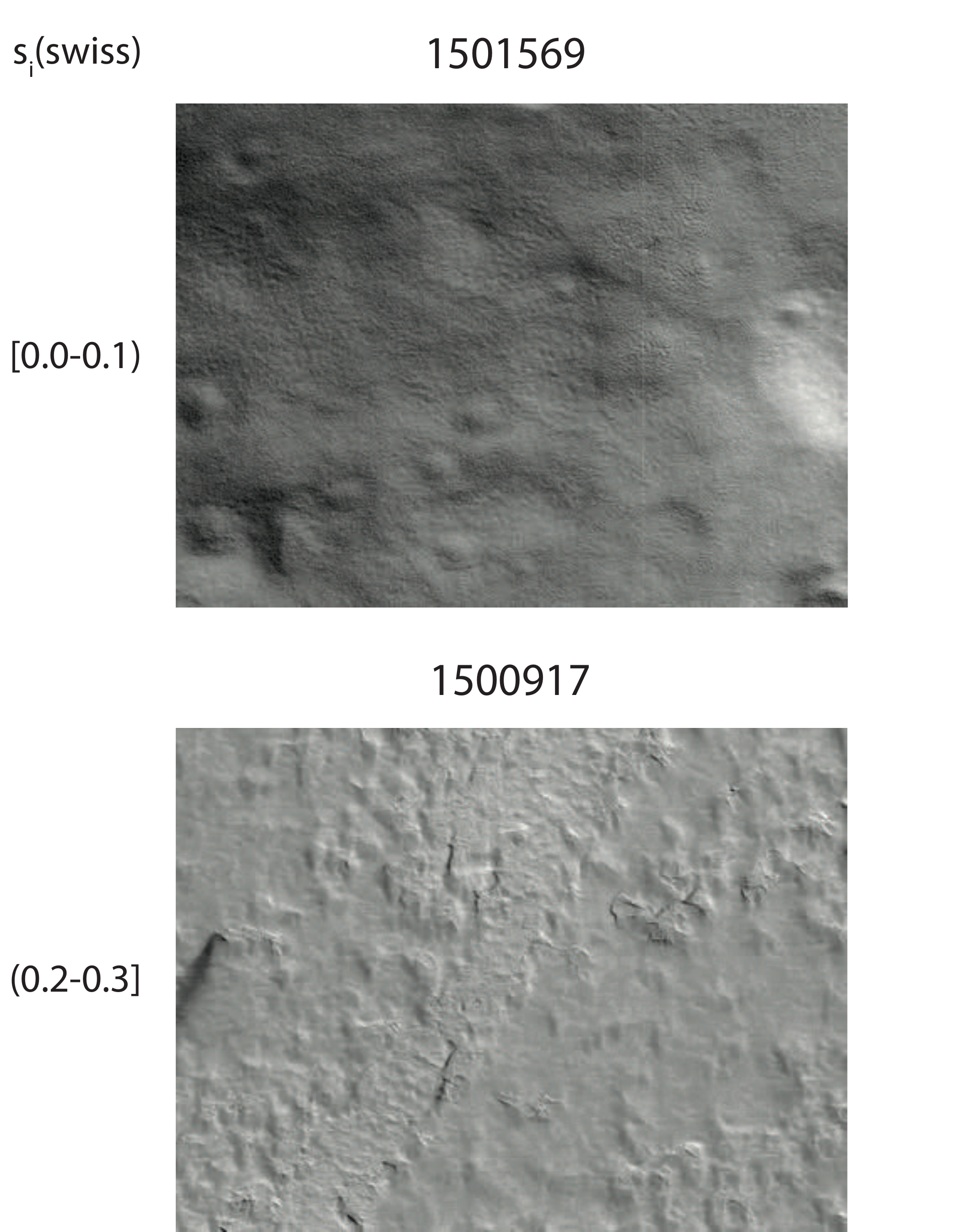}
 \caption{ \label{fig:swiss_score_ex1} Randomly selected sample of subjects per Swiss cheese score ($s_i(\rm Swiss)$) binned with bin size of 0.1. CTX images: P13$\_$006199$\_$1040 (top) and D13$\_$032311$\_$0999 (bottom). Each P4T subject image is $\sim$4.8$\times3.6$ km. }
 \end{center}
 \end{figure}

\begin{figure}
\begin{center}
\includegraphics[width=1.0\columnwidth]{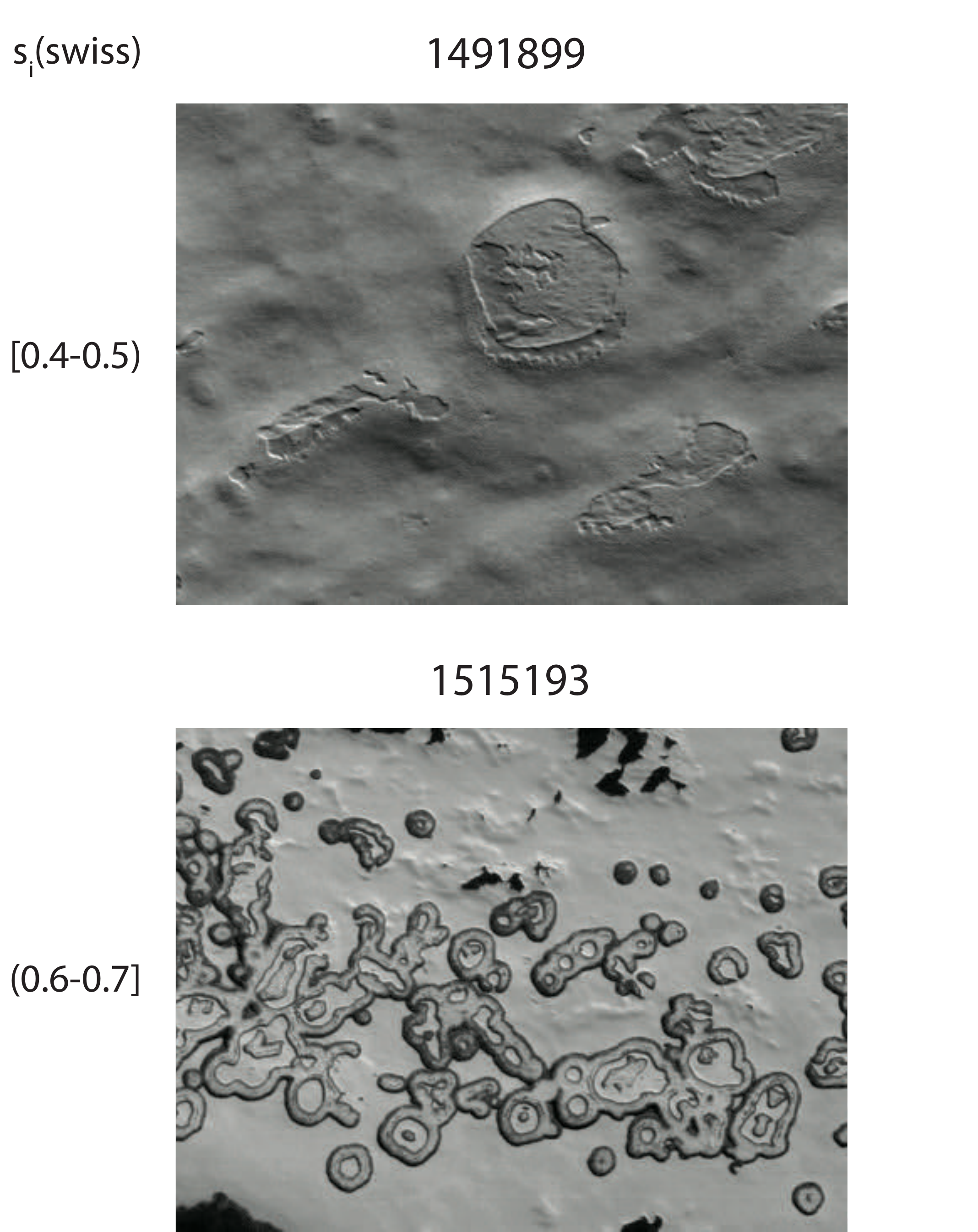}
 \caption{ \label{fig:swiss_score_ex2} Randomly selected sample of subjects per Swiss cheese score  ($s_i(\rm Swiss)$) binned with bin size of 0.1 continued. CTX images: G14$\_$023524$\_$0999 (top) and P13$\_$006234$\_$1008(bottom). Each P4T subject image is $\sim$4.8$\times3.6$ km. }
 \end{center}
 \end{figure}

\begin{figure}
\begin{center}
\includegraphics[width=1.0\columnwidth]{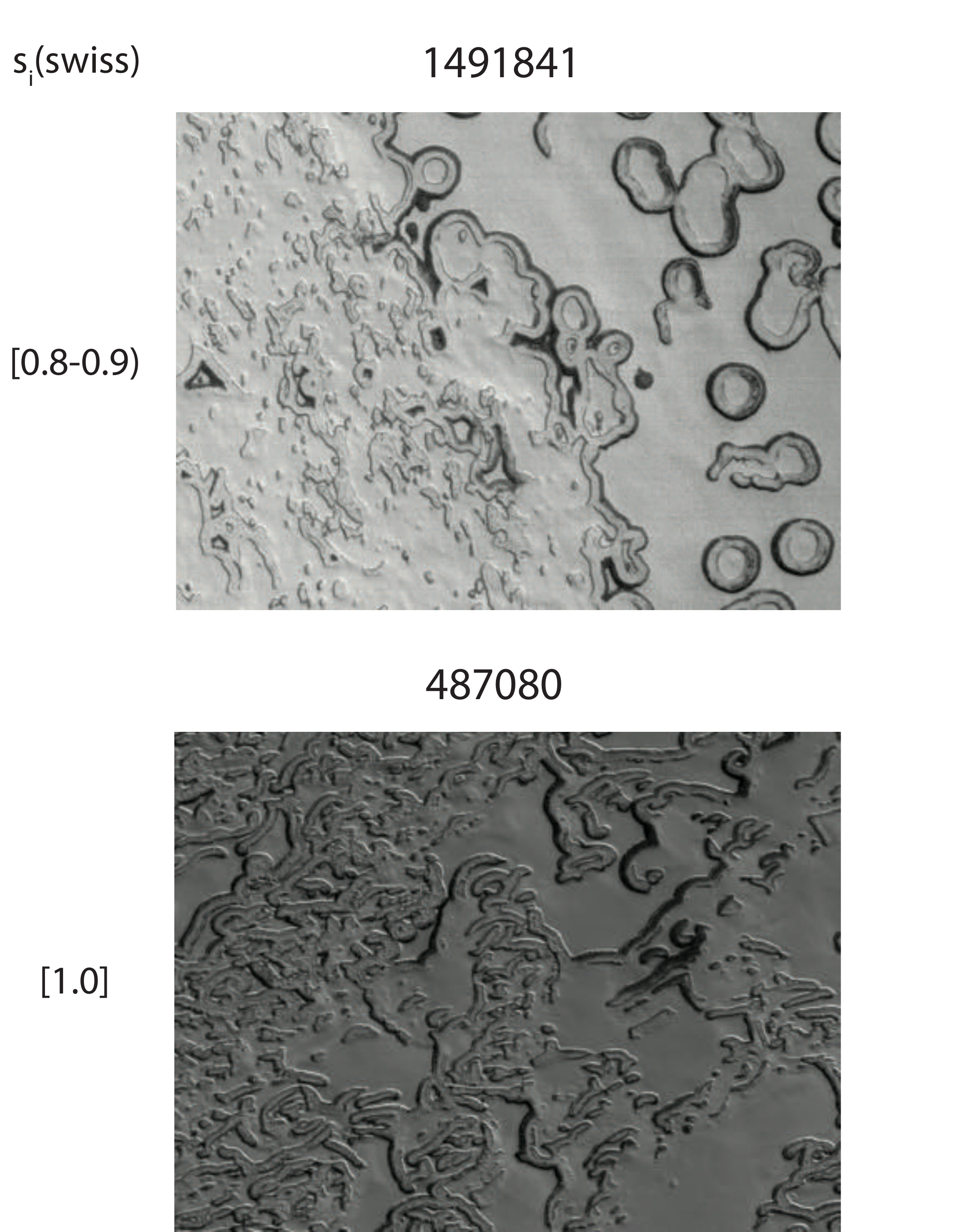}
 \caption{ \label{fig:swiss_score_ex3} Randomly selected sample of subjects per Swiss cheese score  ($s_i(\rm Swiss)$) binned with bin size of 0.1 continued. CTX images: G14$\_$023835$\_$0906(top) and P13$\_$006229$\_$0951 (bottom). Each P4T subject image is $\sim$4.8$\times3.6$ km.}
 \end{center}
 \end{figure}

We then use the final subject scores, $s_i(\rm spider)$ and $s_i(\rm Swiss)$, to identify the locations of spiders and Swiss cheese terrain in the surveyed CTX images. Figure \ref{fig:scores} plots the cumulative distribution for the final calculated scores. Table \ref{tab:scores_bin} reports the binned distribution in subject scores, and  Table \ref{tab:scores} provides the final score values for each subject used in this study. Figures \ref{fig:spider_score_ex1}-\ref{fig:spider_score_ex3} contain a representative sample of subjects for $s_i(\rm spider)$ randomly selected in bins of 0.1. Figures \ref{fig:swiss_score_ex1}-\ref{fig:swiss_score_ex3} show the same for $s_i(\rm Swiss)$. It is readily apparent the closer the subject score is to 1, the more consensus amongst the weighted user vote and thus a higher likelihood of the features of interest (spiders or Swiss cheese terrain) being present in  a given subject image. Of the 20,122 subjects classified by P4T volunteers, 3$\%$ (591) have $s_i(\rm spider)$ $>$ 0.5 and 9$\%$ (1767) have $s_i(\rm Swiss)$ $>$ 0.5. 

We set a detection threshold for $s_i(\rm spider)$ and $s_i(\rm Swiss)$ above which we define a clean sample of subjects identified as having spiders or Swiss cheese terrain present. We set this value such that the number of false positives is sufficiently small while retaining the largest number of true identifications. We determine this detection limit based on the expert assessment by the P4T science team for a  small fraction of the subject data used in this study. A similar validation process has been applied to crowd-sourced crater counting \citep{2014Icar..234..109R, 2016Icar..271...30B}. A subset comprised of 1,009 subjects (505 for Swiss cheese terrain identification and 504 for spider identification),  corresponding in total to 5$\%$ of the total subjects reviewed by volunteers, were used to create the gold standard dataset. The subjects were divided into ten bins based on $s_i(\rm spider)$ and $s_i(\rm Swiss)$. The gold standard subjects were randomly selected from these bins  in order to find the subject score where false positives begin to overwhelm positive identifications of spiders and Swiss cheese terrain. Table \ref{tab:scores_bin} details the number of subjects per $s_i(\rm spider)$ and $s_i(\rm Swiss)$ bins randomly selected for the gold standard review. The P4T subject ids used in the expert  gold standard review are provided in Supplemental Tables 2 and 3. 

Two members of the science team (MES and GP) each independently reviewed the gold standard subjects using the same web interface on the P4T website as used by the volunteers.  The gold standard assessments are available in Supplemental Table 4. Figure \ref{fig:gold_standard} plots for the gold standard dataset, the fraction of positive identification of spider and Swiss cheese detections as a function of score for each expert reviewer. The error bars represent the Poissonian 68$\%$ confidence limit on the positive identifications in each bin, as prescribed by \cite{1991ApJ...374..344K}.   There is strong agreement in expert assessments for Swiss cheese terrain, but less agreement for spider assessments. This may be due to differing levels of difficulty between the two identification tasks. 

\subsection{Clean Spider and Swiss Cheese Terrain Sample}

We set our detection threshold for $s_i(\rm spider)$ and $s_i(\rm Swiss)$ at the score bin where 20$\%$ of the gold standard data subjects reviewed are false positives as  deemed by both expert reviewers. Therefore, we select our `clean' sample of spider detections as subjects with $s_i(\rm spider) \ge 0.6 $, and our `clean' Swiss cheese sample is comprised of subjects with $s_i(\rm Swiss) \ge 0.7 $. Applying these cuts, 390 (1.9$\%$) of the P4T subjects have $s_i(\rm spider) \ge$ 0.6, and 1,537 (7.6$\%$) have  $s_i(\rm Swiss) \ge 0.7 $.  We use the clean samples in the rest of our analysis.

The clean spider and clean swiss cheese identifications do not represent a complete sample in our search region. Instead these are high confidence identifications where false positives in the sample are minimized. Thus a location within the searched CTX images that is not part of our clean samples, may still have spider-shaped araneiforms or Swiss cheese terrain present that are  not easily identified by P4T which includes cases where the surface may not be as visible in the CTX image (see Section \ref{sec:dataset}). Thus our analysis only focuses on what we can glean from a sample of confident identifications with approximately less than 20$\%$ false positives. We provide a catalog of the subjects and  associated properties that comprise the clean spider and Swiss cheese samples in Tables \ref{tab:clean spiders} and \ref{tab:clean swiss_cheese} respectively. 

\FloatBarrier
\begin{threeparttable}[h]
\small
\caption{P4T Subject Information and $s_i{(\textrm spider)}$ and $s_i{(\textrm Swiss)}$ Scores \label{tab:scores} }
 \footnotesize
\begin{tabular}{llllllll}
\hline
\hline
 &  & &  & Center & Center &   &    \\
Subject ID & $s_i{(\textrm spider)}$ & $s_i{(\textrm Swiss)}$ & Parent CTX Image &  Latitude &  Longitude & CTX X$^*$  &  CTX  Y$^*$  \\
 &  & &  & (degrees) & (degrees) &  position &   position \\
\hline
83672 &	0.025 &	0.000	 &	G14$\_$023507$\_$1029$\_$XN$\_$77S158W	 &	-77.80	 &	200.96  & 400  & 300  \\
483673 &	0.074 &	0.000 &	G14$\_$023507$\_$1029$\_$XN$\_$77S158W	 &	-77.74 &	200.90 & 400 & 900 \\
483674 & 0.000 &	0.000 &	G14$\_$023507$\_$1029$\_$XN$\_$77S158W	 &	-77.68 &	200.84  & 400  &  1500 \\
483675 &	0.122 &	0.016 &	G14$\_$023507$\_$1029$\_$XN$\_$77S158W	 &	-77.62 &	200.78  & 400 & 2100  \\
483676 &	0.167 &	0.000 &	G14$\_$023507$\_$1029$\_$XN$\_$77S158W	 &	-77.56 &	200.72  &  400 & 2700 \\
483677 &	0.131 &	0.000 &	G14$\_$023507$\_$1029$\_$XN$\_$77S158W	 &	-77.50 &	200.66  & 400 & 3300 \\
483678 &	0.237 &	0.015 &	G14$\_$023507$\_$1029$\_$XN$\_$77S158W	 &	-77.45 &	200.60  &  400 &  3900 \\
483679 &	0.066 &	0.057 &	G14$\_$023507$\_$1029$\_$XN$\_$77S158W	 &	-77.39 &	200.54  & 400 & 4500  \\
483680 &	0.069 &	0.015 &	G14$\_$023507$\_$1029$\_$XN$\_$77S158W	 &	-77.33 &	200.48  & 400  & 5100 \\
\hline
\hline
\end{tabular}
\begin{tablenotes}
\item[$^*$] CTX X and CTX Y are the pixel location of the subject's center in the full frame CTX ISIS generated cube. 
\item[] This table in its entirety can be found in the online Supplemental and at \url{https://www.zooniverse.org/projects/mschwamb/planet-four-terrains/about/results}  A portion is shown here for guidance regarding its form and content.  We include the full CTX filename here; the first 15 characters are the unique CTX observation identifier. 
\end{tablenotes}
\end{threeparttable}
\FloatBarrier

\FloatBarrier
\begin{table}
\caption{ \label{tab:scores_bin} Distribution of P4T Scores for All Subjects and Subjects Used in the Gold Standard Review }
\begin{tabular}{l|rr|rr}
\hline
\hline
bins & $s_i{(\textrm spider)}$ & $s_i{(\textrm Swiss)}$ & gold spider & gold Swiss \\
\hline
0 $\le s_i{()} < $ 0.1 & 16301 & 15885 & 30 & 28 \\
0.1 $\le s_i{()} < $ 0.2 & 1846 & 1533 & 40  & 38 \\
0.2 $\le s_i{()} < $ 0.3 & 713 & 483 & 50  & 50 \\
0.3 $\le s_i{()} < $ 0.4 & 383  & 290 & 60 & 60 \\
0.4 $\le s_i{()} < $ 0.5 & 288 & 164 & 60  & 60\\
0.5 $\le s_i{()} < $ 0.6 & 201 & 135 & 60  & 59 \\
0.6 $\le s_i{()} < $ 0.7 & 150 & 95 & 60 & 60 \\
0.7 $\le s_i{()} < $ 0.8 & 117 & 114 & 60 & 60  \\
0.8 $\le s_i{()} < $ 0.9 & 88 & 353 & 49 & 50 \\
0.9 $\le s_i{()} \le $ 1.0 & 35 & 1070 & 35 & 40 \\ 
\cline{1-5}
Total & 20122 & 20122 & 504 & 510  \\
\hline
\hline
\end{tabular}
\end{table}
\FloatBarrier

\begin{threeparttable}[h]
\caption{Clean Spider Sample \label{tab:clean spiders}}
\begin{tabular}{lllll}
\hline
\hline
Subject ID & $s_i{(\textrm spider)}$ & Parent CTX lmage & Center  Latitude & Center Longitude \\
& & & (degrees) & (degrees) \\
\hline
489994	& 1.000 &	P12$\_$005747$\_$1035$\_$XI$\_$76S195W &	        -76.76 &	165.32 \\
491505	& 1.000 &	D14$\_$032675$\_$0924$\_$XN$\_$87S253W &	        -87.55 &	110.18 \\
1045043	& 1.000 &	P13$\_$006151$\_$0974$\_$XN$\_$82S055W &        -82.13 &	303.34 \\
1320810	& 1.000 &	P12$\_$005839$\_$0994$\_$XI$\_$80S170W & 	        -81.10 &	189.69 \\
1510306	& 0.953 &	G13$\_$023338$\_$1043$\_$XI$\_$75S229W &	        -75.34 &	131.48 \\
491434	& 0.953 &	P13$\_$006173$\_$0934$\_$XN$\_$86S263W & 	-86.73 &	100.61 \\
491859	& 0.951 &	P13$\_$006148$\_$1028$\_$XN$\_$77S342W & 	-79.08 &	19.48 \\
1491749	& 0.951 &	P13$\_$006197$\_$0930$\_$XN$\_$87S187W & 	-86.91 &	172.28 \\
1058665	& 0.951 &	D14$\_$032593$\_$1037$\_$XN$\_$76S174W & 	-76.20 &	185.99 \\\hline
\hline
\end{tabular}
\begin{tablenotes}
\item[] This table in its entirety can be found in the online Supplemental and at \url{https://www.zooniverse.org/projects/mschwamb/planet-four-terrains/about/results}  A portion is shown here for guidance regarding its form and content.  We include the full CTX filename here; the first 15 characters are the unique CTX observation identifier. 

\end{tablenotes}
\end{threeparttable}

\begin{threeparttable}
\caption{Clean Swiss Cheese Terrain Sample \label{tab:clean swiss_cheese}}
\begin{tabular}{lllll}
\hline
\hline
Subject ID & $s_i{(\textrm Swiss)}$ & Parent CTX Image & Center Latitude & Center Longitude \\
& & & (degrees) & (degrees) \\
\hline
1320528	& 1.000 &	P13$\_$006005$\_$0947$\_$XN$\_$85S015W &      -86.32	  &  355.10 \\
1320615	& 1.000 &	P13$\_$006005$\_$0947$\_$XN$\_$85S015W &	     -85.39	&    345.54 \\
1320728	& 1.000 &	P13$\_$006005$\_$0947$\_$XN$\_$85S015W &	      -86.34	&    359.96 \\
1491818	& 1.000 &	G14$\_$023835$\_$0906$\_$XN$\_$89S050W &      -89.36	 &    342.33 \\
1491849	& 1.000 &	G14$\_$023835$\_$0906$\_$XN$\_$89S050W &	-89.14 &	300.68 \\
1040008	& 1.000 &	G14$\_$023833$\_$0926$\_$XN$\_$87S017W &	-87.75 &	346.66 \\
1040047	& 1.000 &	G14$\_$023833$\_$0926$\_$XN$\_$87S017W &	-87.38   &  342.37 \\
1040089	& 1.000 &	P13$\_$006123$\_$0953$\_$XN$\_$84S001W &  	-86.26 &	9.46  \\
1040272	& 1.000 &	P13$\_$006123$\_$0953$\_$XN$\_$84S001W &         -85.75 & 	5.97 \\
\hline
\hline
\end{tabular}
\begin{tablenotes}
\item[] This table in its entirety can be found in the online Supplemental and at \url{https://www.zooniverse.org/projects/mschwamb/planet-four-terrains/about/results}  A portion is shown here for guidance regarding its form and content. We include the full CTX filename here; the first 15 characters are the unique CTX observation identifier. 
\end{tablenotes}
\end{threeparttable}

   \begin{figure}
\begin{center}
\includegraphics[width=0.8\columnwidth]{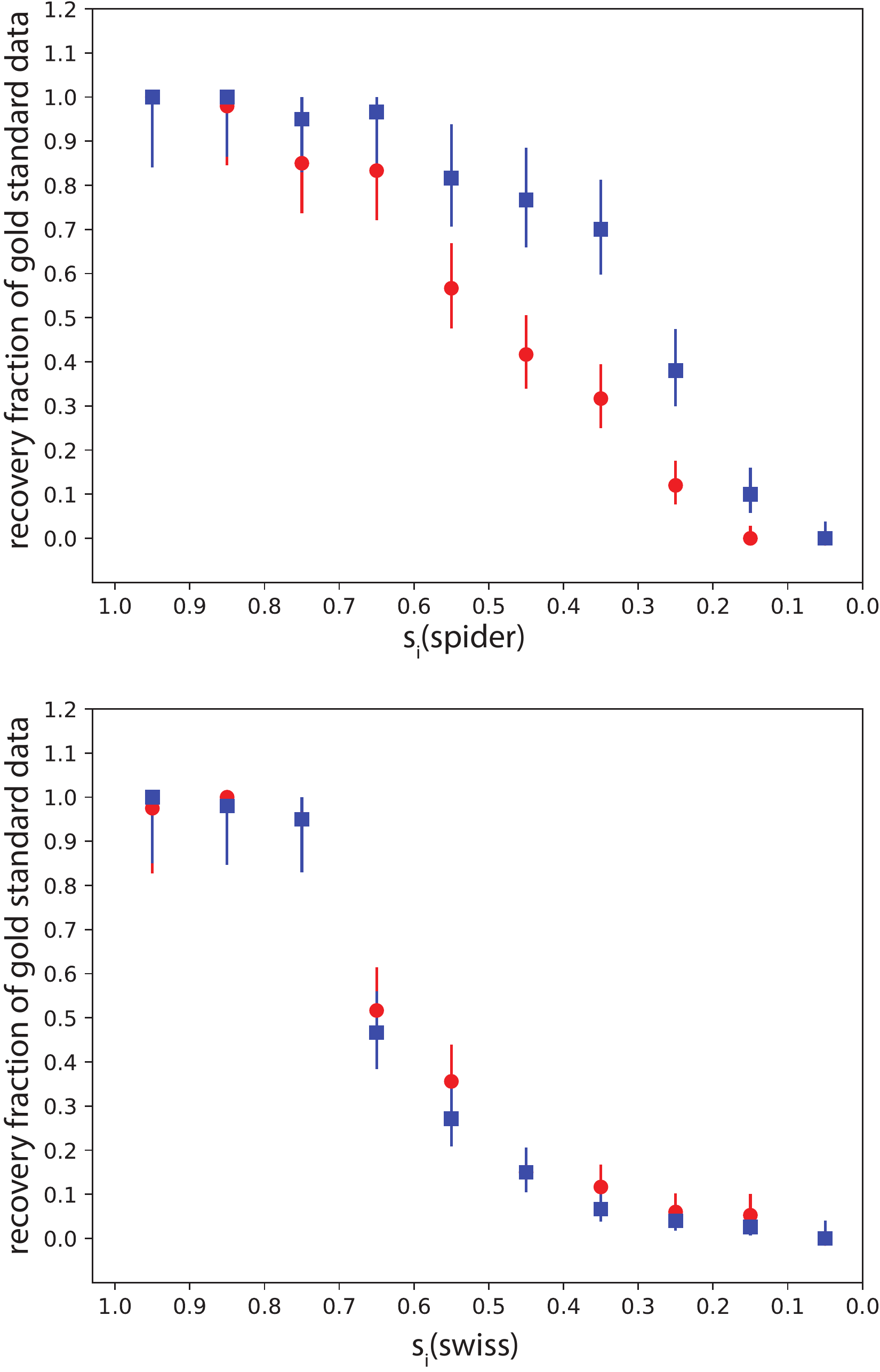}
 \caption{ \label{fig:gold_standard} Detection efficiency/recovery rate of  subjects with spider and Swiss cheese features per  subject score ($s_i(\rm spider)$ and $s_i(\rm Swiss)$) in the gold standard dataset  bin size of 0.1. The assessments from the two expert reviewers, MES and GP) are plotted separately as red circles and blue squares respectively. Error bars represent the Poissonian 1-$\sigma$ uncertainty.} 
 \end{center}
 \end{figure}
 
\section{Spider Distribution}
\label{sec:spider_distribution}
\begin{figure}
\begin{center}
\includegraphics[width=1.0\columnwidth]{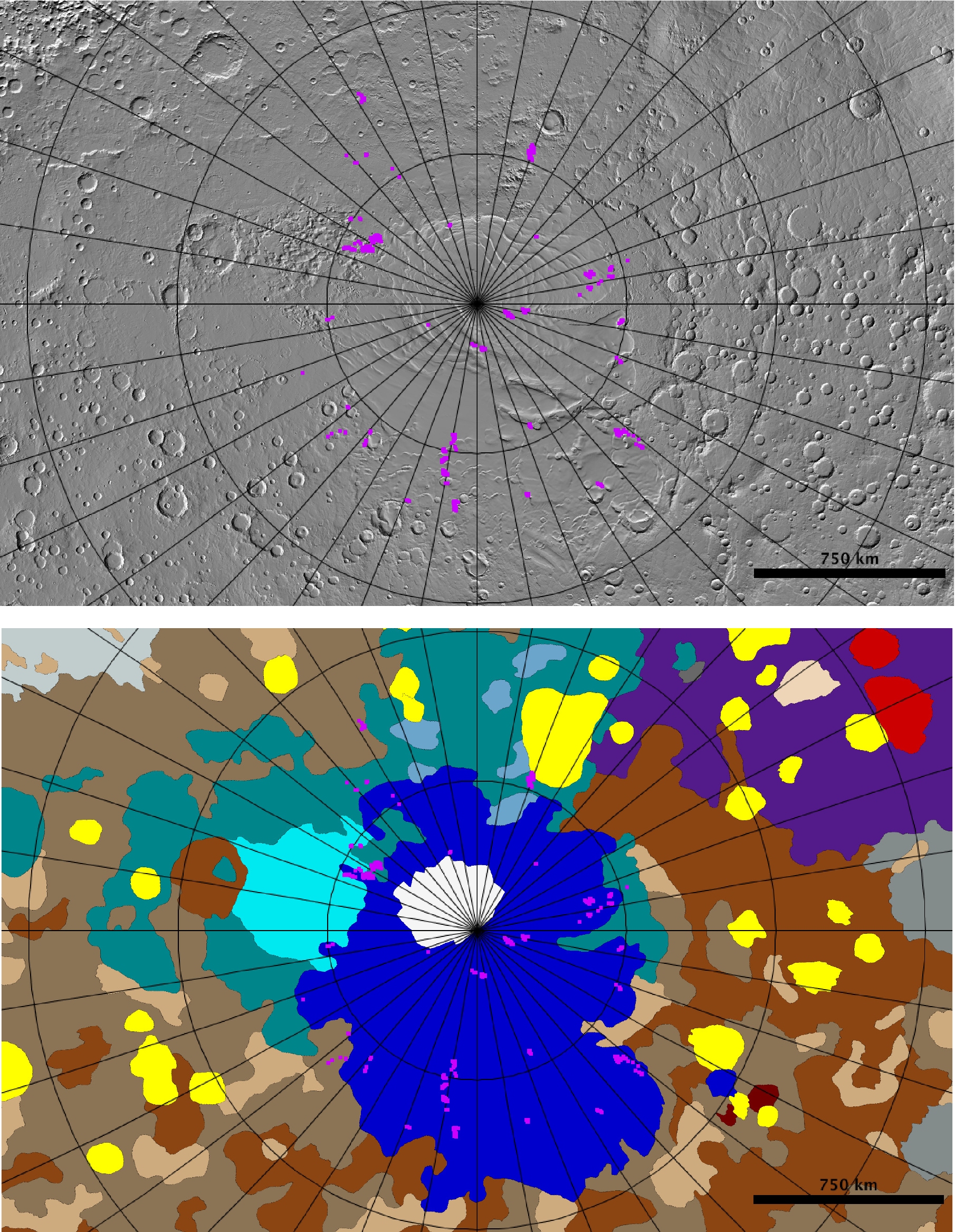}
 \caption{ \label{fig:spider_dist} Top: Clean spider distribution overlaid on a MOLA shaded elevation map \citep{1992JGR....97.7781Z,2001JGR...10623689S}  (top) and over on the geologic map from \cite{Tanaka:2014wd} (bottom).  For both plots, latitude and longitude lines are plotted every 10  degrees. A legend for the geologic map is provided in \ref{ref:appendix}.} 
 \end{center}
 \end{figure}

\begin{figure}
\begin{center}
\includegraphics[width=1.0\columnwidth]{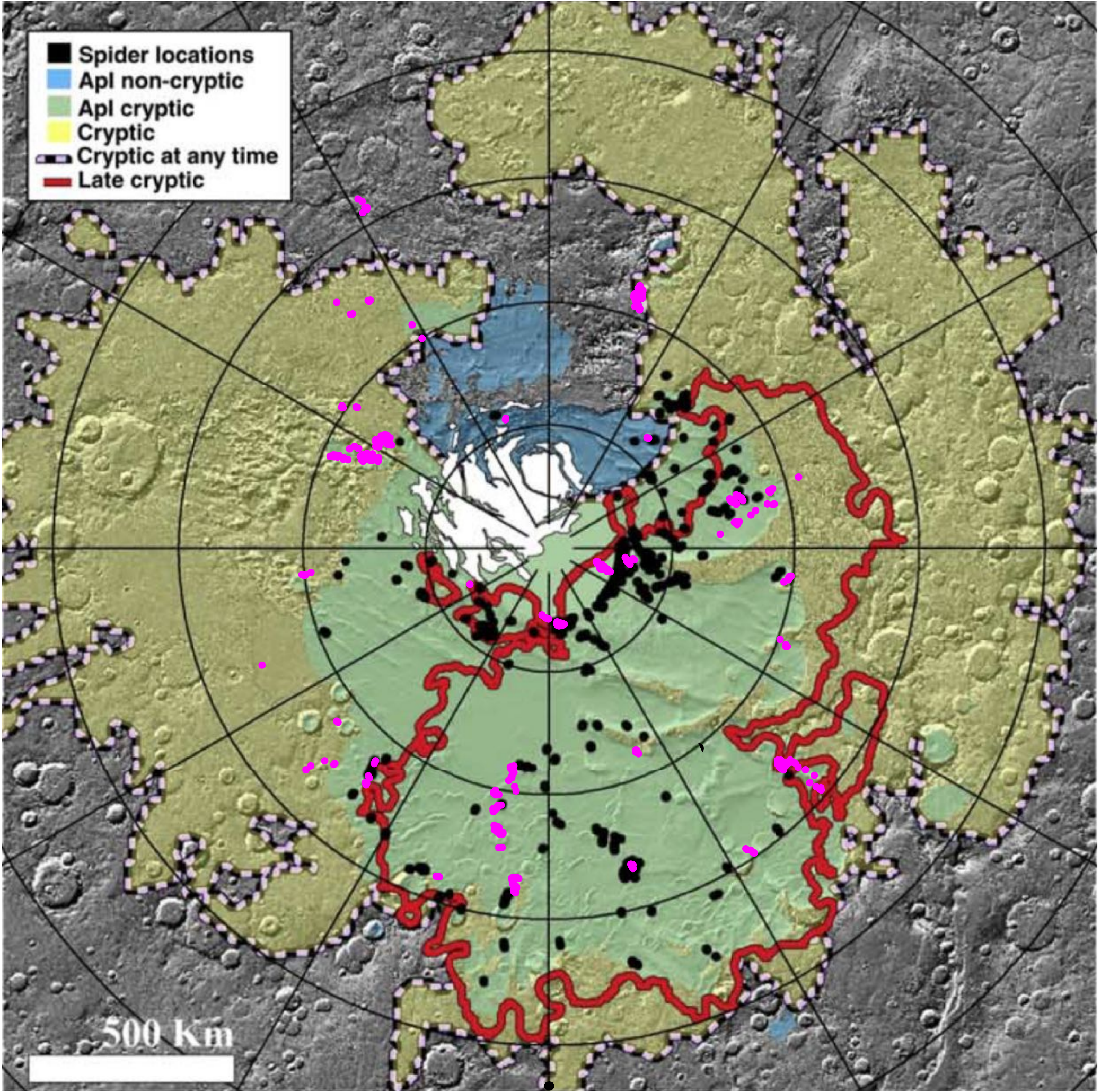}
 \caption{ \label{fig:2003spider_dist} Comparison of P4T clean spider distribution with that of \cite{2003JGRE..108.5084P}. We overlay our distribution on Figure 2a  of \cite{2003JGRE..108.5084P}. Plotted in magenta is the clean spider sample from this work.  The black points are the spider identifications from \cite{2003JGRE..108.5084P}. The black and white dashed line encloses polar areas that \cite{2003JGRE..108.5084P} found  exhibited cryptic behavior (and thus the presence of  semi-translucent slab ice CO$_2$ ) at any point between Ls = 180$^\circ$ and Ls = 250$^\circ$. The red solid line \cite{2003JGRE..108.5084P} identifies as cryptic terrain from Ls = 190$^\circ$ onward.An older definition for the SPLD  (Amazonian polar polar layered deposit [APL]) [Kolb et al., 2003] is shown in green and blue based on the albedo of the regions between Ls = 180$^\circ$ and Ls = 250$^\circ$) regions. Green represents SPLD areas that exhibited cryptic behavior, and blue are parts of the SPLD that did not exhibit cryptic characteristics as identified by  \cite{2003JGRE..108.5084P}. A visual comparison of the \cite{2003JGRE..108.5084P} spider identifications find that the locations of nearly all are within the outline of SPLD in the more modern geologic mapping by \cite{Tanaka:2014wd}. Latitude lines are plotted every 5 degrees and longitude lines are plotted every 30 degrees.} 
 \end{center}
 \end{figure}

\begin{figure}
\begin{center}
\includegraphics[width=0.8\columnwidth]{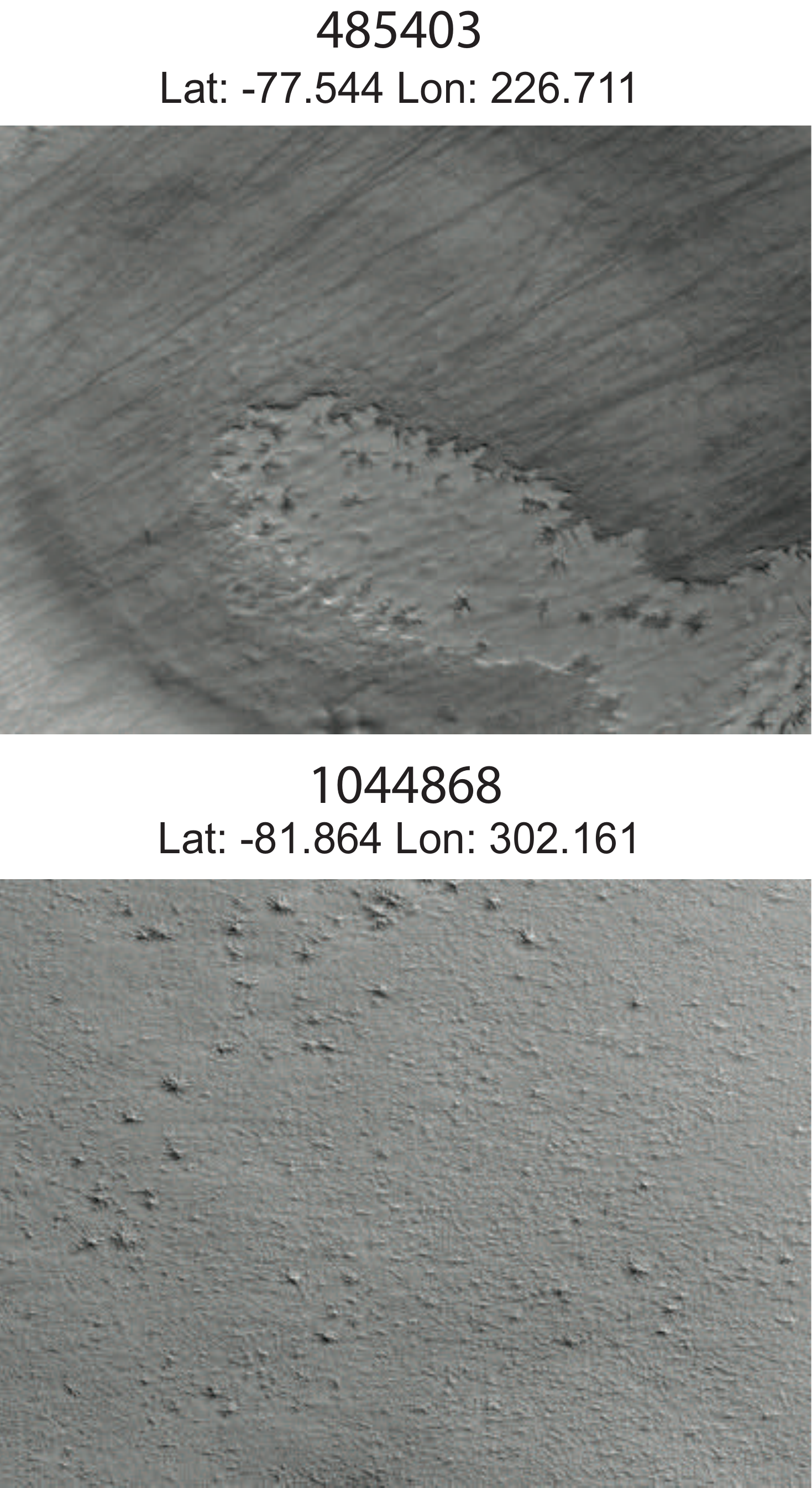}
 \caption{ \label{fig: off_spld1_1} Representative sample of each of the regions identified in the clean spider sample as outside of the SPLD unit. P4T subjects derived from CTX images P13$\_$006151$\_$0974 and D14$\_$032733$\_$1028.Each P4T subject image is $\sim$4.8$\times3.6$ km.} 
 \end{center}
 \end{figure}

\begin{figure}
\begin{center}
\includegraphics[width=0.8\columnwidth]{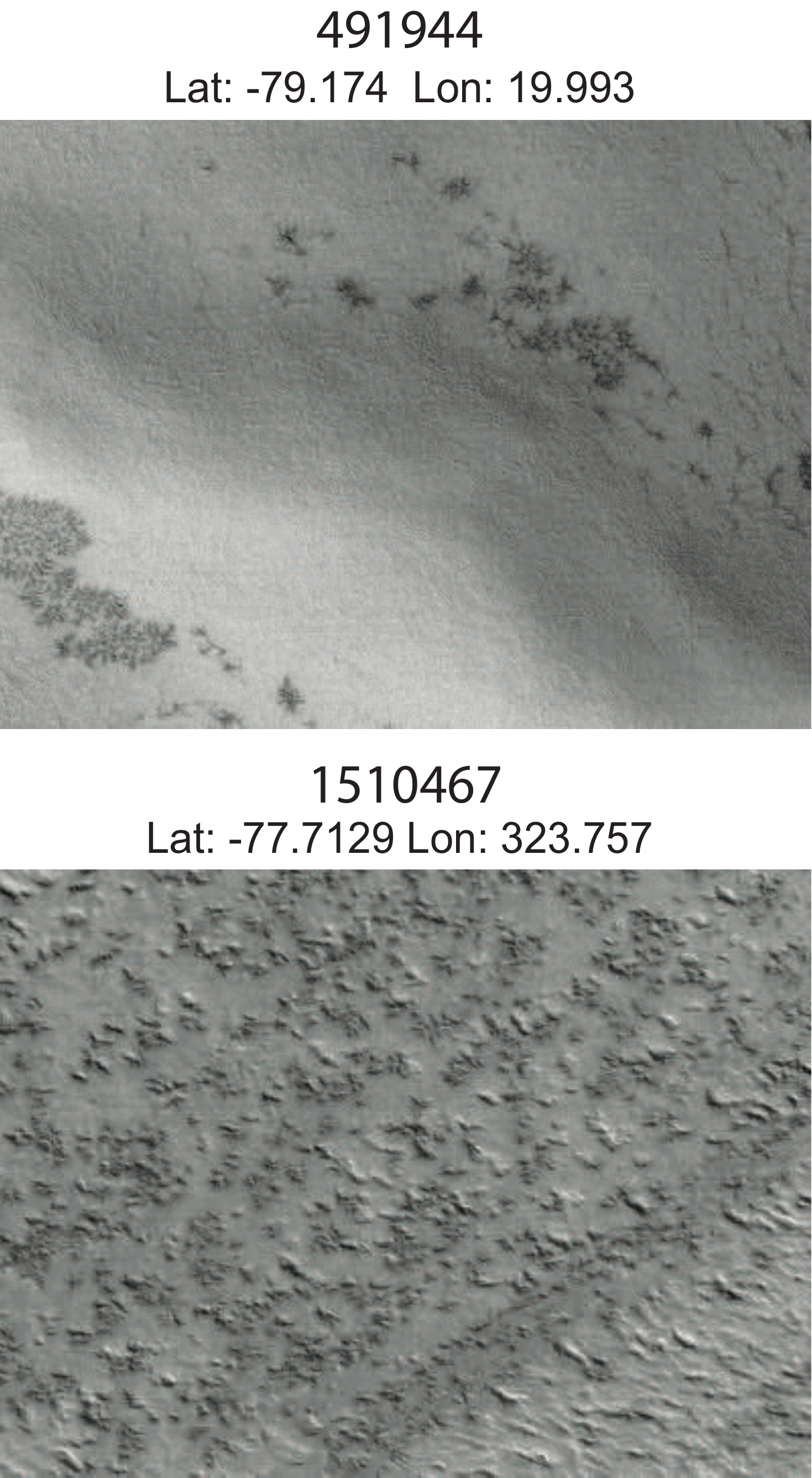}
 \caption{ \label{fig: off_spld1_2} Representative sample of each of the regions identified in the clean spider sample as outside of the SPLD unit, continued. P4T subjects derived from CTX images G14$\_$023506$\_$1036 and P13$\_$006151$\_$0974.Each P4T subject image is $\sim$4.8$\times3.6$ km.} 
 \end{center}
 \end{figure}

\begin{figure}
\begin{center}
\includegraphics[width=0.8\columnwidth]{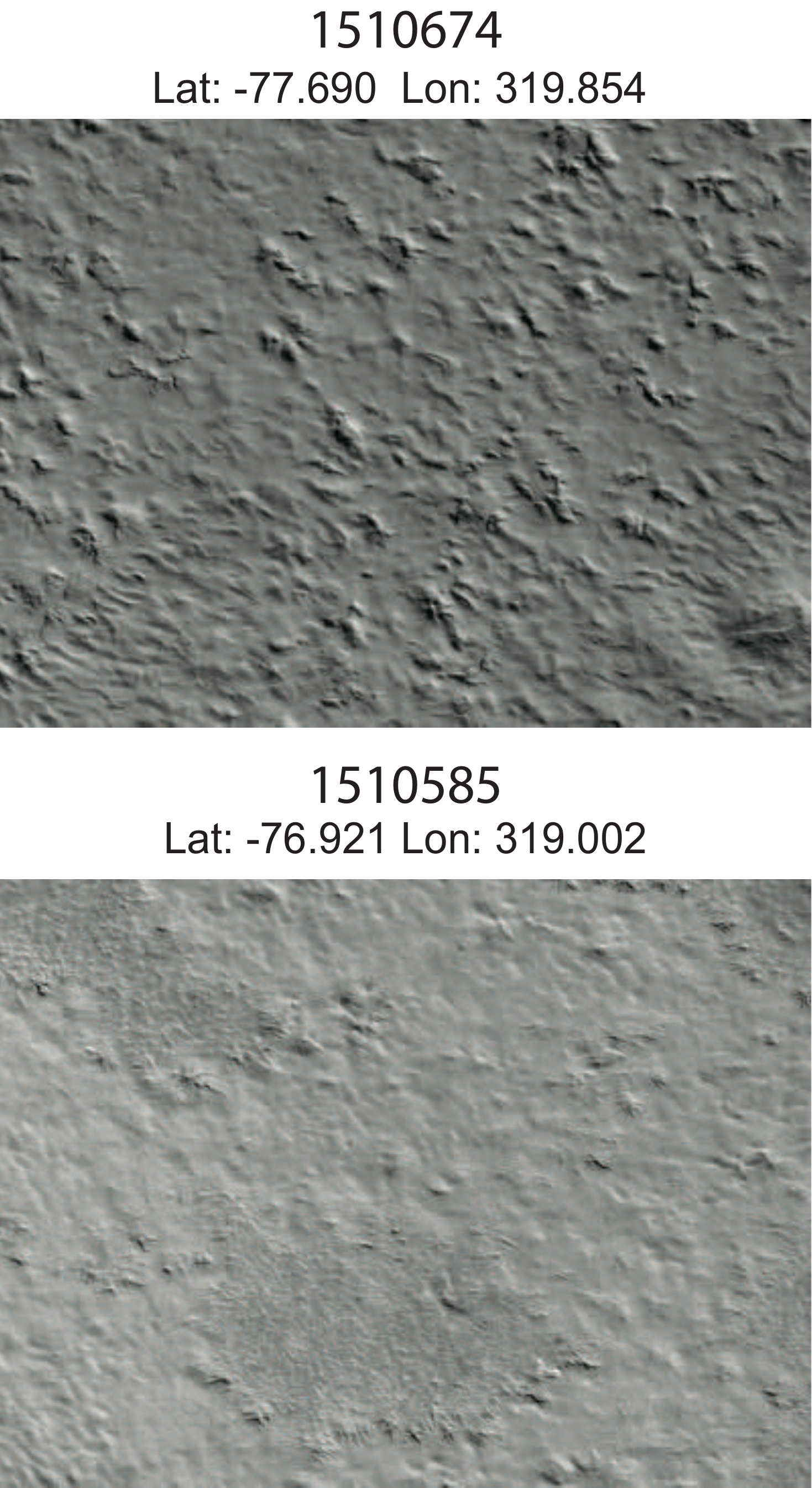}
 \caption{ \label{fig: off_spld1_3} Representative sample of each of the regions identified in the clean spider sample as outside of the SPLD unit, continued. P4T subjects derived from CTX images P13$\_$006148$\_$1028 and P13$\_$006283$\_$1003. Each P4T subject image is $\sim$4.8$\times3.6$ km.} 
 \end{center}
 \end{figure}

\begin{figure}
\begin{center}
\includegraphics[width=0.8\columnwidth]{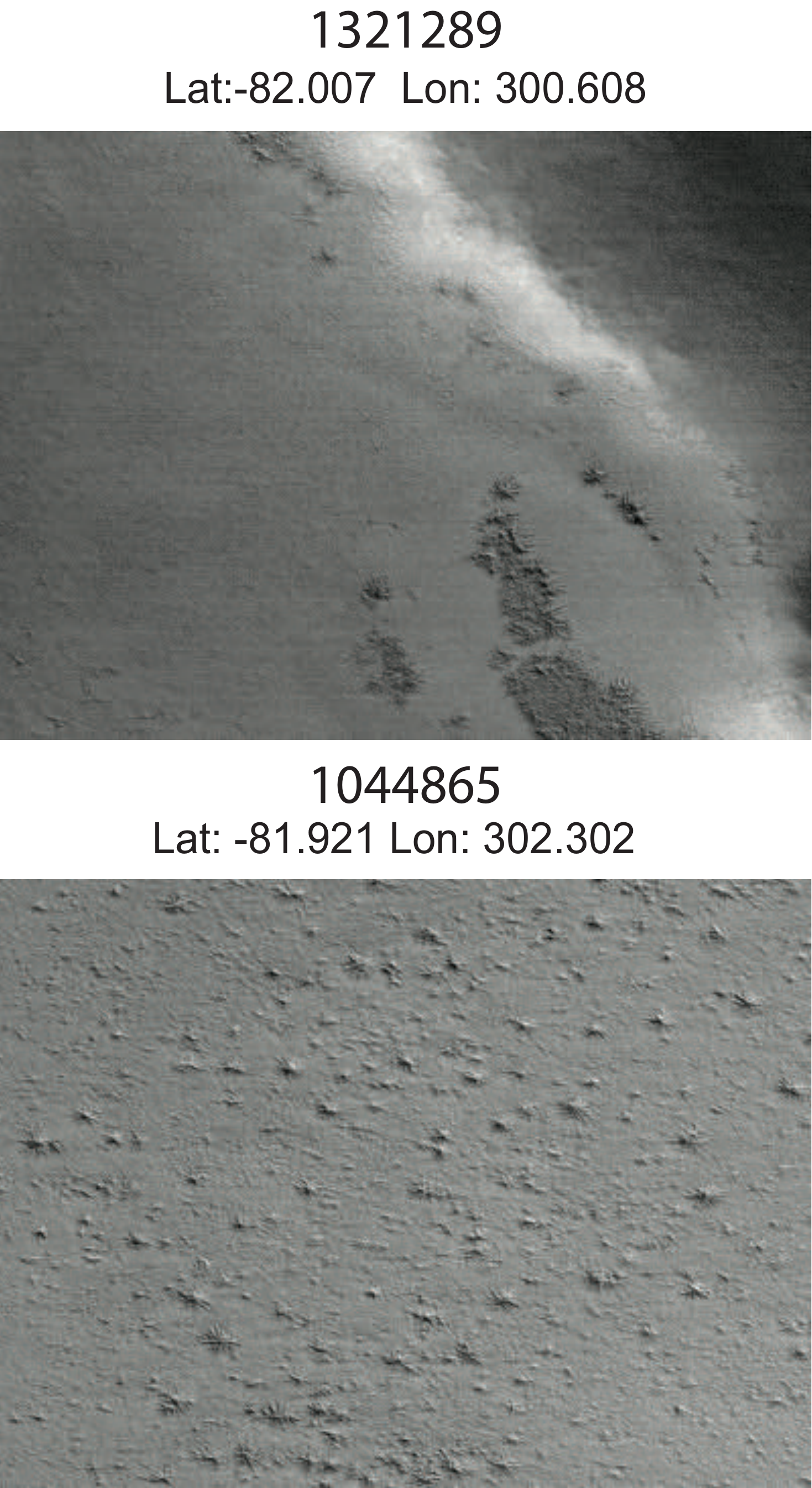}
 \caption{ \label{fig: off_spld1_4} Representative sample of each of the regions identified in the clean spider sample as outside of the SPLD unit, continued. P4T subjects derived from CTX images  P13$\_$006151$\_$0974 and P13$\_$006282$\_$1046. Each P4T subject image is $\sim$4.8$\times3.6$ km.} 
 \end{center}
 \end{figure}

\begin{figure}
\begin{center}
\includegraphics[width=0.8\columnwidth]{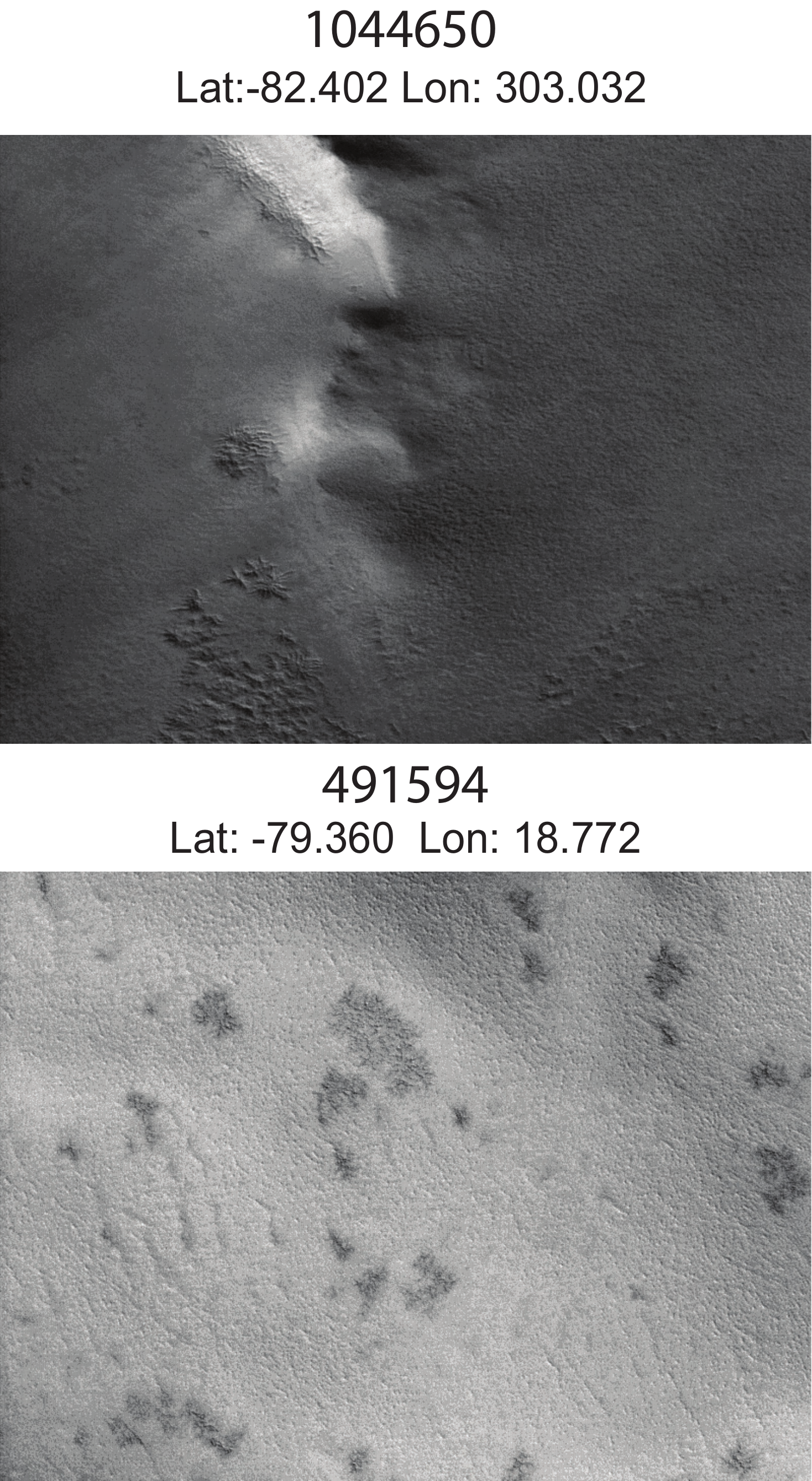}
 \caption{ \label{fig: off_spld2_1} Representative sample of each of the regions identified in the clean spider sample as outside of the SPLD unit, continued. Planet Four: Terrains subjects derived from CTX images P13$\_$006151$\_$0974 and  P13$\_$006148$\_$1028 .Each P4T subject image is $\sim$4.8$\times3.6$ km.} 
 \end{center}
 \end{figure}

\begin{figure}
\begin{center}
\includegraphics[width=0.8\columnwidth]{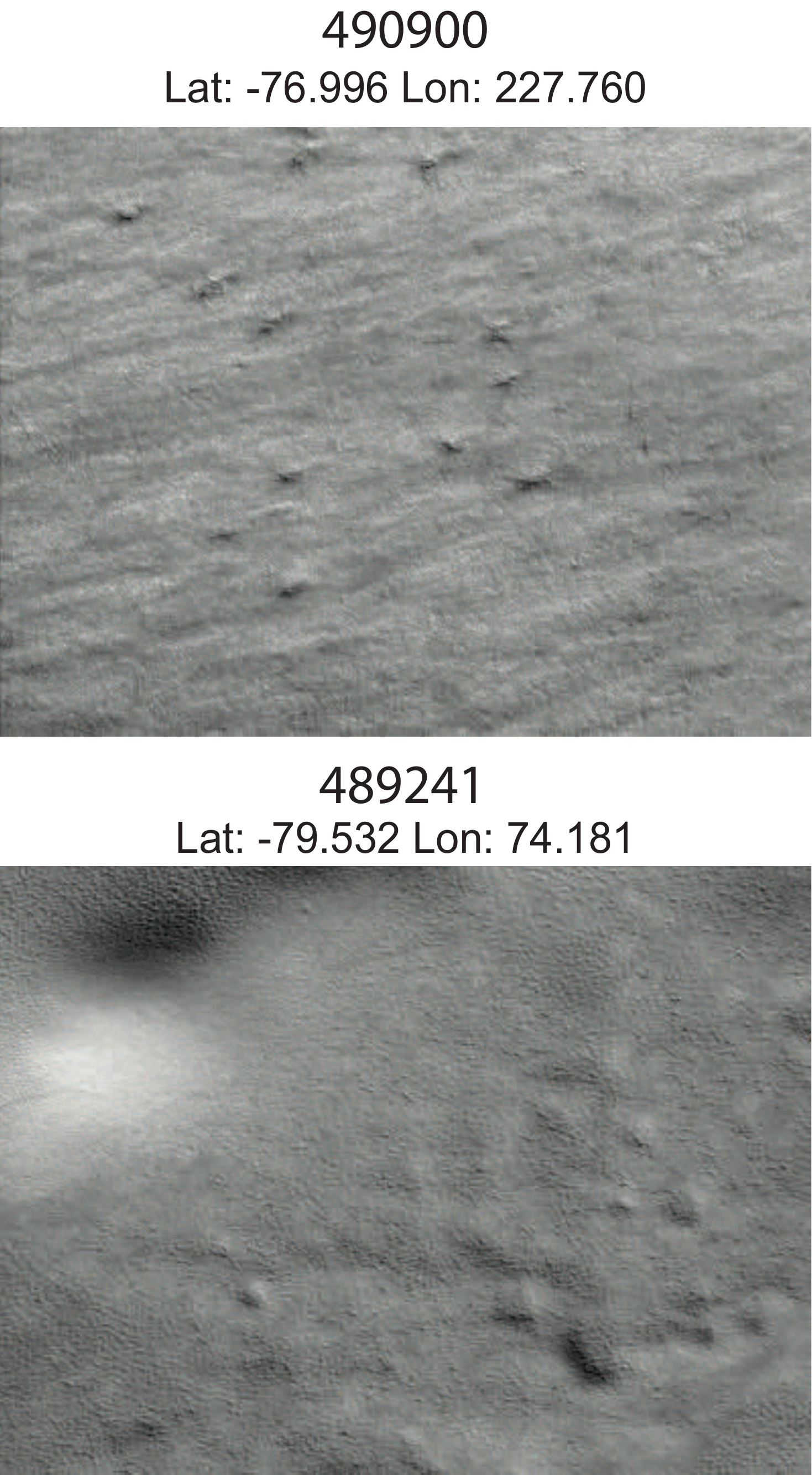}
 \caption{ \label{fig: off_spld2_2} Representative sample of each of the regions identified in the clean spider sample as outside of the SPLD unit, continued. Planet Four: Terrains subjects derived from CTX images  P13$\_$006148$\_$1028 and G14$\_$023634$\_$103. Each P4T subject image is $\sim$4.8$\times3.6$ km.} 
 \end{center}
 \end{figure}

\begin{figure}
\begin{center}
\includegraphics[width=0.8\columnwidth]{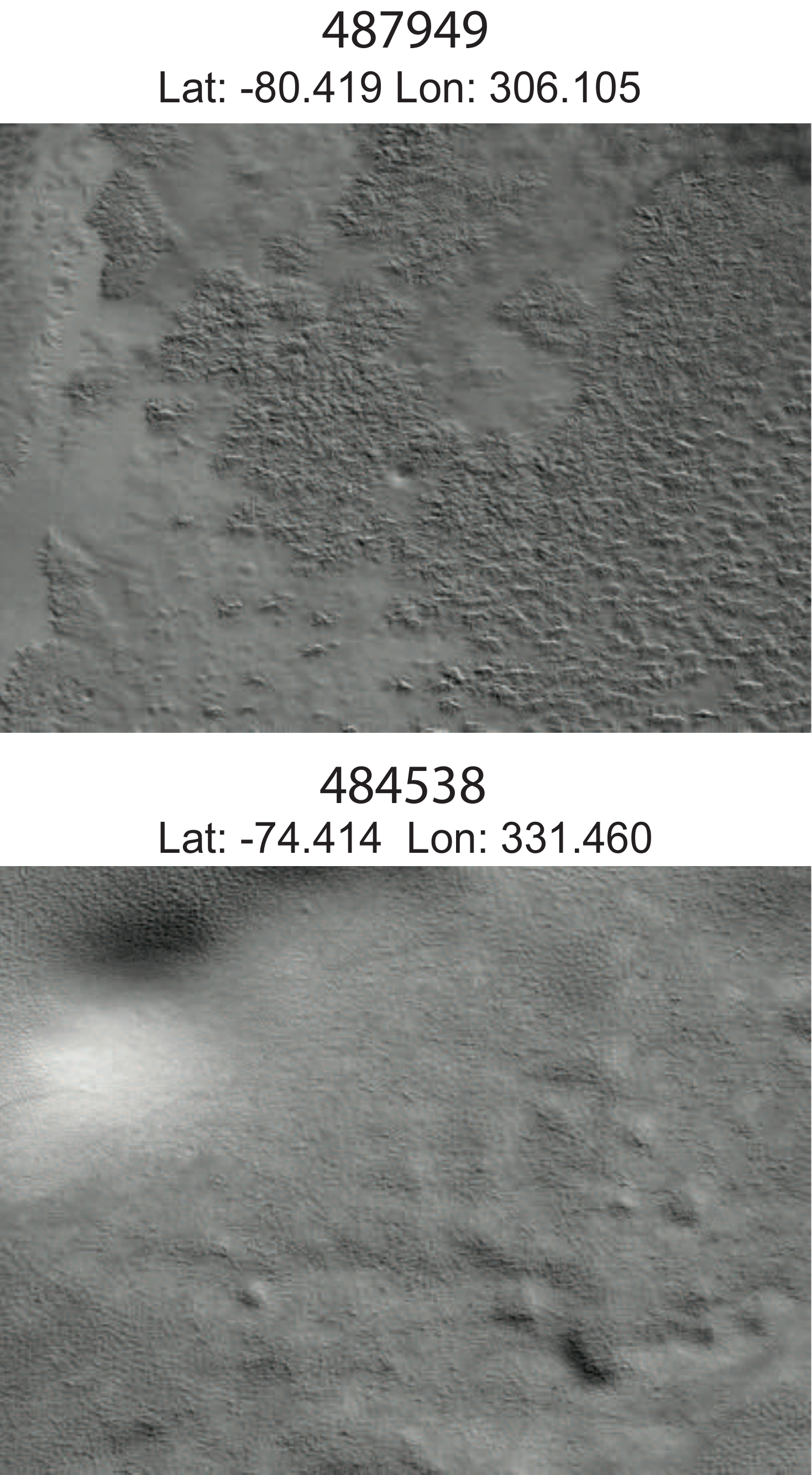}
 \caption{ \label{fig: off_spld2_3} Representative sample of each of the regions identified in the clean spider sample as outside of the SPLD unit, continued. Planet Four: Terrains subjects derived from CTX images P13$\_$006151$\_$0974 and P13$\_$006283$\_$1003. Each P4T subject image is $\sim$4.8$\times3.6$ km.} 
 \end{center}
 \end{figure}

Our clean spider sample is comprised of 390 subjects, and their locations are plotted in Figure \ref{fig:spider_dist} overlaid on a MOLA elevation map \citep{1992JGR....97.7781Z,2001JGR...10623689S} and the geologic map from \cite{Tanaka:2014wd}. Using MOC Narrow Angle observations,  \cite{2003JGRE..108.5084P} previously mapped the distribution of spiders in the South Polar region. They found a correlation between areneiforms and CO$_2$ slab ice, which is thought to be required for the CO$_2$ jet process. Additionally,  \cite{2003JGRE..108.5084P} found that areneiforms  were restricted to the SPLD. All but one grouping of their araneiform regions found in the \cite{2003JGRE..108.5084P} study were located within the definition of the SPLD at the time. The outlier region (located at  approximately latitude=-82$^{\circ}$, longitude=35$^{\circ}$) has similar surface characteristics to the SPLD, and the more modern \cite{Tanaka:2014wd} definition does include this location within the SPLD. There is no published table of \cite{2003JGRE..108.5084P}'s positive identifications, only the plotted distribution in Figure 2 from their paper. In order to compare our clean spider sample to that observed by \cite{2003JGRE..108.5084P}, in Figure \ref{fig:2003spider_dist} we overlay our clean spider distribution on top of \cite{2003JGRE..108.5084P}'s Figure 2a. Like \cite{2003JGRE..108.5084P}, we find the spiders are concentrated on the SPLD with $75\%$ of the clean sample located on the \cite{Tanaka:2014wd}  defined SPLD, but we also have 96 identifications located outside of the SPLD. We list these locations and subject details in Table \ref{tab:spiders_off_spld} and display a set of representative subject images of these regions in Figures \ref{fig: off_spld1_1}-\ref{fig: off_spld2_3}. With the gold standard classifications providing a $\sim$20$\%$ value for the false positive rate, we are confident that the majority of the subjects identified in the clean sample as outside of the SPLD are  valid. In order to verify the detection of araneiform features outside of the SPLD, we have also obtained high resolution imaging with HiRISE. We present those observations in Section \ref{sec:HiRISEobs}. With this discovery, we find for the first time that there are surfaces other than the SPLD on the Martian South Polar region that are conducive to the growth of araneiforms. 

 We further investigated why the araneiform features outside of the SPLD in our clean spider sample were not found by \cite{2003JGRE..108.5084P}. The MOC NA camera has better resolution ($\sim$1-2 m per pixel) but less coverage per image than CTX does. One would have expected these features to be more clearly visible if present in the MOC NA images surveyed by \cite{2003JGRE..108.5084P}. Examining the coverage of the MOC NA  observations used by \cite{2003JGRE..108.5084P}, we find that the locations listed in Table \ref{tab:spiders_off_spld} are either outside the search area or in regions with less than $20\%$ coverage. Thus, the likely reason why these araneiforms features were first identified by P4T is that these areas were covered in the CTX observations and not in the  \cite{2003JGRE..108.5084P} reviewed observations. Bolstering this argument is the araneiform identifications from \cite{Portyankina2005} which surveyed MOC NA observations from September 1997 to March 2004, including reviewing newer observations than what was available to  \cite{2003JGRE..108.5084P}. \cite{Portyankina2005} did not compare their araneiform identifications to the outline of the SPLD or the locations of other geologic units. Although \cite{Portyankina2005}  do not resolve araneiform identifications  to subimage positions, comparing their distribution, we find that there are several identifications outside of the SPLD (E0701468, R0902433, R0902028, R0901662, R0800241, R0904104, R0801805, R0903024, R0903639, R0701390, R0801195, M1101070, R0601143, R0903607, R0901019, R0601842, R0902403, R0903025, M1102723, M0905981, M1103774,M1000442, M1301816, and M0900454). With a visual inspection, we find many of these off SPLD MOC observations identified as containing araneiforms by \cite{Portyankina2005} are heavily covered with dark seasonal fans, making it more difficult for a positive identification. In our P4T search, the majority of the CTX observations are  ice free off the SPRC, allowing for clearer identification of spiders and other types of araneiforms. 
 
\begin{small}
\begin{longtable}{cccc}
\caption[short]{Clean Spider Sample Outside of the SPLD \label{tab:spiders_off_spld}} \\ 

\hline
\hline \\[-2ex]
 Subject ID &   Center Latitude & Center Longitude  \\
& (degrees)  & (degrees)   \\
  \hline 
\endfirsthead

\multicolumn{3}{c}{{\tablename} \thetable{} -- Continued} \\[0.5ex]
  \hline \hline \\[-2ex]
 Subject ID &   Center Latitude & Center Longitude  \\
& (degrees)  & (degrees)   \\
   \hline 
 \endhead

 \hline
  \multicolumn{3}{l}{{Continued on Next Page\ldots}} \\
\endfoot

  \\[-1.8ex] \hline \hline

\endlastfoot

1044650	&	-82.4	 &	303.03	\\
1044654	&	-82.34	&	302.88	\\
1044658	&	-82.29	&	302.73	\\
1044835	&	-82.27	&	303.19	\\
1045035	&	-82.24	&	303.64	\\
1044662	&	-82.23	&	302.58	\\
1058053	&	-82.22	&	304.11	\\
1044841	&	-82.21	&	303.03	\\
1058095	&	-82.20	  &	304.56	\\
1058054	&	-82.16	&	303.95	\\
1058096	&	-82.14	&	304.40	\\
1045043	&	-82.13	&	303.34	\\
1058055	&	-82.11	&	303.79	\\
1044851	&	-82.09	&	302.73	\\
1058139	&	-82.06	&	304.69	\\
1321289	&	-82.01	&	300.61	\\
1058140	&	-82.00	&	304.53	\\
1044676	&	-82.00	&	302.00	\\
1044860	&	-81.98	&	302.44	\\
1058099	&	-81.97	&	303.93	\\
1045055	&	-81.96	&	302.89	\\
1321290	&	-81.95	&	300.45	\\
1058141	&	-81.94	&	304.37	\\
1044681	&	-81.94	&	301.86	\\
1058058	&	-81.93	&	303.34	\\
1044865	&	-81.92	&	302.30	\\
1058100	&	-81.91	&	303.78	\\
1045059	&	-81.9	0 &	302.75	\\
1321291	&	-81.89	&	300.29	\\
1044687	&	-81.88	&	301.72	\\
1058059	&	-81.88	&	303.19	\\
1044868	&	-81.86	&	302.16	\\
1058101	&	-81.85	&	303.63	\\
1044690	&	-81.83	&	301.59	\\
1044697	&	-81.77	&	301.45	\\
1044878	&	-81.75	&	301.89	\\
1321091	&	-81.33	&	297.46	\\
1321181	&	-81.14	&	297.50	\\
487949	&	-80.42	&	306.11	\\
488331	&	-80.41	&	292.65	\\
487950	&	-80.36	&	306.00	\\
487697	&	-79.95	&	304.11	\\
487698	&	-79.89	&	304.02	\\
487785	&	-79.87	&	304.38	\\
491934	&	-79.76	&	20.89	\\
491935	&	-79.70	&	20.79	\\
491675	&	-79.7	0  &	19.62	\\
491849	&	-79.66	&	20.34	\\
491850	&	-79.6	&	20.25	\\
489241	&	-79.53	&	74.18	\\
491679	&	-79.46	&	19.28	\\
491766	&	-79.44	&	19.64	\\
491594	&	-79.36	&	18.77	\\
491768	&	-79.33	&	19.47	\\
491595	&	-79.3	&	18.69	\\
491682	&	-79.29	&	19.04	\\
492029	&	-79.27	&	20.51	\\
491769	&	-79.27	&	19.38	\\
491856	&	-79.25	&	19.73	\\
491596	&	-79.24	&	18.61	\\
491943	&	-79.23	&	20.08	\\
491683	&	-79.23	&	18.96	\\
492030	&	-79.22	&	20.42	\\
491857	&	-79.19	&	19.65	\\
491944	&	-79.18	&	19.99	\\
492031	&	-79.16	&	20.33	\\
491771	&	-79.15	&	19.22	\\
491858	&	-79.13	&	19.57	\\
491945	&	-79.12	&	19.91	\\
492032	&	-79.10 	&	20.25	\\
491859	&	-79.08	&	19.48	\\
491946	&	-79.06	&	19.82	\\
491947	&	-79.00	&	19.74	\\
492034	&	-78.98	&	20.08	\\
491040	&	-78.98	&	230.66	\\
491774	&	-78.97	&	18.98	\\
491127	&	-78.96	&	231.00	\\
491948	&	-78.94	&	19.66	\\
491863	&	-78.84	&	19.16	\\
485155	&	-77.77	&	225.05	\\
1510467	&	-77.71	&	323.76	\\
485156	&	-77.71	&	225.02	\\
1510416	&	-77.7	&	324.06	\\
1510674	&	-77.69	&	319.85	\\
1510468	&	-77.65	&	323.69	\\
1510690	&	-77.65	&	319.48	\\
485403	&	-77.54	&	226.71	\\
490900	&	-77.00	&	227.76	\\
1510585	&	-76.92	&	319.00	\\
490991	&	-76.74	&	227.82	\\
484275	&	-74.57	&	330.81	\\
484536	&	-74.53	&	331.55	\\
484624	&	-74.46	&	331.75	\\
484538	&	-74.41	&	331.46	\\
484629	&	-74.16	&	331.53	\\
484545	&	-74.00	&	331.17	\\
\end{longtable}
\end{small}

\section{Comparison to the Swiss Cheese Distribution}

\begin{figure}
\begin{center}
\includegraphics[width=0.9\columnwidth]{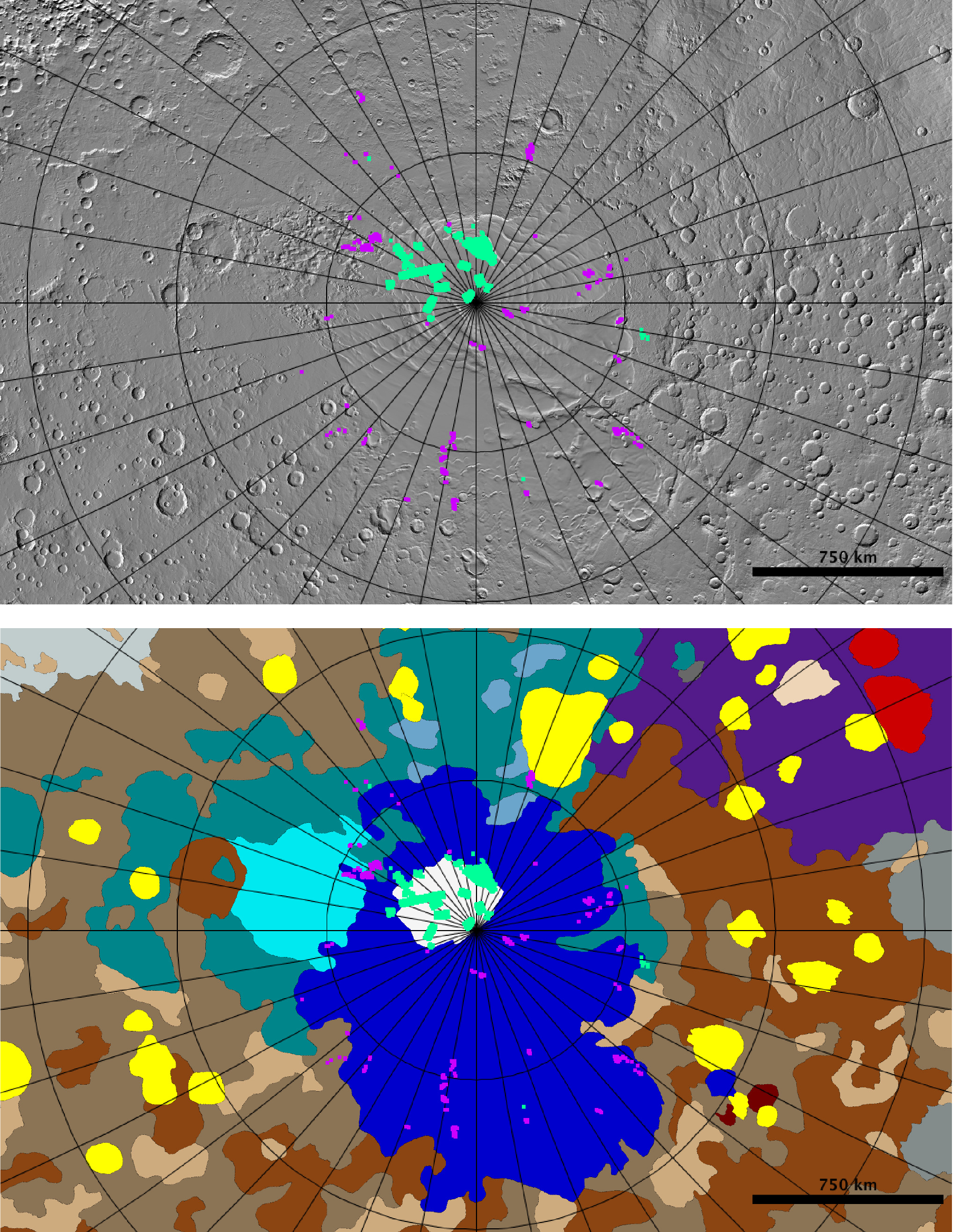}
 \caption{ \label{fig:pft_both} Distribution of the clean spider sample and clean Swiss cheese sample on the Martian South Polar region overlaid on a MOLA shaded elevation map \citep{1992JGR....97.7781Z,2001JGR...10623689S}  (top) and over on the geologic map from \cite{Tanaka:2014wd} (bottom).  For both plots, latitude and longitude lines are plotted every 10  degrees. Swiss cheese terrain are identified as the mint green square markers. Spiders are identified as magenta squares. The zero meridian is pointing straight up. A legend for the geologic map is provided in \ref{ref:appendix}. } 
 \end{center}
 \end{figure}

In Figure \ref{fig:pft_both}, we compare the locations of  our clean spider sample to our clean Swiss cheese terrain distribution. There is a clear separation between the locations of the spiders and Swiss cheese terrain in our clean samples. As expected, Swiss cheese terrain is concentrated on the SPRC, the only location on the Martian South polar region that has exposed carbon dioxide ice present at all times of the Martian year. Of the 1,537 subjects identified as containing Swiss cheese terrain, only 10 are located outside of the SPRC region latitudes northward of -80$^\circ$ (subject identifiers: 1070921, 1046056, 1058252, 1058338, 1058430, 1058511, 1058604, 1510352, 485480, and 1510512). In our clean Swiss cheese and spider samples. we find no subjects identified as containing both Swiss cheese terrain and spiders. Our coverage of the South Polar region is moderate with 11$\%$ surface area covered, but this result is consistent with the CO$_2$ slab ice hypothesis for the production of araneiforms, as the locations identified fall in regions that have been observed to darken in albedo at some point during the thawing of the seasonal CO$_2$ ice sheet. \cite{2017Icar..286...69B} have recently identified  dark fans emanating from exclusively on the sides of mesas in the SPRC. Given the restriction locations of seasonal fan activity mesa walls, this may suggest that araneiforms should not form on the SPRC.  

\begin{figure}
\begin{center}
\includegraphics[width=1.0\columnwidth]{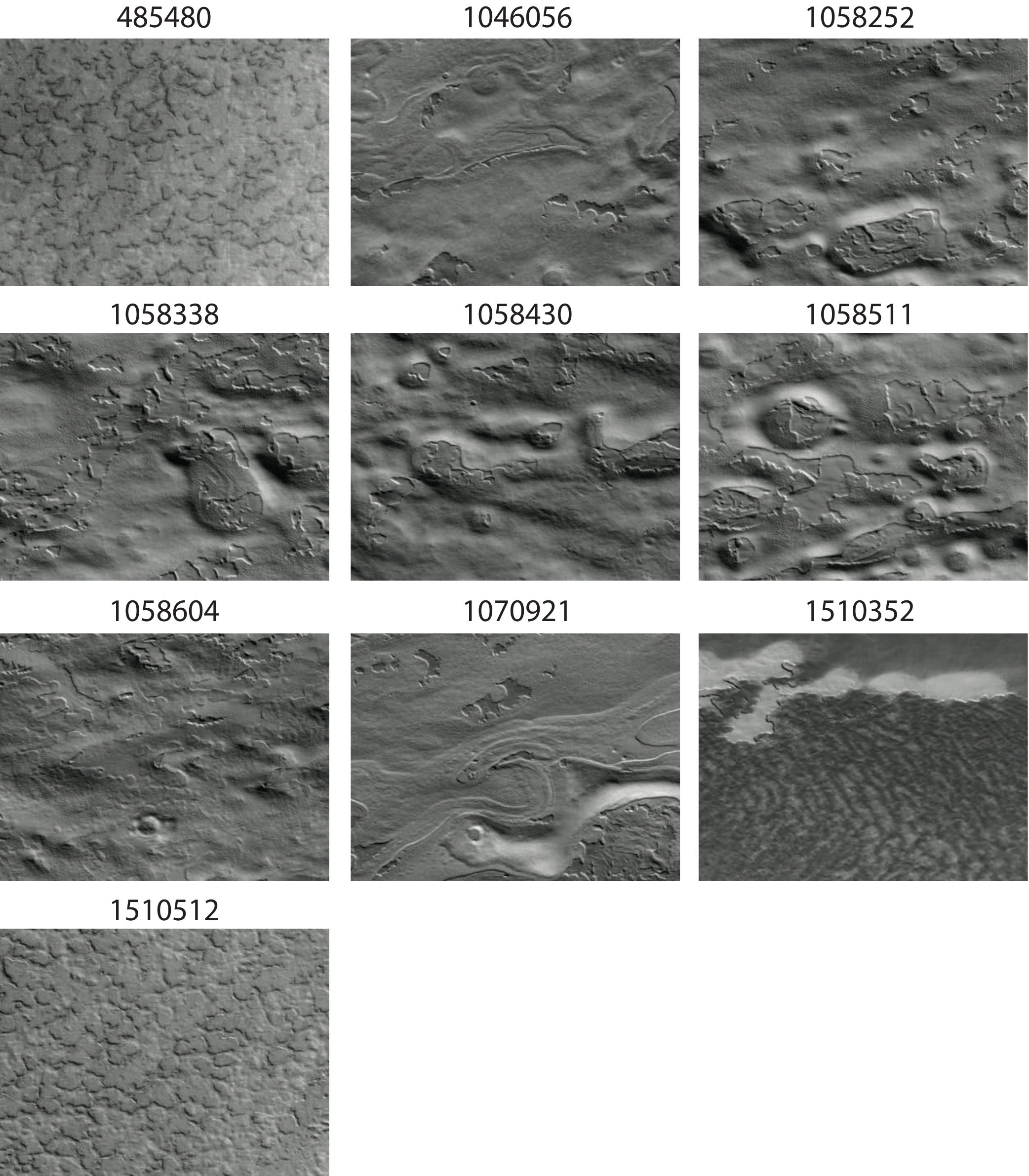}
 \caption{ \label{fig:outside_swiss} The subjects comprising the Swiss cheese clean sample outside of the SPRC northward of -80$^\circ$ N. Expert review would not classify these images as Swiss cheese, features in these images have some morphologically similarities to Swiss cheese. These 10 images are subimages of CTX frames: G13$\_$023432$\_$1026, P13$\_$006290$\_$1017, P13$\_$005958$\_$1030, G14$\_$023687$\_$1031, and D14$\_$032733$\_$1028. Each P4T subject image is $\sim$4.8$\times3.6$ km. } 
 \end{center}
 \end{figure}

The dissected mesas, linear troughs, moats, and quasi-circular pits characteristic of the SPRC form where there is carbon dioxide ice that persists through the southern summer \citep[e.g.][]{2005Icar..174..535T}. The 10 subject images outside of the SPRC  boundary represent 0.65$\%$ of our clean Swiss cheese sample. Thus, our results are generally consistent with the previously mapped extent of the SPRC \citep[e.g.][]{2016Icar..268..118T}, but indicate some newly discovered regions where carbon dioxide ice likely persists through the southern summer. Of the 10 deviant subjects from the clean Swiss cheese sample lying outside of the SPRC area (Figure  \ref{fig:outside_swiss}), 6 are clustered around 78.487$^{\circ}$ and 101.61$^{\circ}$ E (CTX image: P13$\_$006290$\_$1017), in the periglacially-deformed thin mantling deposit topping the Hesperian polar unit  \citep{Tanaka:2014wd}.

\section{HiRISE Observations and Interpretations of Candidate Araneiform Locations Outside of the SPLD}
\label{sec:HiRISEobs}
\begin{figure}
\begin{center}
\includegraphics[width=1.0\columnwidth]{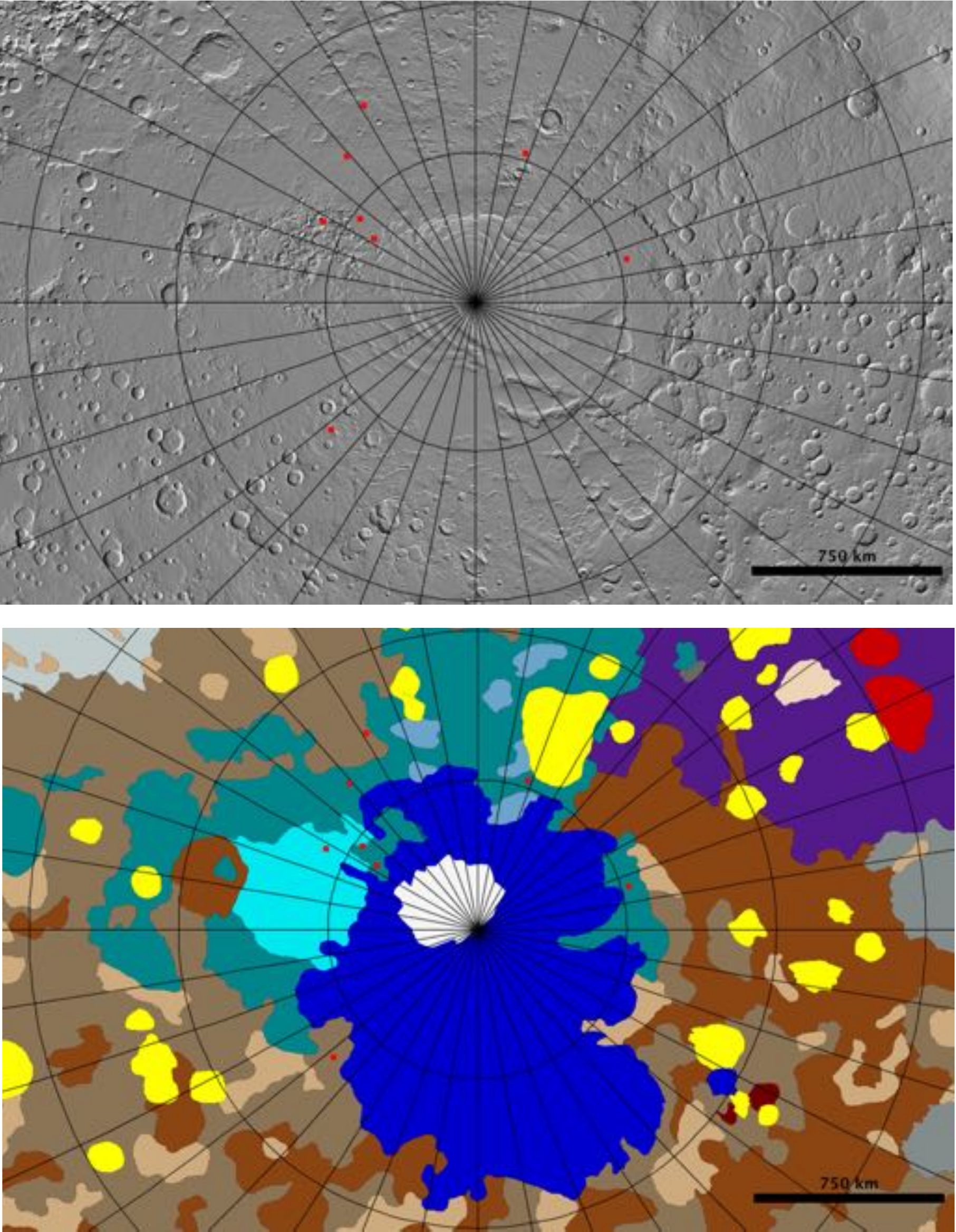}
 \caption{ \label{fig:hirise_targets} Follow-up HiRISE imaging target locations are shown as the red squares overlaid on top of the MOLA shaded elevation map \citep{1992JGR....97.7781Z,2001JGR...10623689S}  (top) and  geologic map (bottom) from \cite{Tanaka:2014wd}. The plot marker is not to scale of the HiRISE image area. For both plots, latitude and longitude lines are plotted every 10  degrees. The zero meridian is pointing straight up. The zero meridian is pointing straight up. A legend for the geologic map is provided in \ref{ref:appendix}. } 
 \end{center}
 \end{figure}

To confirm the araneiform identifications outside of the SPLD (presented in Section \ref{sec:spider_distribution}), HiRISE targeted and imaged near these locations in the Southern spring and summer of Mars Year 33. Additionally, we visually inspected all observations outside of the SPLD with  $s_i{(\textrm spider)}$  greater than 0.5, and the most promising candidates were also added to the HiRISE target database during this time period. With 12-18 times finer spatial resolution than CTX, HiRISE can reveal detail not visible in the original CTX images including whether or not fans are present at the same locations of the candidate spider channels. We present these HiRISE observations obtained to date and discuss each region below. Table \ref{tab:HiRISE_targets}  lists the HiRISE targets imaged so far, the relevant observations, and the informal names given to each HiRISE target. Figure \ref{fig:hirise_targets} plots the HiRISE pointings overlaid on a MOLA elevation map \citep{1992JGR....97.7781Z,2001JGR...10623689S}  and the geologic map from \cite{Tanaka:2014wd}.  Overall, seasonal fans were visible in the early spring observations at these locations, confirming  CO$_2$  jet activity consistent with other araneiform locations on the Martian South Polar region \citep{2010Icar..205..283H}. Although these observations are not ice free,  dendritic-like channels with similar morphologies to previous HiRISE observations of araneiforms  \citep[e.g][]{2010Icar..205..283H} are visible.  Thus, we positively identify araneiforms outside of the SPLD.  In this Section, we briefly summarize the characteristics of the araneiform morphologies, if present at each HiRISE target location and present our interpretations based on the high resolution imagery. 

\begin{table}
\caption{Candidate Araneiform Locations Outside of the SPLD Targeted by HiRISE \label{tab:HiRISE_targets}}
\begin{tabular}{cccc}
\hline
\hline
Latitude &  Longitude & Informal Name &  HiRISE Image IDs \\
(degrees) & (degrees) & &  \\
\hline
-74.99 & 330.98  &  Hempstead &  ESP$\_$046948$\_$1050 \\
-77.07 & 319.00 & Bethpage &  ESP$\_$046421$\_$1030/ESP$\_$046777$\_$1030$/$ESP$\_$048056$\_$1030  \\
-77.15 & 227.76 & Stony Brook & ESP$\_$046398$\_$1030  \\ 
 -78.54 & 298.10 &  Shoram & ESP$\_$046712$\_$1015  \\
-79.46  & 	74.34 &  Montauk &  ESP$\_$046562$\_$1005  \\ 
-79.49 & 18.77 &  Hauppauge & ESP$\_$046564$\_$1005  \\ 
-80.53 & 306.10 & Dix Hills & ESP$\_$046448$\_$0995$/$ESP$\_$046870$\_$0995/ESP$\_$047938$\_$0995 \\
 -82.02 & 302.31 & Patchogue & ESP$\_$046567$\_$0980 \\
\hline
\hline
 \end{tabular}
\end{table}

\subsection{Boulders and Fans: Hempstead and Shoram}

\begin{figure}
\begin{center}
\includegraphics[width=0.5\columnwidth]{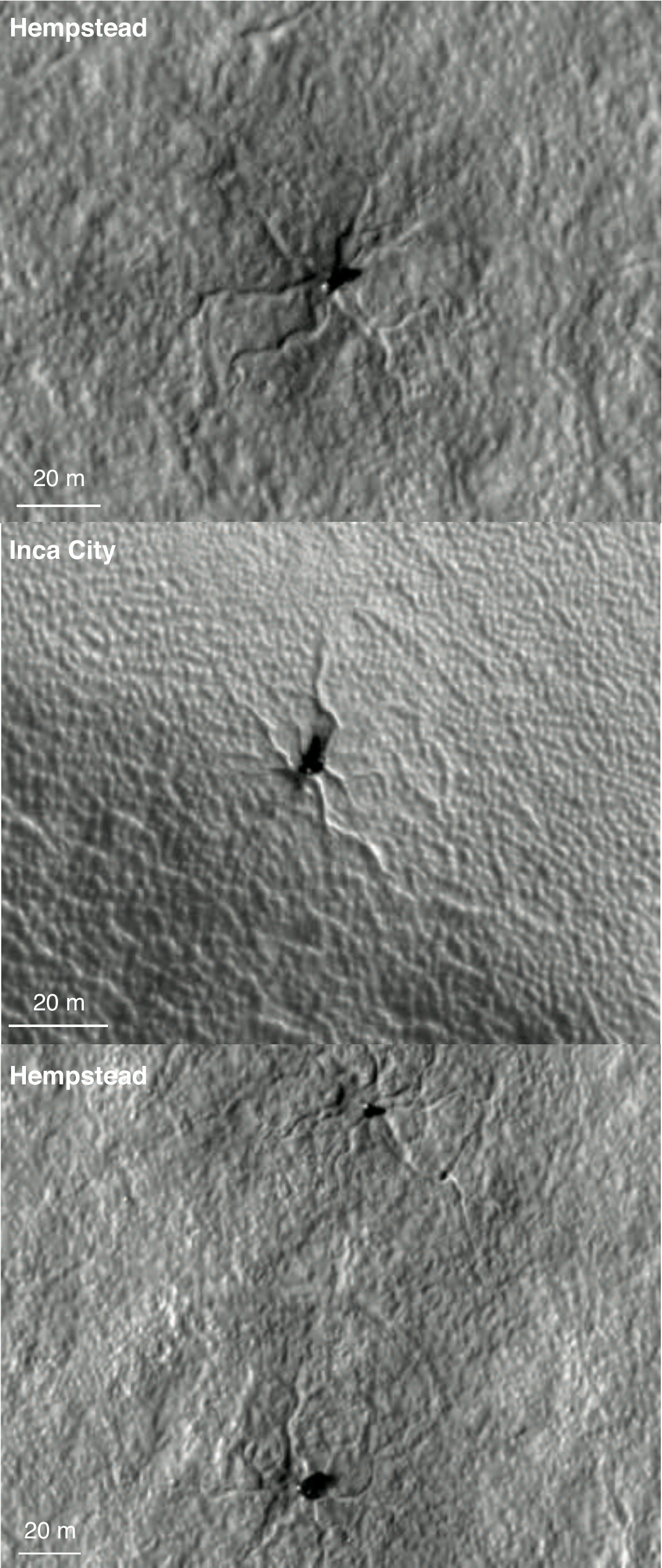}
 \caption{ \label{fig:Hempstead} Hempstead spiders (ESP$\_$046948$\_$1050;L$_{\rm s}$=196.2$^\circ$) compared to an Inca City spider (ESP$\_$029741$\_$0985;L$_{\rm s}$=216.2$^\circ$).  At the center of each pit is a boulder in all three cutouts. Sinuous channels characteristic of araneiform morphology converge at each boulder/pit location in the cutouts. All three cutouts  also show a dark seasonal fan emerging from the center originating at the boulder location indicative of active CO$_2$ outgassing at these locations.} 
 \end{center}
 \end{figure}

\begin{figure}
\begin{center}
\includegraphics[width=1.0\columnwidth]{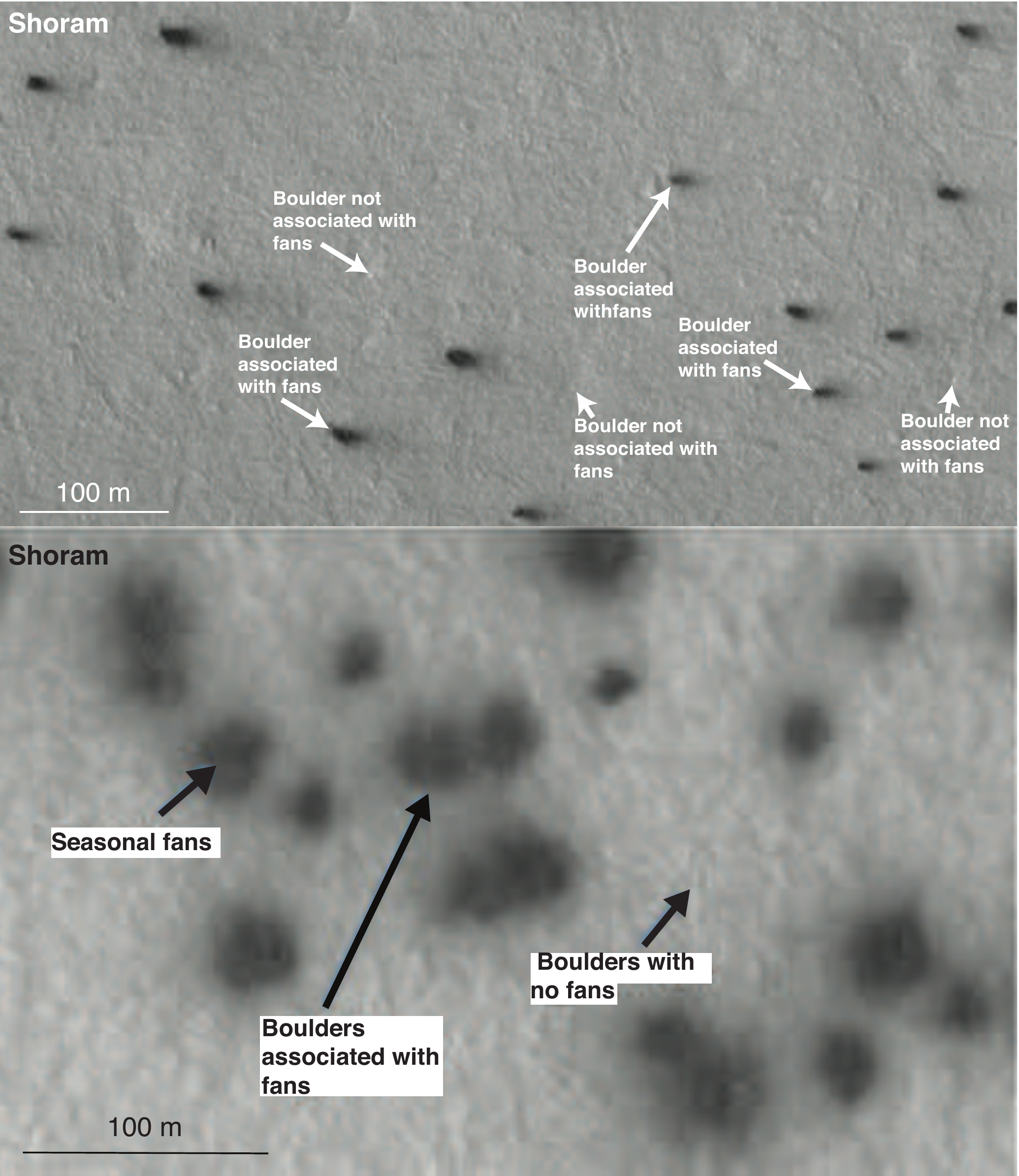}
 \caption{ \label{fig:shoram_boulders}  Boulders associated with seasonal activity in Shoram are shown in the top panel, a cutout from HiRISE image ESP$\_$046712$\_$1015, acquired at L$_{\rm s}$=185.5$^\circ$.  The fine particles deposited on the seasonal ice layer may land in a fan-shaped deposit if the ambient winds are blowing in a particular direction, or just in a directionless blotch if not.  Both panels show that seasonal activity at this t L$_{\rm s}$ is associated with the locations of some boulders but not all.}
 \end{center}
 \end{figure}

At -74.81$^{\circ}$  latitude Hempstead is the furthest site from the Martian South Polar region that is outside of the SPLD. Spiders are primarily centered around boulders, the  bright points at the center of the spiders. The region informally nicknamed `Inca City' (latitude = -81.3$^{\circ}$ N, longitude= 295.7$^{\circ}$) was previously the only known location with araneiforms and boulders. Fans emanate from the boulders in Hempstead, similar to Inca City. See Figure \ref{fig:Hempstead}  for a comparison with ESP$\_$029741$\_$0985 from Inca City. Like Inca City, the boulders appear to spur the generation of the carbon dioxide jets.  As proposed by \cite{2010Icar..205..296T}, the boulders are surrounded and covered by the seasonal ice sheet.  The boulders provide a discontinuity in thermal inertia and an additional source of thermal heat that enables the surrounding ice to sublimate and crack more easily providing an outlet for the trapped carbon dioxide gas.  As we see at Shoram (Figure \ref{fig:shoram_boulders}) it is not always a simple story, however. In Shoram there are also small boulders,  but they are not always associated with fans. 

 \subsection{Boundaries and Interfaces: Bethpage}
 
\begin{figure}
\begin{center}
\includegraphics[width=1.0\columnwidth]{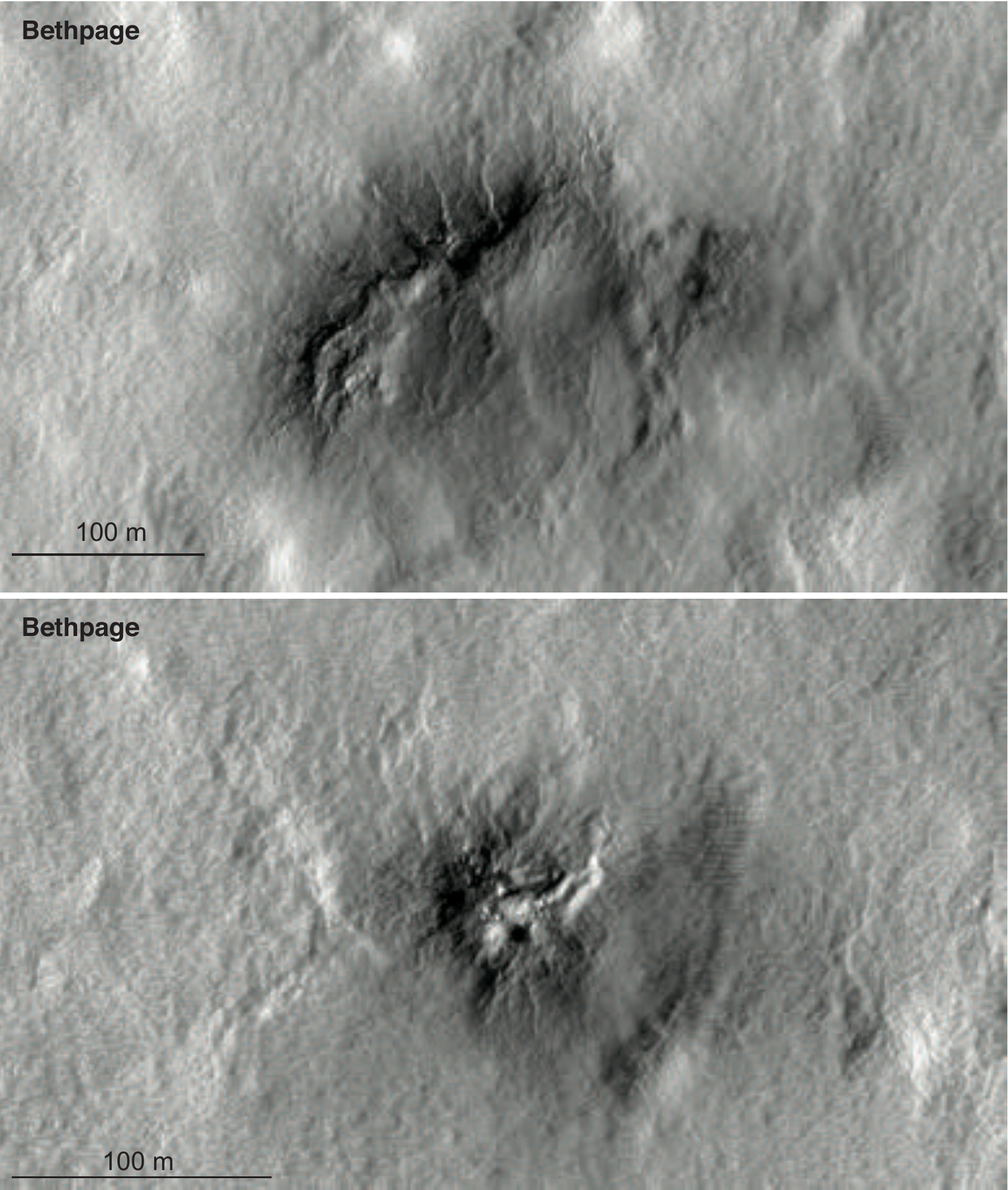}
 \caption{ \label{fig:Bethpage} Broad troughs with araneiform-like channels in Bethpage observed by HiRISE (ESP$\_$046777$\_$1030; L$_{\rm s}$=188.4$^\circ$).The fans are less distinct in the middle and lower panels, examples of broad troughs and araneiform channels at this time in the Southern spring season, but the seasonal fans deposits are still visible as a darker area in the vicinity of the channels. } 
 \end{center}
 \end{figure}

\begin{figure}
\begin{center}
\includegraphics[width=1.0\columnwidth]{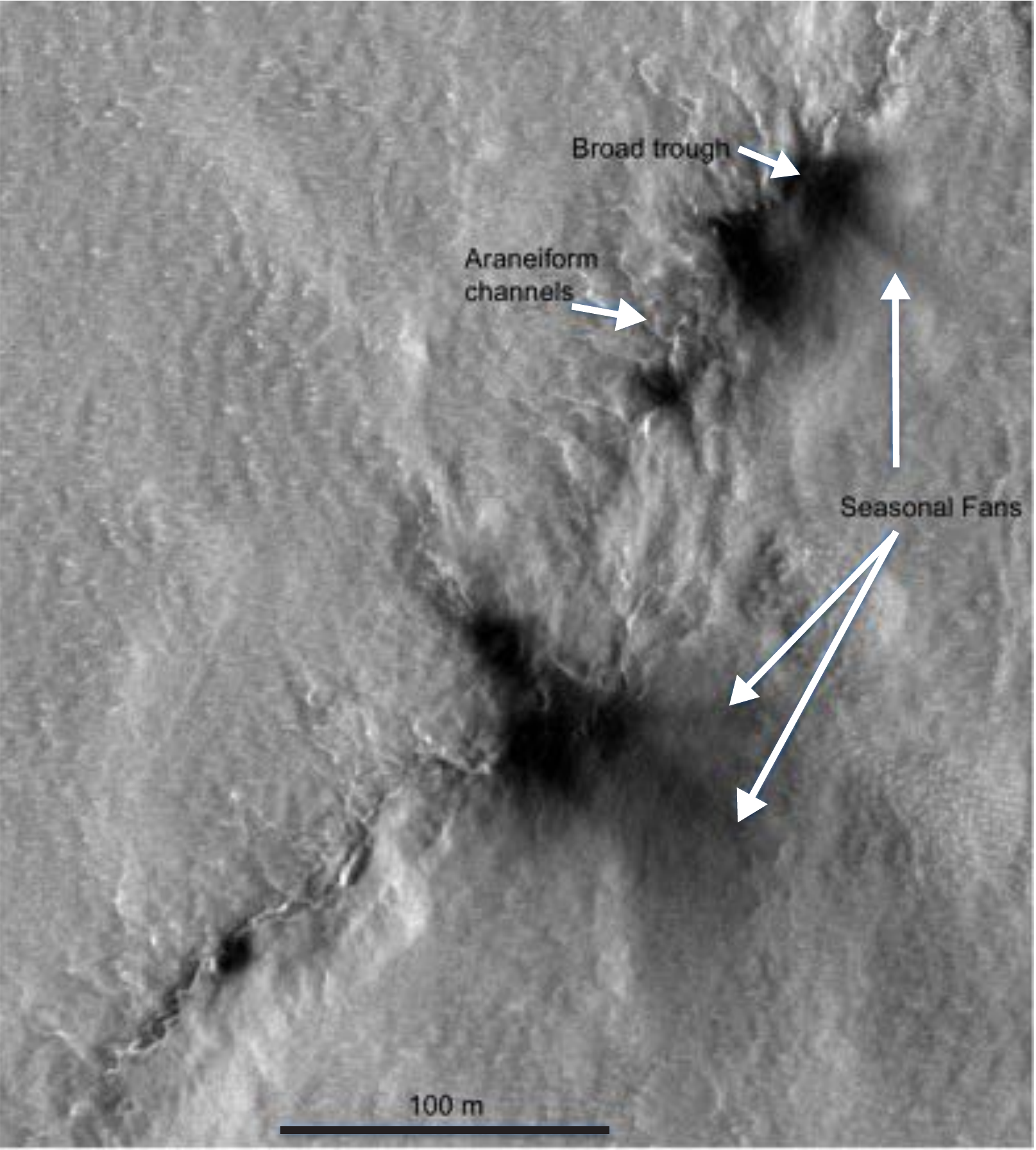}
 \caption{ \label{fig:Bethpage_fans} Araneiforms in Bethpage. Dark seasonal fans associated with channels in this Bethpage cutout taken from HiRISE image ESP$\_$046421$\_$1030, acquired at L$_{\rm s}$= 172.7$^\circ$). }
 \end{center}
 \end{figure}

\begin{figure}
\begin{center}
\includegraphics[width=1.0\columnwidth]{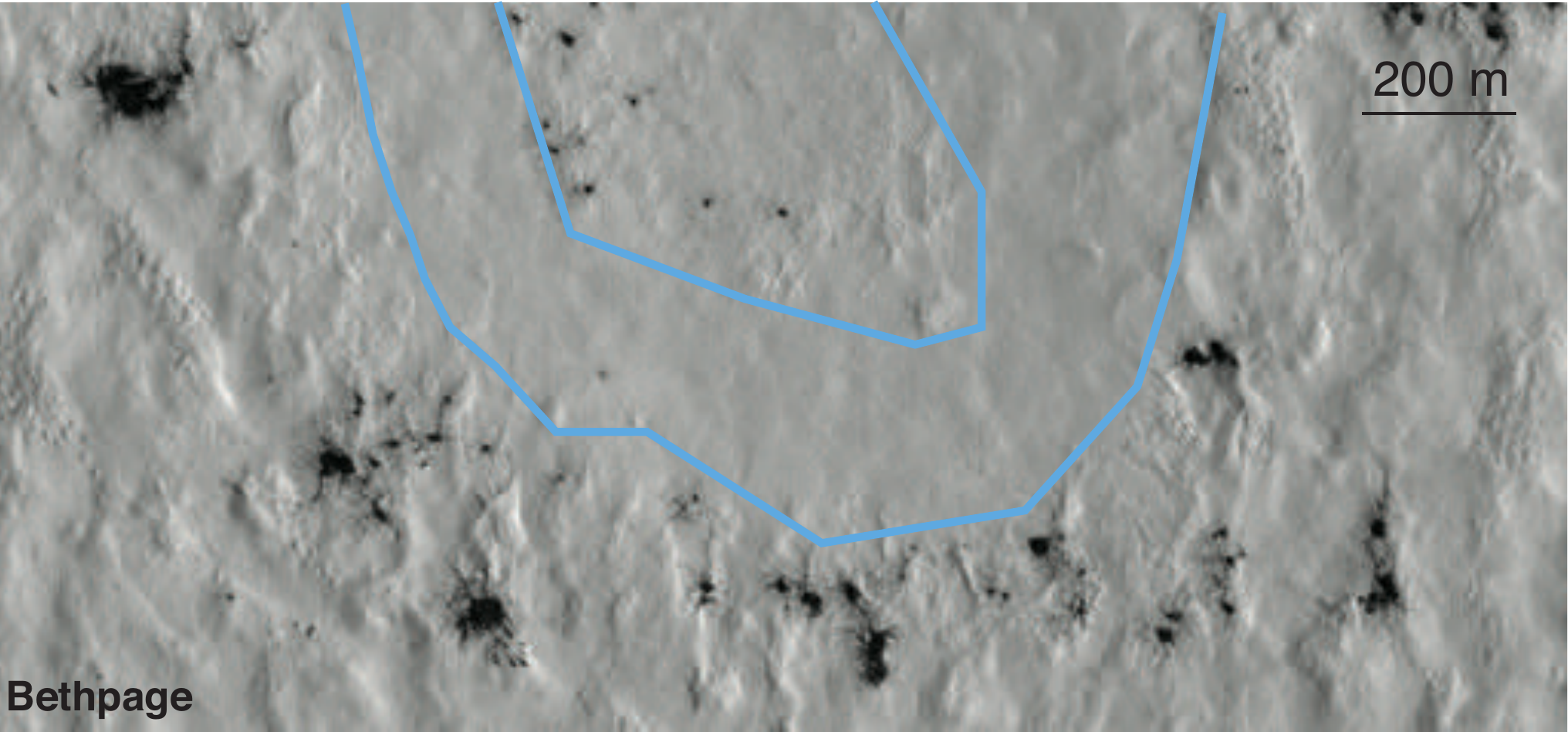}
 \caption{ \label{fig:Bethpage_2} Broad troughs with araneiform like channels in Bethpage observed by HiRISE (ESP$\_$048056$\_$1030;  L$_{\rm s}$=249.5$^\circ$). At the time the image was taken the surface in this region was covered by a semi-translucent seasonal layer of CO$_2$ ice.  The black patches in the image are seasonal fan material. Seasonal jet activity was found occurring near the boundaries of relatively smooth terrain and more hummocky terrain.  Light blue solid lines are drawn at the approximate outline of a smooth region surrounded by hummocky terrain in this HiRISE cutout.  } 
 \end{center}
 \end{figure}

Bethpage has narrow channels radiating from broad troughs.  A representative close-up is shown in Figure \ref{fig:Bethpage}.  Dark dust on the top of the seasonal ice surrounds the channels as highlighted in Figure \ref{fig:Bethpage_fans}. The appearance of the seasonal fans is consistent with active CO$_2$ jet processes on-going in this region, thus we infer that at least the narrow channels are formed by erosion from seasonal CO$_2$ gas escape entraining loose material from the surface under the seasonal ice.  As shown in Figure \ref{fig:Bethpage_2} the broad troughs and narrow channels are found in hummocky terrain and along the boundary of smoother terrain, consistent with the hypothesis that CO$_2$ jet processes exploit the weaknesses in the regolith to form sinuous araneiform channels.  

\subsection{Short Channels and Boulders: Stony Brook}
\begin{figure}
\begin{center}
\includegraphics[width=1.0\columnwidth]{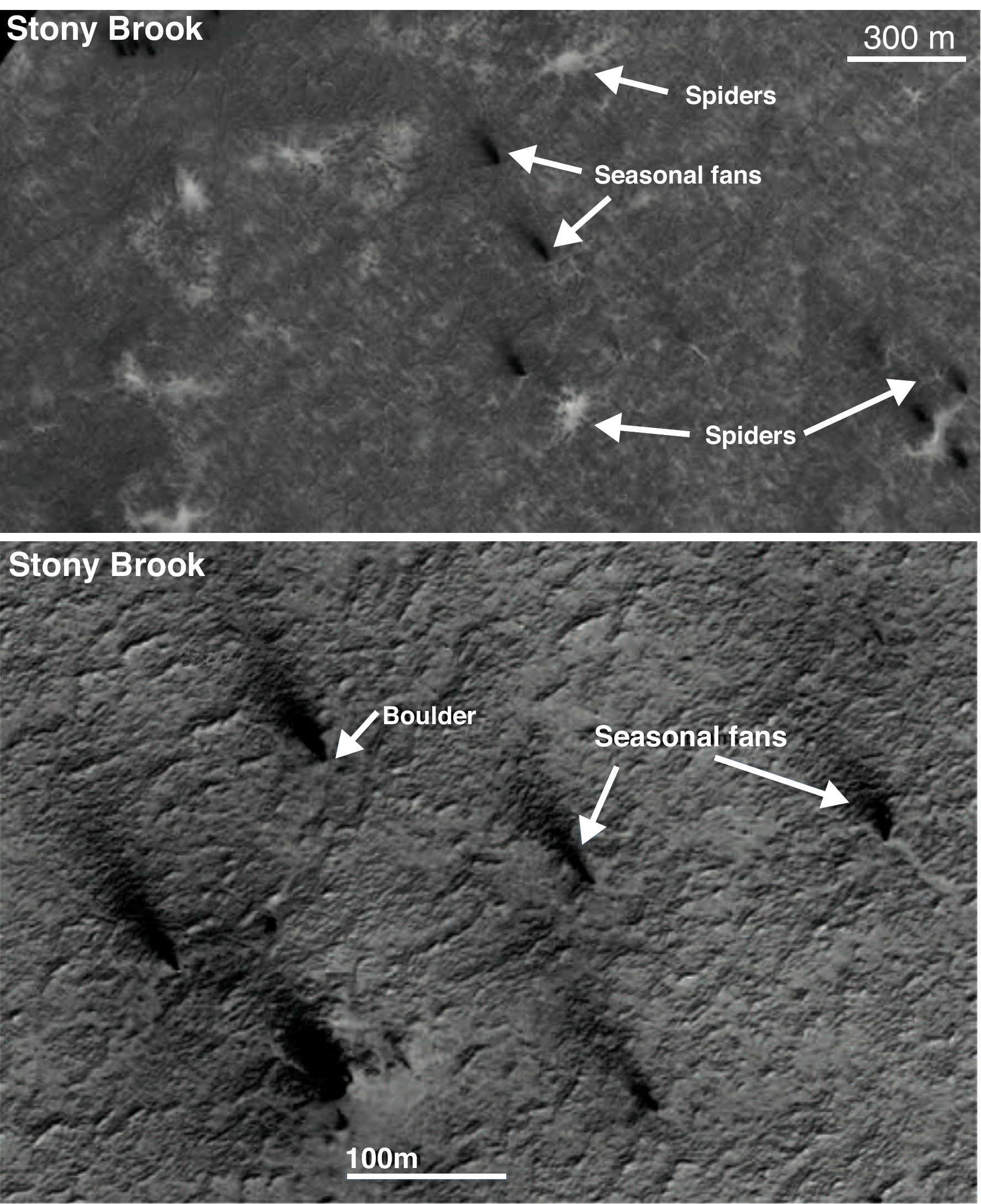}
 \caption{ \label{fig:stonybrook}  Stony Brook: Frosted spiders and dark seasonal fans are visible in the top two panel, a cutout rom HiRISE image ESP$\_$046398$\_$1030 taken at L$_{\rm s}$=171.7$^\circ$.  The lower panel shows seasonal fans, visible in ESP$\_$047308$\_$1030, acquired at L$_{\rm s}$=213$^\circ$. } 
 \end{center}
 \end{figure}

At Stony Brook the terrain is rough on the scale of meters.  There are short channels but with a few exceptions they do not show the branching, dendritic characteristics or the tight sinuosity of the araneiform terrains previously reported in the literature \citep[e.g.][]{2010Icar..205..283H}. HiRISE image ESP$\_$046398$\_$1030, shown in Figure \ref{fig:stonybrook}, displays seasonal fans (some coming from boulders and some from short grooves) and a few well-developed spiders. Frosted spider outlines are also visible in the region as well. 

 \subsection{Connected Aaraneiforms: Montauk, Happauge, and Shoram}
 
\begin{figure}[!]
\begin{center}
\includegraphics[width=1.0\columnwidth]{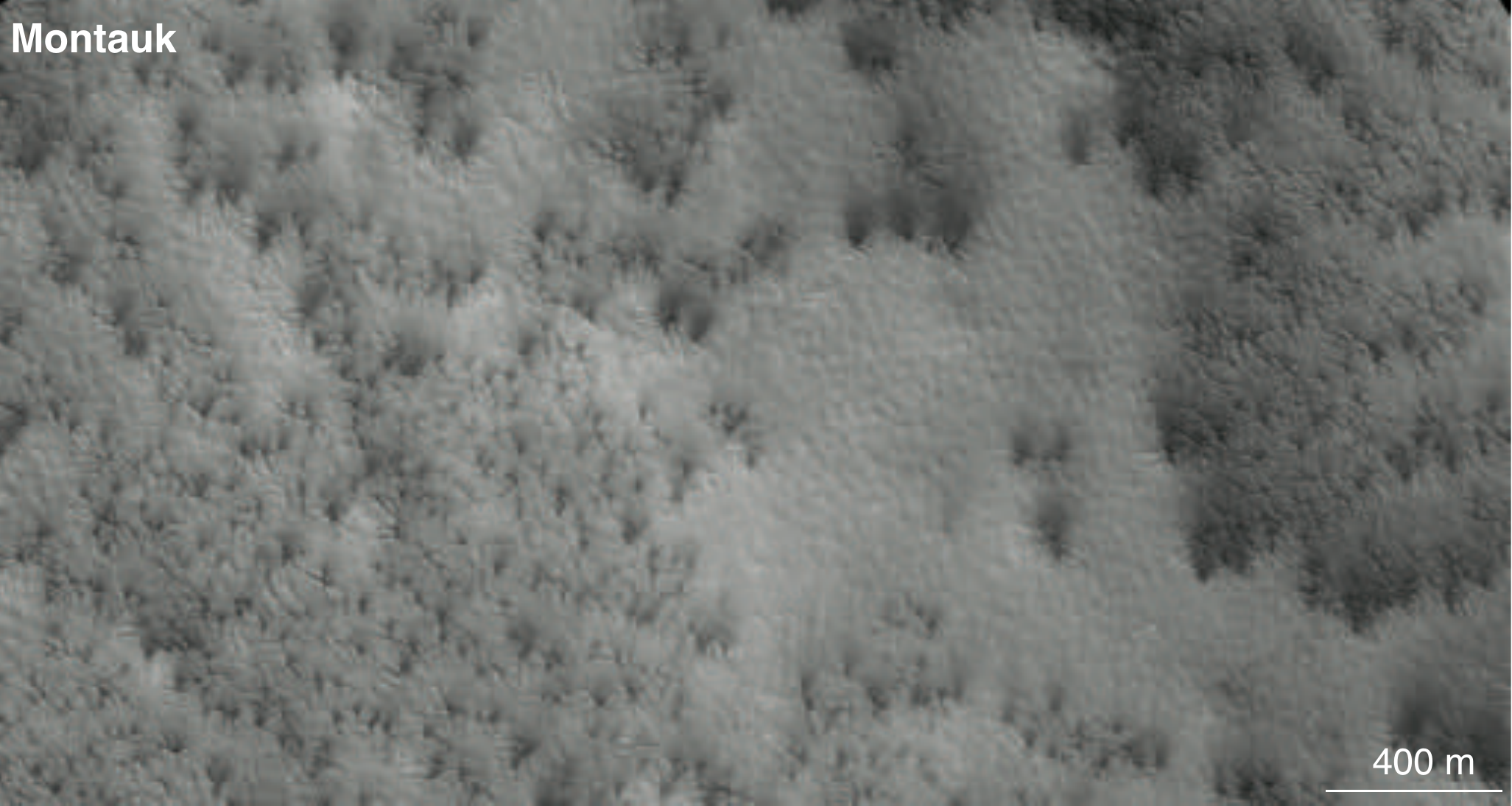}
 \caption{ \label{fig:Montauk} Connected araneiforms in Montauk (ESP$\_$046562$\_$1005; L$_{\rm s}$=178.8$^\circ$). In this subimage, branching channels are visible with no central organization. Dark seasonal fans are also present indicating active carbon dioxide CO$_2$ jets have formed this region.} 
 \end{center}
 \end{figure}

\begin{figure}[!]
\begin{center}
\includegraphics[width=1.0\columnwidth]{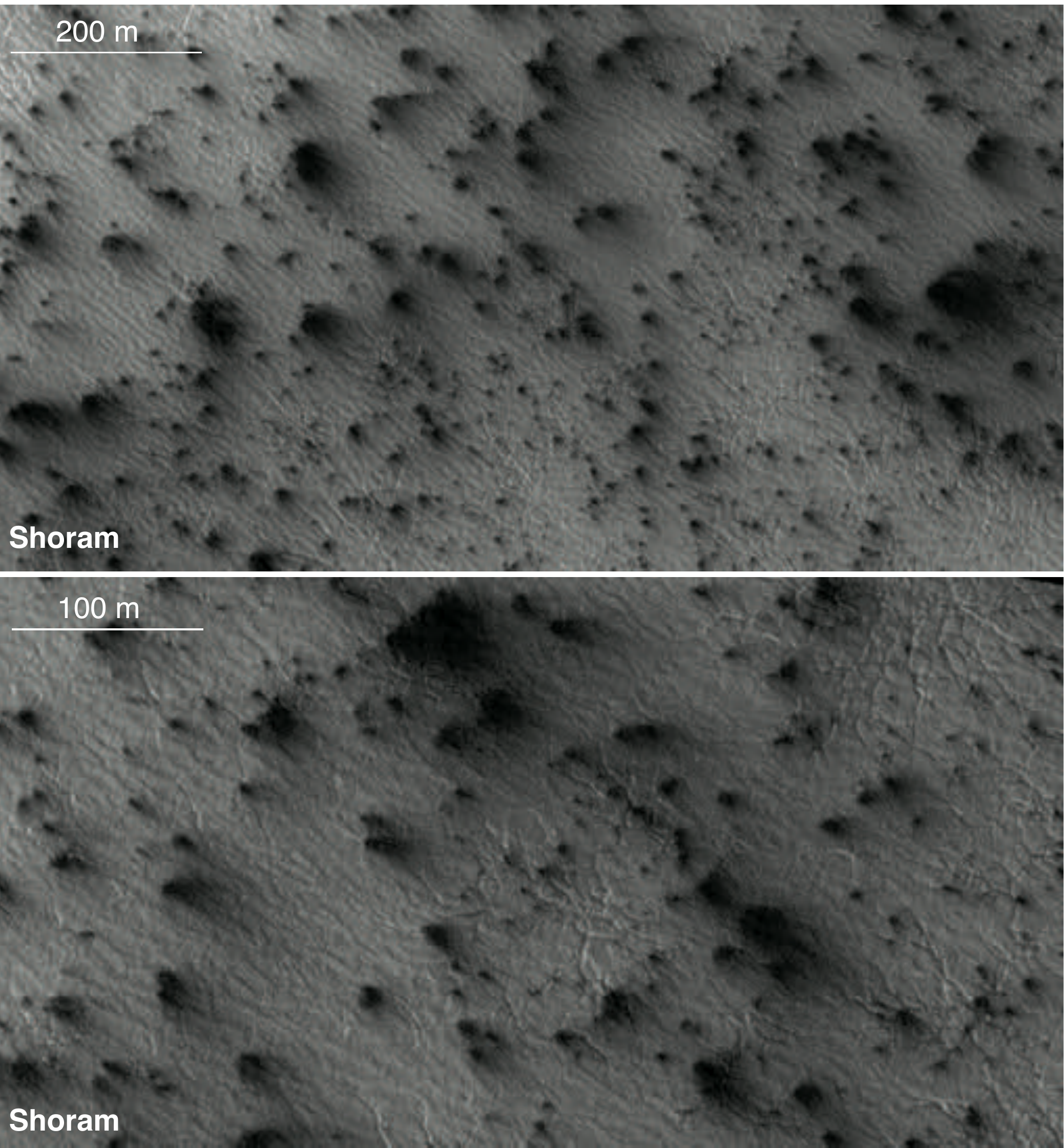}
 \caption{ \label{fig:shoram}Sample of connected araneiform features identified in HiRISE observations of Shoram (ESP$\_$046712$\_$1015;L$_{\rm s}$=185.5$^\circ$). The sinuous channels appear to  not be centralized around a single point or surface pit.  Dark seasonal fans present at the top of the semi-translucent seasonal CO$_2$ ice sheet are  also visible at this locale. Many of the fans originate at locations above  the dendritic araneiform channels.} 
 \end{center}
 \end{figure}

\begin{figure}[!]
\begin{center}
\includegraphics[width=1.0\columnwidth]{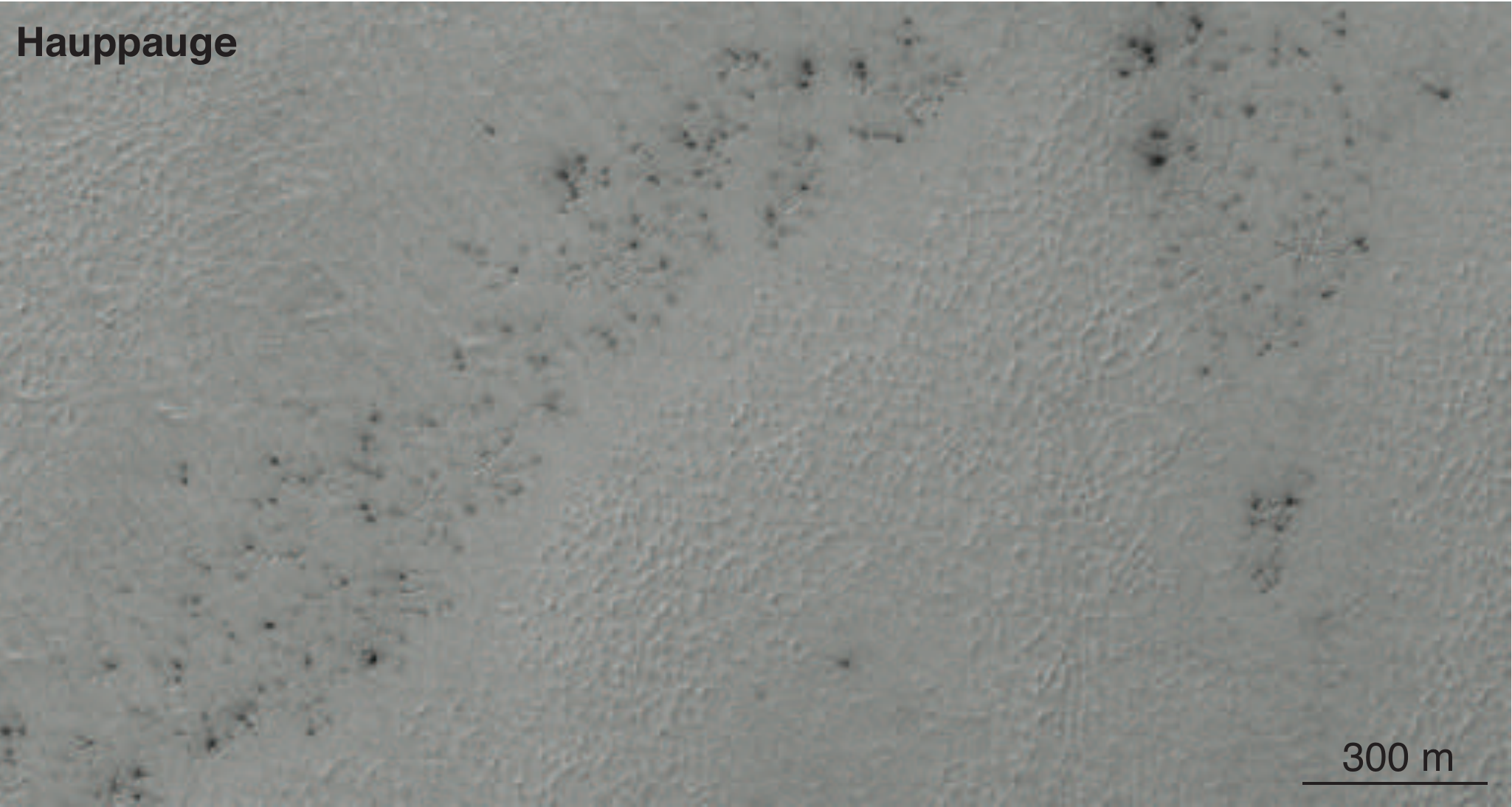}
 \caption{ \label{fig:Hauppague} Araneiform morphology in Hauppauge imaged by HiRISE (ESP$\_$046564$\_$1005; L$_{\rm s}$=178.9$^\circ$). Dark seasonal fans can be seen on top of the semi-translucent CO$_2$ ice  at locations where dendritic channels  are present in the surface below. The terrain in between the araneiforms is stippled in texture.} 
 \end{center}
 \end{figure}

Montauk has well-developed araneiforms as shown in Figure \ref{fig:Montauk}. Separated spiders are visible, but the vast majority of the araneiforms visible in these HiRISE observations are connected spiders, where dendritic channels are not emanating from a point but rather the channels form complex and interlinked networks of araneiform channels. This has a morphology similar to what  \cite{2010Icar..205..283H} describes as Connected Araneiform Morphology. Along with having boulders, Shoram also has pockets of connected araneiform features.  At L$_{\rm s}$=185.5$^\circ$, fans can be seen originating from many of the dendritic channels as shown in Figure \ref{fig:shoram}.  Additionally, Happauge (see Figure \ref{fig:Hauppague}) also exhibits connected araneiforms. The ground surrounding the spiders in Hauppage also has a stippled texture although this may not be related to the seasonal carbon dioxide jets.

 \subsection{Spiders on Possible Eject Blankets: Montauk and Happauge}

\begin{figure}[!]
\begin{center}
\includegraphics[width=1.0\columnwidth]{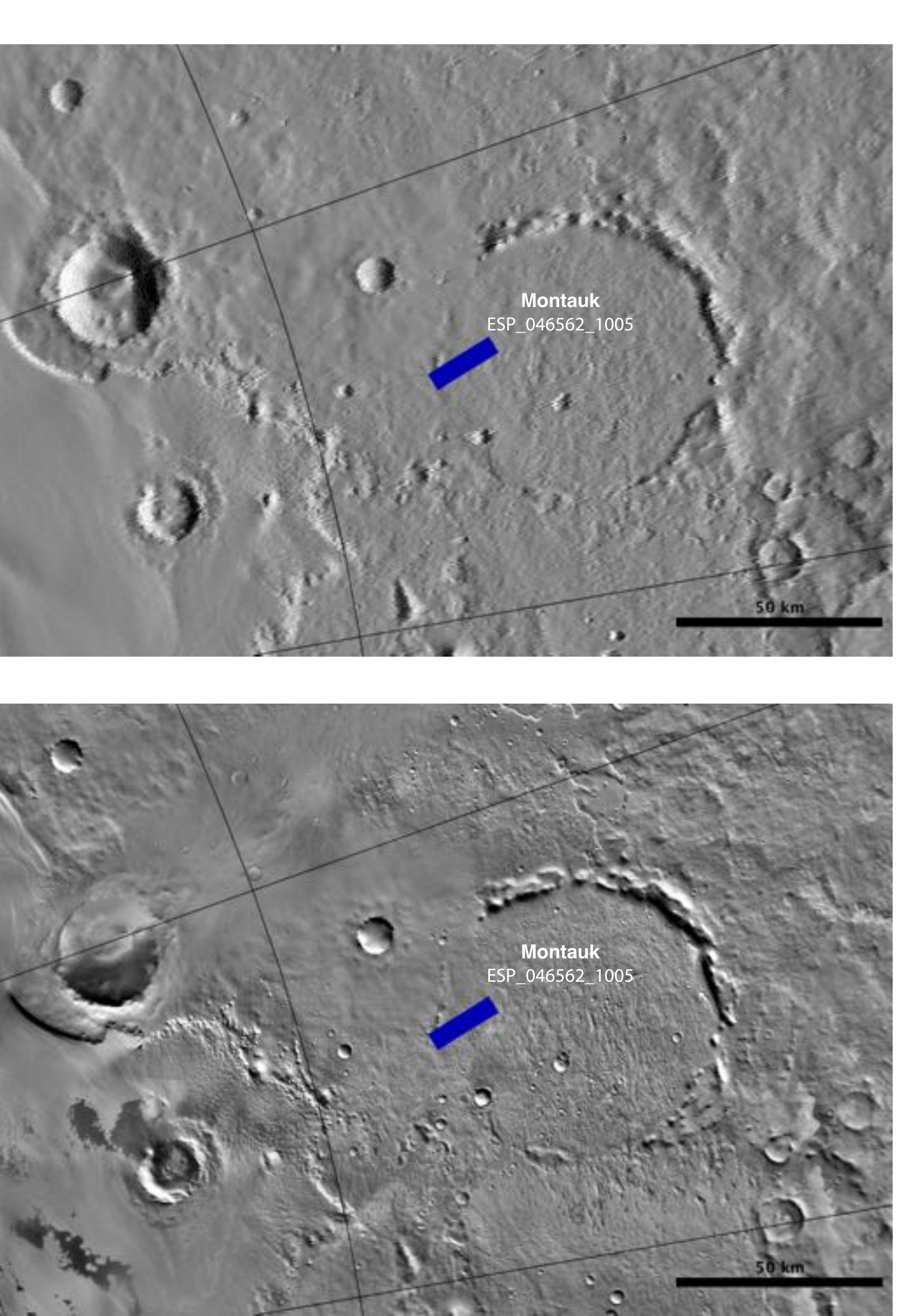}
 \caption{ \label{fig:Montauk_crater} Proximity of  Montauk in relation to a nearby crater rim and possible ejecta blanket.  The top shows the extent of HiRISE observation ESP$\_$046562$\_$1005 (L$_{\rm s}$=178.8$^\circ$) overlaid on a MOLA shaded elevation map \citep{1992JGR....97.7781Z,2001JGR...10623689S}.   The bottom image displays  the  HiRISE target overlaid on top of the THEMIS DAY IR 100m global map mosaic \citep{2011JGRE..11610008E,2014LPICo1791.1141H}.  For both figures, latitude and longitude lines are plotted every 10  degrees. The zero meridian is pointing straight up. } 
 \end{center}
 \end{figure}

\begin{figure}[!]
\begin{center}
\includegraphics[width=1.0\columnwidth]{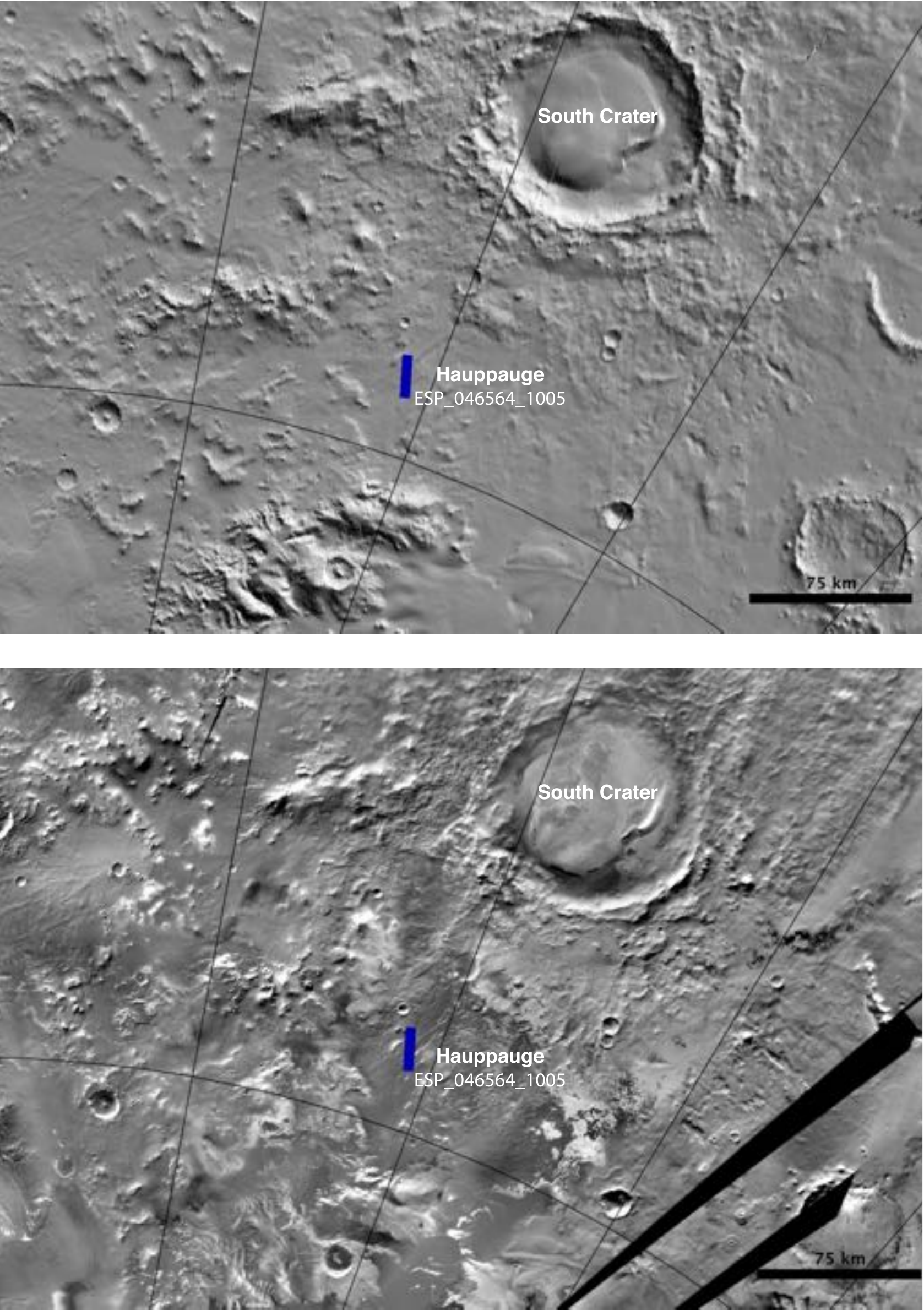}
 \caption{ \label{fig:Hauppague_south_crater} Proximity of Hauppauge to South Crater.  The top shows the extent of HiRISE observation ESP$\_$046562$\_$1005 (L$_{\rm s}$=178.8$^\circ$) overlaid on a MOLA shaded elevation map \citep{1992JGR....97.7781Z,2001JGR...10623689S}.   The bottom image displays  the  HiRISE target overlaid on top of the THEMIS DAY IR 100m global map mosaic \citep{2011JGRE..11610008E,2014LPICo1791.1141H}.  For both figures, latitude and longitude lines are plotted every 10  degrees. The zero meridian is pointing straight up. Hauppage appears in the THEMIS observations to be on a darker unit, but it is not clear if this can be identified as part of South Crater's ejecta blanket. } 
 \end{center}
 \end{figure}

Montauk is located on the rim of a partially eroded un-named crater (see Figure \ref{fig:Montauk_crater}) and Hauppauge is located near a large crater named `South Crater' as shown in Figure \ref{fig:Hauppague_south_crater}. These two regions may reside on the crater ejecta blankets.  The layering or unconsolidated nature of the surface inherent in an ejecta blanket may provide a conducive environment for spider development, similar to the SPLD, where araneiforms were originally  found to reside \citep[see ][]{2003JGRE..108.5084P}. Identification of crater ejecta is challenging. Only fresh or moderately eroded craters will have ejecta blankets surrounding the crater rim identifiable in orbital images.  Following  \citep{2012JGRE..117.5004R}, we examined  Thermal Emission Imaging System (THEMIS )Day Infrared  IR) 100m per pixel global map mosaic \citep{2011JGRE..11610008E,2014LPICo1791.1141H} of the area surrounding Montauk and Happauge (see  Figures  \ref{fig:Montauk_crater} and \ref{fig:Hauppague_south_crater}). Hauppage appears in the THEMIS observations to be on a darker unit, but it is not clear if this can be identified as part of South Crater's ejecta blanket. From our analysis we cannot confirm that these locations outside of the SPLD coincide with ejecta blankets. Further high resolution observations and detections of more araneiforms outside of the SPLD in close proximity to crater rims would bolster this hypothesis.    
 
\subsection{Polygons and Spiders: Dix Hills}

\begin{figure}[!]
\begin{center}
\includegraphics[width=1.0\columnwidth]{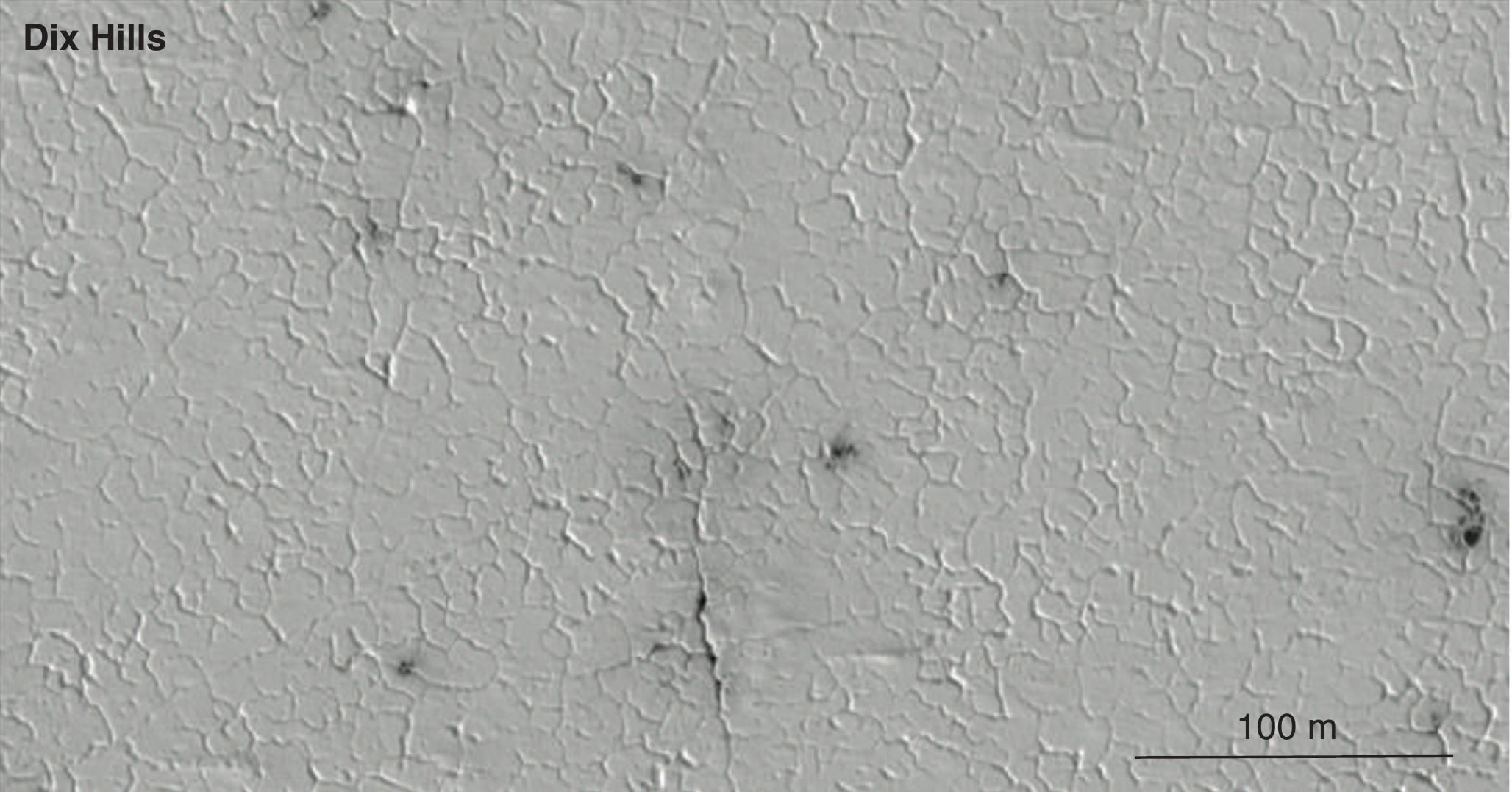}
 \caption{ \label{fig:DixHills_polygons1} Polygonal channels in the ground are visible at Dix Hills in HiRISE observation  ESP$\_$047938$\_$0995 acquired at L$_{\rm s}$=243.7$^\circ$.  Seasonal fans (dark patches) associated with the channels are evidence for seasonal sub-ice gas flow along the polygonal grooves. } 
 \end{center}
 \end{figure}

\begin{figure}[!]
\begin{center}
\includegraphics[width=1.0\columnwidth]{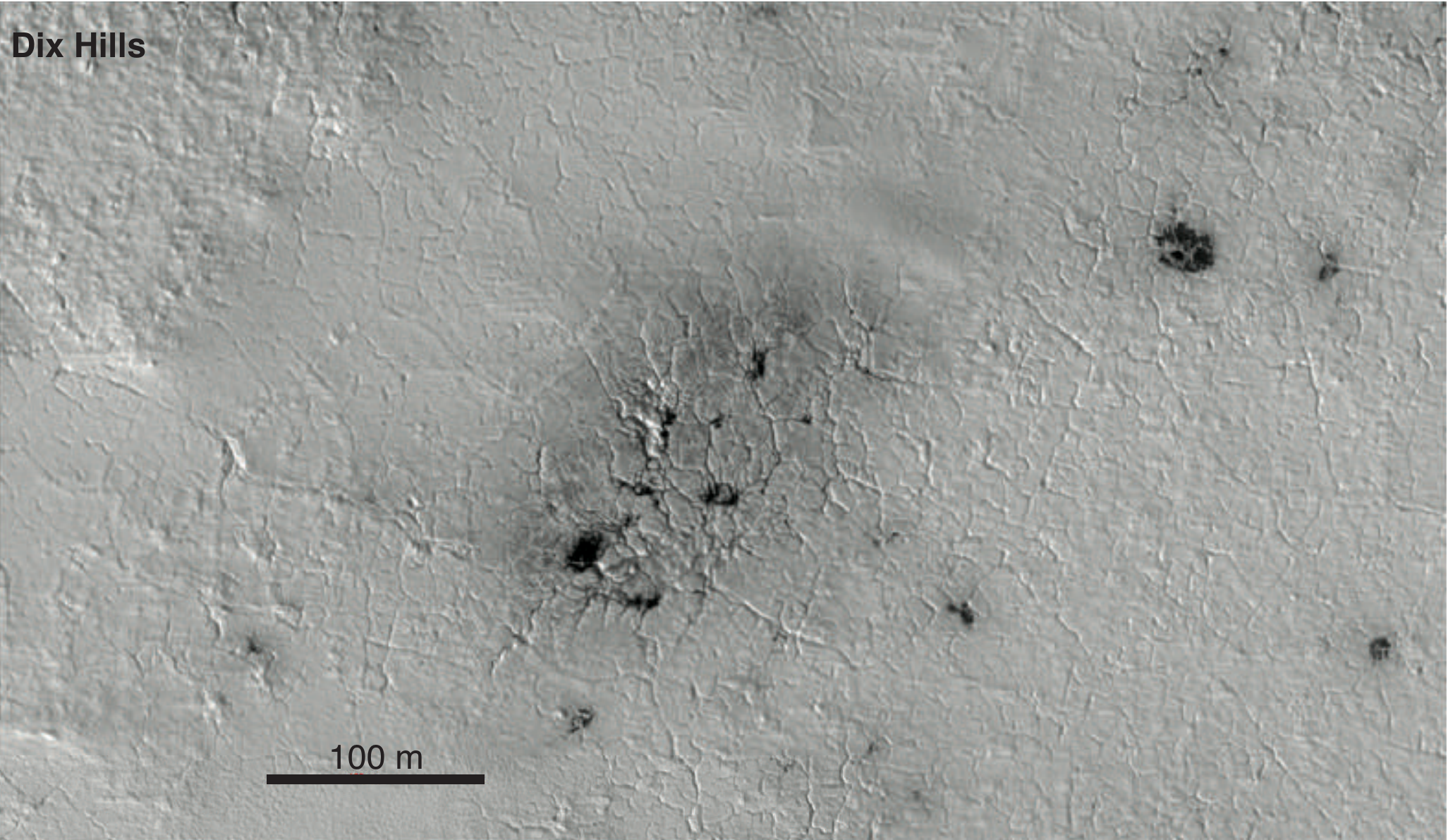}
 \caption{ \label{fig:DixHills_polygons2}  Araneiform terrain located along the polygonal grooves is shown in this cutout of HiRISE image ESP$\_$047938$\_$0995 (L$_{\rm s}$=243.7$^\circ$) of Dix Hills.  Seasonal fan (dark black patches in the image) are associated with some of the araneiform  channels, indicating active CO$_2$ jets at these locations.} 
 
 \end{center}
 \end{figure}

Extensive polygonal cracks are visible in the surface at Dix Hills. Polygonal cracks are one of many well-documented types of patterned ground formed by the cyclic freezing and thawing of underlying water ice-cemented ground, and the subsequent cracking of ice-rich material produces shallow grooves\citep{2005Icar..174..336M}. The association of fans with these grooves leads us to infer that they provide routes for the flow of the trapped pressurized gas under the CO$_2$ seasonal cap.  As shown in Figure \ref{fig:DixHills_polygons1} and Figure \ref{fig:DixHills_polygons2} more sinuous araneiform channels are also present at these locales. Fans are visible in later observations of Dix Hills where the polygonal channels have pits, are wider, and/or develop secondary branching channels, as shown in Figure \ref{fig:DixHills_polygons2}. 

\subsection{Sinuous Channels: Patchogue}

\begin{figure}[!]
\begin{center}
\includegraphics[width=1.0\columnwidth]{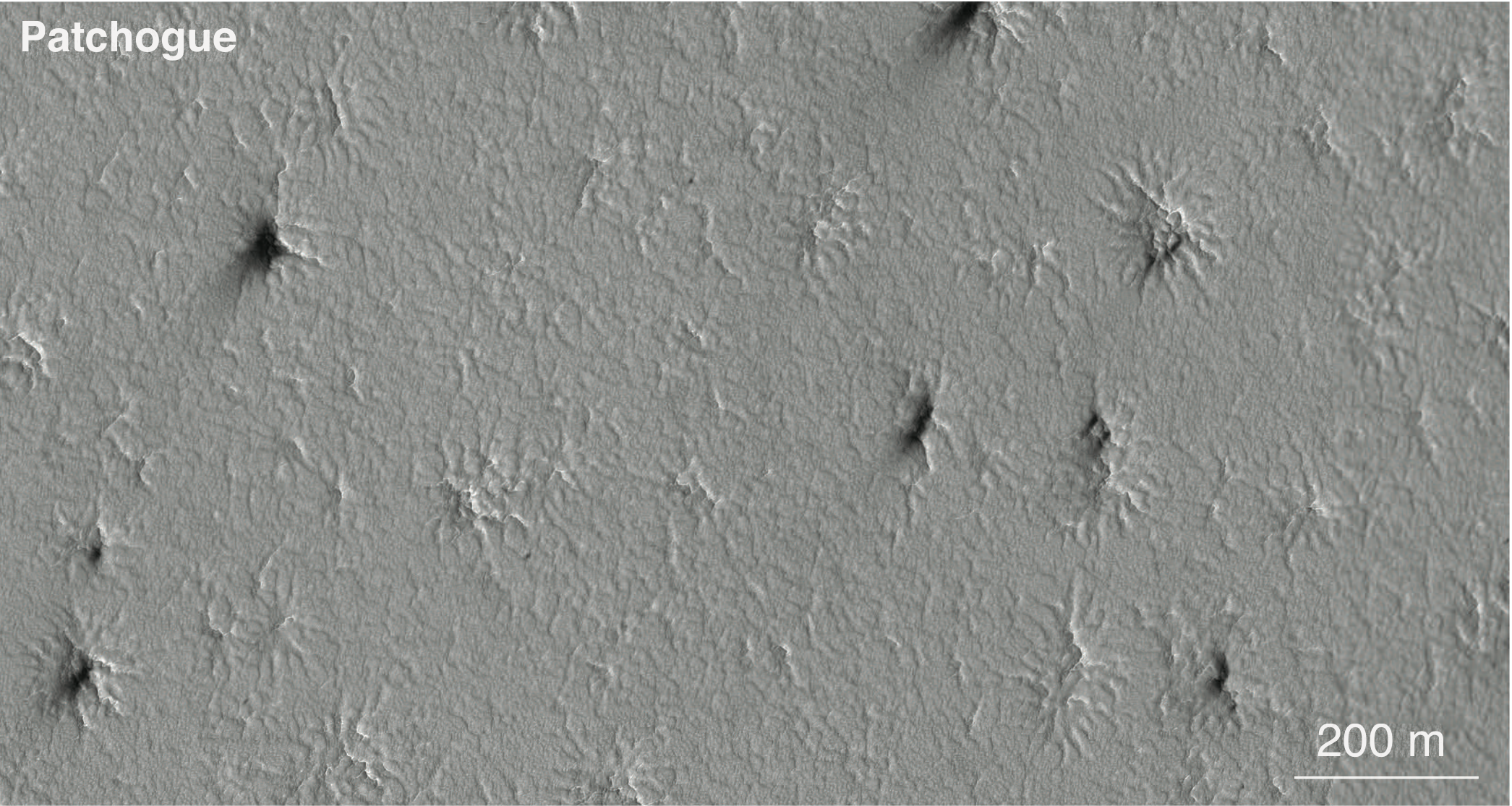}
 \caption{ \label{fig:patchogue} Sample of spider araneiform features identified in HiRISE observations of Patchogue (ESP$\_$046567$\_$0980; L$_{\rm s}$=179.1$^\circ$).  Distinct spiders are visible with sinuous channels emanating from a central pit. At this  L$_{\rm s}$, some of the isolated spiders also exhibit seasonal fans at the locations above on the top of the seasonal ice sheet.} 
 \end{center}
 \end{figure}

Patchogue has both sinuous channels and distinct well-separated spiders, with distinctly visible central pits and radiating channels, see Figure \ref{fig:patchogue}.  The spiders' morphology is the classic spider with radially-organized channels, that have been observed by HiRISE on other locales on the SPLD \citep{2010Icar..205..283H}. Patchogue is near Inca City, but unlike Hempstead, Shoram, and Inca City, no boulders were visible.

\section{Implications for the CO$_2$ Jet Model}

The majority of the P4T clean spider sample ( $75\%$) resides on the SPLD, where the carbon dioxide jet model has been previously put forth for the formation of araneiforms \citep{2000mpse.conf...93K, 2003JGRE..108.5084P,2006Natur.442..793K, 2007JGRE..112.8005K,  2008JGRE..113.6005P, 2010Icar..205..296T, 2010Icar..205..311P, 2011Icar..213..131P,  2011GeoRL..38.8203T}. High resolution HiRISE imaging confirms the identification of several locations outside of the SPLD with araneiforms  found by P4T.  All of the HiRISE observations of the locales outside of the SPLD with confirmed araneiform features visible, also exhibited seasonal fans, suggesting that active  CO$_2$ jets were present during the season. Additionally, all but one of these HiRISE target locations have already been identified by \cite{2003JGRE..108.5084P} as becoming cryptic (with temperatures near the CO$_2$ frost sublimation temperature but with an albedo lower than that of typical CO$_2$ frost) for a period of time between L$_S$ = 180$^{\circ}$ and  L$_S$ = 250$^{\circ}$, indicative of semi-translucent CO$_2$ slab ice required for the CO$_2$ jet formation. For Hempstead, there was no albedo information available to \cite{2003JGRE..108.5084P}, but \cite{2015Icar..251..164P} identify the presence of the seasonal ice cap extending beyond that location to -50$^{\circ}$ over multiple Mars years. Thus our clean spider distribution and new araneiform discoveries outside of the SPLD are consistent with the preferred CO$_2$ jet model and slab ice formation model. 

Our outside of the SPLD spider identifications are located in  Amazonian and Hesperian polar units and the Early Noachian highland units outside of the SPLD. With our current coverage of  $\sim$11$\%$ of the South Polar region southward of 75$^{\circ}$, we cannot make a definitive statement regarding whether these are the only geologic units with araneiforms within the Martian South Polar region. A larger search for araneiforms in CTX and MOC NA observations are needed. The presence araneiforms in the new areas identified in this study suggests that the regolith in these areas can be easily weakened or is loosely conglomerated like that of the SPLD, enabling the trapped carbon dioxide gas to slowly carve channels into the surface over time. Further study of the surface properties may illuminate the link between the SPLD and these regions for araneiform formation. Two of the HiRISE target locations (Hauppague and Montauk) are close to craters and may reside on associated ejecta blankets. This may indicate that crater ejecta blankets play an important role in the formation of araneiforms outside of the SPLD as an alternative source for loosely conglomerated materials to excavate. Further HiRISE observations and samples of araneiforms outside the SPLD are needed before a definitive conclusion can be made.  

\section{Conclusions}

Through the effort of over 10,000 volunteers, P4T has mapped the distribution of spider araneiforms in 90 CTX observations of the Martian South Pole. This paper presents results from our analysis of 303,192 km$^2$ of CTX observations  between -75$^\circ$  and the South Pole, mapping locations of araneiforms and Swiss cheese terrain on the Martian South Pole. We present the first identification of araneiforms present at  locations outside the SPLD. This discovery is consistent with the general paradigm where the CO$_2$ jet process carves channels into the surface by the trapped CO$_2$ gas exploiting weaknesses in the conglomeration of the soil layer below  \citep{2000mpse.conf...93K, 2003JGRE..108.5084P,2006Natur.442..793K, 2007JGRE..112.8005K,  2008JGRE..113.6005P, 2010Icar..205..296T, 2010Icar..205..311P, 2011Icar..213..131P,  2011GeoRL..38.8203T}. 

Although the spiders are found primarily on the SPLD, the source of the non-uniformity of the spider distribution requires further study. The morphology of the aranieforms and rate of the erosion process to carve channels into the surfaces outside the SPLD is likely a combination of material strength of the surface, surface slopes,  and the amount of time the seasonal CO$_2$ ice sheet is present at each location. We find that several of our newly identified araneiform locales are on or close to crater ejecta blankets. With our limited sample size, we can not definitively conclude that araneiform formation outside of the SPLD prefers disturbed regolith such as crater ejecta blankets. A larger survey of the South Polar region is required. Further high resolution imaging of these new spider locations coupled with additional surveys of the South Polar region will further elucidate the formation of these features and the CO$_2$ seasonal jet process. 

\section{Future Prospects}
The success of P4T in  providing timely targets for follow-up high resolution imaging highlights citizen science and crowd sourcing as a potential future avenue for efficiently identifying planetary surface features in planetary mission datasets and feeding back into mission planning. The area searched by P4T presented in this Paper was constrained in order to have both the citizen science review and full analysis of the classifications completed before the beginning of the South spring Equinox in Mars Year 33 and the  start of HiRISE's seasonal processes monitoring campaign. With the completion of this first set of 90 CTX observations, additional CTX images have continued to be available on the P4T website for review; expanding the coverage wider beyond -75$^{\circ}$ latitude. Classification of these images is currently underway. We expect in future works to be able to produce a sample of baby spider and channel network locations from P4T classifications. P4T will be able to further examine the abundance of spiders outside the SPLD and the correlations with surface properties that can shed light on how the CO$_2$ jet process is able to carve channels into the SPLD and these other regions of the South Pole.  More examples of araneiform locations outside of the SPLD will enable a better analysis of the prevalence of crater ejecta blankets as suitable environments from araneiform formation. 

\section*{Acknowledgments}

The data presented in this Paper are the result of the efforts of the P4T volunteers, without whom this work would not have been possible. Their contributions are individually acknowledged at:  \newline  \url{http://p4tauthors.planetfour.org}. We also thank our P4T Talk moderators Andy Martin and John Keegan for their time and efforts helping the P4T volunteer community. 

This publication uses data generated via the Zooniverse.org (\url{https://www.zooniverse.org}) platform, development of which is funded by generous support, including a Global Impact Award from Google, and by a grant from the Alfred P. Sloan Foundation. The authors thank the Zooniverse team for early access to their Project Builder Platform. 

MES was supported by Gemini Observatory, which is operated by the Association of Universities for Research in Astronomy, Inc., on behalf of the international Gemini partnership of Argentina, Brazil, Canada, Chile, and the United States of America.  MES was also supported in part by an Academia Sinica Postdoctoral Fellowship. KMA and AP were supported in part by NASA Grant Nr. 14-SSW14 2-0237. The authors also thank Chris Lintott (University of Oxford), who had to decline authorship on this Paper. We thank him for his efforts contributing to the development of the Planet Four: Terrains website and for his useful discussions. The authors also thank Chris Snyder, Chris Schaller, and Serina Diniega for useful discussions. The authors also thank Peter Buhler and an anonymous reviewer for constructive reviews that improved the manuscript. 

The Mars Reconnaissance Orbiter mission is operated at the Jet Propulsion Laboratory, California Institute of Technology, under contracts with NASA. This research has made use of the USGS Integrated Software for Imagers and Spectrometers (ISIS) and of NASA's Astrophysics Data System. We thank the Context Camera (CTX) team  for making their data publicly available through NASA's Planetary Data System (PDS). CTX is operated by Malin Space Science Systems. This research also  made use of Astropy, a community-developed core Python package for Astronomy \citep{2013A&A...558A..33A}. This work made use of the JAVA Mission-planning and Analysis for Remote Sensing (JMARS) \citep{2009AGUFMIN22A..06C} and the MRO Context Camera Image Explorer \url{http://viewer.mars.asu.edu/viewer/ctx#T=0}. 

\appendix
\section{Geologic Map Legend}
\label{ref:appendix}
\begin{figure}[!]
\begin{center}
\includegraphics[width=1.0\columnwidth]{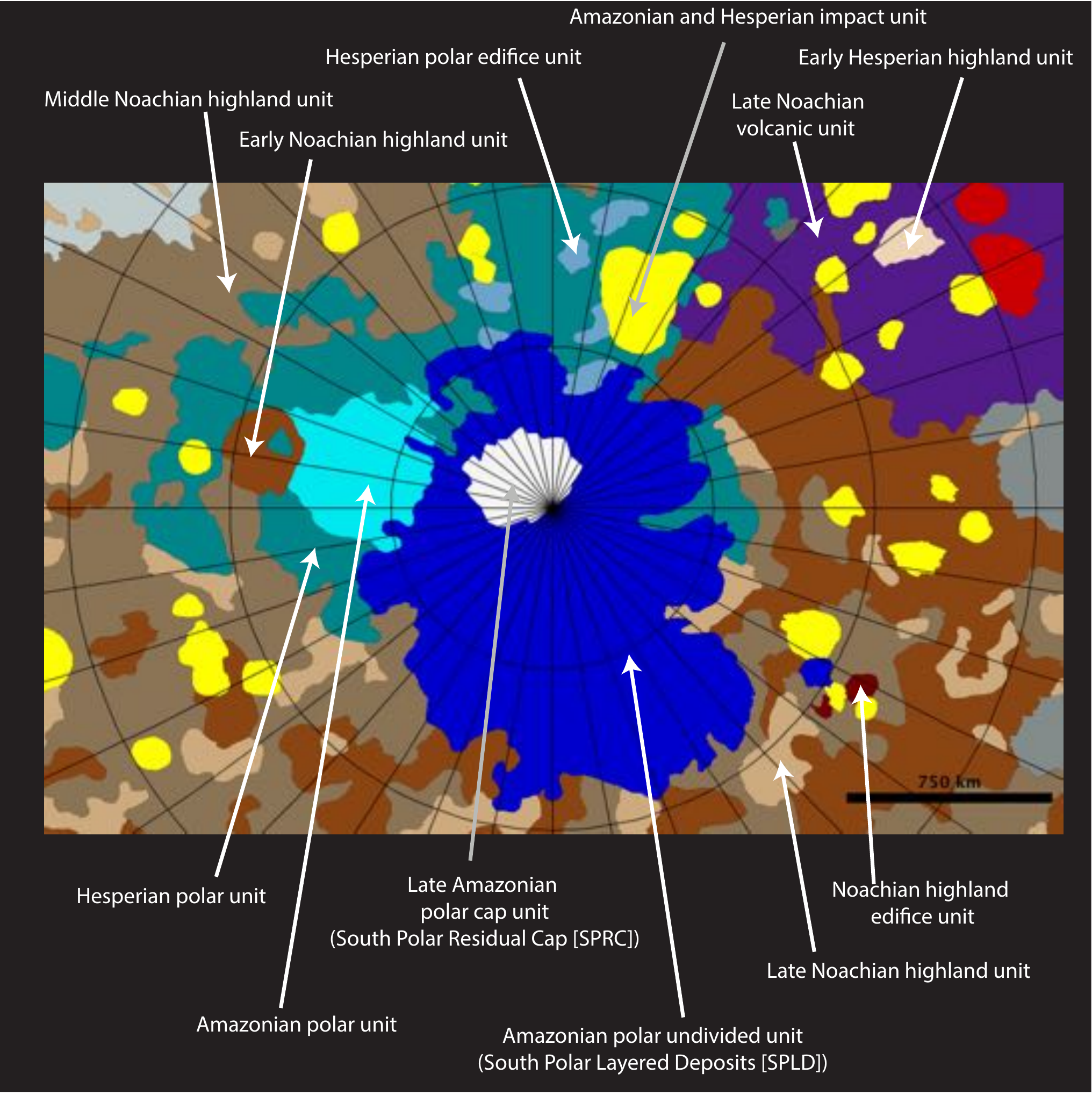}
 \caption{ \label{fig:legend}  Main geologic units comprising the South Polar region of Mars as identified by \cite{Tanaka:2014wd}. Latitude and longitude lines are plotted every 10  degrees. The zero meridian is pointing straight up.} 
 \end{center}
 \end{figure}

We provide in Figure \ref{fig:legend} a legend identifying the main geologic units of the South Polar region in the geologic map from \cite{Tanaka:2014wd} used in this work. We refer the reader to \cite{Tanaka:2014wd}  for further details about the characteristics of each geologic unit. We note that the Amazonian polar undivided unit is also known as the South Polar Layered Deposits (SPLD), and the  Late Amazonian polar cap unit is also referred to as the  South Polar Residual Cap (SPRC). 

\section*{References}

\bibliographystyle{model2-names.bst}\biboptions{authoryear}

\begin{thebibliography}{84}
\expandafter\ifx\csname natexlab\endcsname\relax\def\natexlab#1{#1}\fi
\providecommand{\url}[1]{\texttt{#1}}
\providecommand{\href}[2]{#2}
\providecommand{\path}[1]{#1}
\providecommand{\DOIprefix}{doi:}
\providecommand{\ArXivprefix}{arXiv:}
\providecommand{\URLprefix}{URL: }
\providecommand{\Pubmedprefix}{pmid:}
\providecommand{\doi}[1]{\href{http://dx.doi.org/#1}{\path{#1}}}
\providecommand{\Pubmed}[1]{\href{pmid:#1}{\path{#1}}}
\providecommand{\bibinfo}[2]{#2}
\ifx\xfnm\relax \def\xfnm[#1]{\unskip,\space#1}\fi
\bibitem[{{Aharonson} et~al.(2004){Aharonson}, {Zuber}, {Smith}, {Neumann},
  {Feldman} and {Prettyman}}]{2004JGRE..109.5004A}
\bibinfo{author}{{Aharonson}, O.}, \bibinfo{author}{{Zuber}, M.T.},
  \bibinfo{author}{{Smith}, D.E.}, \bibinfo{author}{{Neumann}, G.A.},
  \bibinfo{author}{{Feldman}, W.C.}, \bibinfo{author}{{Prettyman}, T.H.},
  \bibinfo{year}{2004}.
\newblock \bibinfo{title}{{Depth, distribution, and density of CO$_{2}$
  deposition on Mars}}.
\newblock \bibinfo{journal}{Journal of Geophysical Research (Planets)}
  \bibinfo{volume}{109}, \bibinfo{pages}{E05004}.
\newblock \DOIprefix\doi{10.1029/2003JE002223}.
\bibitem[{{Anderson} et~al.(2004){Anderson}, {Sides}, {Soltesz}, {Sucharski}
  and {Becker}}]{2004LPI....35.2039A}
\bibinfo{author}{{Anderson}, J.A.}, \bibinfo{author}{{Sides}, S.C.},
  \bibinfo{author}{{Soltesz}, D.L.}, \bibinfo{author}{{Sucharski}, T.L.},
  \bibinfo{author}{{Becker}, K.J.}, \bibinfo{year}{2004}.
\newblock \bibinfo{title}{{Modernization of the Integrated Software for Imagers
  and Spectrometers}}, in: \bibinfo{editor}{{Mackwell}, S.},
  \bibinfo{editor}{{Stansbery}, E.} (Eds.), \bibinfo{booktitle}{Lunar and
  Planetary Science Conference}, p. \bibinfo{pages}{2039}.
\bibitem[{{Astropy Collaboration} et~al.(2013){Astropy Collaboration},
  {Robitaille}, {Tollerud}, {Greenfield}, {Droettboom}, {Bray}, {Aldcroft},
  {Davis}, {Ginsburg}, {Price-Whelan}, {Kerzendorf}, {Conley}, {Crighton},
  {Barbary}, {Muna}, {Ferguson}, {Grollier}, {Parikh}, {Nair}, {Unther},
  {Deil}, {Woillez}, {Conseil}, {Kramer}, {Turner}, {Singer}, {Fox}, {Weaver},
  {Zabalza}, {Edwards}, {Azalee Bostroem}, {Burke}, {Casey}, {Crawford},
  {Dencheva}, {Ely}, {Jenness}, {Labrie}, {Lim}, {Pierfederici}, {Pontzen},
  {Ptak}, {Refsdal}, {Servillat} and {Streicher}}]{2013A&A...558A..33A}
\bibinfo{author}{{Astropy Collaboration}}, \bibinfo{author}{{Robitaille},
  T.P.}, \bibinfo{author}{{Tollerud}, E.J.}, \bibinfo{author}{{Greenfield},
  P.}, \bibinfo{author}{{Droettboom}, M.}, \bibinfo{author}{{Bray}, E.},
  \bibinfo{author}{{Aldcroft}, T.}, \bibinfo{author}{{Davis}, M.},
  \bibinfo{author}{{Ginsburg}, A.}, \bibinfo{author}{{Price-Whelan}, A.M.},
  \bibinfo{author}{{Kerzendorf}, W.E.}, \bibinfo{author}{{Conley}, A.},
  \bibinfo{author}{{Crighton}, N.}, \bibinfo{author}{{Barbary}, K.},
  \bibinfo{author}{{Muna}, D.}, \bibinfo{author}{{Ferguson}, H.},
  \bibinfo{author}{{Grollier}, F.}, \bibinfo{author}{{Parikh}, M.M.},
  \bibinfo{author}{{Nair}, P.H.}, \bibinfo{author}{{Unther}, H.M.},
  \bibinfo{author}{{Deil}, C.}, \bibinfo{author}{{Woillez}, J.},
  \bibinfo{author}{{Conseil}, S.}, \bibinfo{author}{{Kramer}, R.},
  \bibinfo{author}{{Turner}, J.E.H.}, \bibinfo{author}{{Singer}, L.},
  \bibinfo{author}{{Fox}, R.}, \bibinfo{author}{{Weaver}, B.A.},
  \bibinfo{author}{{Zabalza}, V.}, \bibinfo{author}{{Edwards}, Z.I.},
  \bibinfo{author}{{Azalee Bostroem}, K.}, \bibinfo{author}{{Burke}, D.J.},
  \bibinfo{author}{{Casey}, A.R.}, \bibinfo{author}{{Crawford}, S.M.},
  \bibinfo{author}{{Dencheva}, N.}, \bibinfo{author}{{Ely}, J.},
  \bibinfo{author}{{Jenness}, T.}, \bibinfo{author}{{Labrie}, K.},
  \bibinfo{author}{{Lim}, P.L.}, \bibinfo{author}{{Pierfederici}, F.},
  \bibinfo{author}{{Pontzen}, A.}, \bibinfo{author}{{Ptak}, A.},
  \bibinfo{author}{{Refsdal}, B.}, \bibinfo{author}{{Servillat}, M.},
  \bibinfo{author}{{Streicher}, O.}, \bibinfo{year}{2013}.
\newblock \bibinfo{title}{{Astropy: A community Python package for astronomy}}.
\newblock \bibinfo{journal}{\aap} \bibinfo{volume}{558}, \bibinfo{pages}{A33}.
\newblock \DOIprefix\doi{10.1051/0004-6361/201322068},
  \href{http://arxiv.org/abs/1307.6212}{\tt arXiv:1307.6212}.
\bibitem[{{Becerra} et~al.(2015){Becerra}, {Byrne} and
  {Brown}}]{2015Icar..251..211B}
\bibinfo{author}{{Becerra}, P.}, \bibinfo{author}{{Byrne}, S.},
  \bibinfo{author}{{Brown}, A.J.}, \bibinfo{year}{2015}.
\newblock \bibinfo{title}{{Transient bright ''halos'' on the South Polar
  Residual Cap of Mars: Implications for mass-balance}}.
\newblock \bibinfo{journal}{\icarus} \bibinfo{volume}{251},
  \bibinfo{pages}{211--225}.
\newblock \DOIprefix\doi{10.1016/j.icarus.2014.04.050}.
\bibitem[{{Becker} et~al.(2007){Becker}, {Anderson}, {Sides}, {Miller},
  {Eliason} and {Keszthelyi}}]{2007LPI....38.1779B}
\bibinfo{author}{{Becker}, K.J.}, \bibinfo{author}{{Anderson}, J.A.},
  \bibinfo{author}{{Sides}, S.C.}, \bibinfo{author}{{Miller}, E.A.},
  \bibinfo{author}{{Eliason}, E.M.}, \bibinfo{author}{{Keszthelyi}, L.P.},
  \bibinfo{year}{2007}.
\newblock \bibinfo{title}{{Processing HiRISE Images Using ISIS3}}, in:
  \bibinfo{booktitle}{Lunar and Planetary Science Conference}, p.
  \bibinfo{pages}{1779}.
\bibitem[{{Benson} and {James}(2005)}]{2005Icar..174..513B}
\bibinfo{author}{{Benson}, J.L.}, \bibinfo{author}{{James}, P.B.},
  \bibinfo{year}{2005}.
\newblock \bibinfo{title}{{Yearly comparisons of the martian polar caps: 1999
  2003 Mars Orbiter Camera observations}}.
\newblock \bibinfo{journal}{\icarus} \bibinfo{volume}{174},
  \bibinfo{pages}{513--523}.
\newblock \DOIprefix\doi{10.1016/j.icarus.2004.08.025}.
\bibitem[{{Bierson} et~al.(2016){Bierson}, {Phillips}, {Smith}, {Wood},
  {Putzig}, {Nunes} and {Byrne}}]{2016GeoRL..43.4172B}
\bibinfo{author}{{Bierson}, C.J.}, \bibinfo{author}{{Phillips}, R.J.},
  \bibinfo{author}{{Smith}, I.B.}, \bibinfo{author}{{Wood}, S.E.},
  \bibinfo{author}{{Putzig}, N.E.}, \bibinfo{author}{{Nunes}, D.},
  \bibinfo{author}{{Byrne}, S.}, \bibinfo{year}{2016}.
\newblock \bibinfo{title}{{Stratigraphy and evolution of the buried CO$_{2}$
  deposit in the Martian south polar cap}}.
\newblock \bibinfo{journal}{\grl} \bibinfo{volume}{43},
  \bibinfo{pages}{4172--4179}.
\newblock \DOIprefix\doi{10.1002/2016GL068457}.
\bibitem[{{Bugiolacchi} et~al.(2016){Bugiolacchi}, {Bamford}, {Tar}, {Thacker},
  {Crawford}, {Joy}, {Grindrod} and {Lintott}}]{2016Icar..271...30B}
\bibinfo{author}{{Bugiolacchi}, R.}, \bibinfo{author}{{Bamford}, S.},
  \bibinfo{author}{{Tar}, P.}, \bibinfo{author}{{Thacker}, N.},
  \bibinfo{author}{{Crawford}, I.A.}, \bibinfo{author}{{Joy}, K.H.},
  \bibinfo{author}{{Grindrod}, P.M.}, \bibinfo{author}{{Lintott}, C.},
  \bibinfo{year}{2016}.
\newblock \bibinfo{title}{{The Moon Zoo citizen science project: Preliminary
  results for the Apollo 17 landing site}}.
\newblock \bibinfo{journal}{\icarus} \bibinfo{volume}{271},
  \bibinfo{pages}{30--48}.
\newblock \DOIprefix\doi{10.1016/j.icarus.2016.01.021},
  \href{http://arxiv.org/abs/1602.01664}{\tt arXiv:1602.01664}.
\bibitem[{{Buhler} et~al.(2017){Buhler}, {Ingersoll}, {Ehlmann}, {Fassett} and
  {Head}}]{2017Icar..286...69B}
\bibinfo{author}{{Buhler}, P.B.}, \bibinfo{author}{{Ingersoll}, A.P.},
  \bibinfo{author}{{Ehlmann}, B.L.}, \bibinfo{author}{{Fassett}, C.I.},
  \bibinfo{author}{{Head}, J.W.}, \bibinfo{year}{2017}.
\newblock \bibinfo{title}{{How the martian residual south polar cap develops
  quasi-circular and heart-shaped pits, troughs, and moats}}.
\newblock \bibinfo{journal}{\icarus} \bibinfo{volume}{286},
  \bibinfo{pages}{69--93}.
\newblock \DOIprefix\doi{10.1016/j.icarus.2017.01.012}.
\bibitem[{{Byrne} and {Ingersoll}(2003)}]{2003Sci...299.1051B}
\bibinfo{author}{{Byrne}, S.}, \bibinfo{author}{{Ingersoll}, A.P.},
  \bibinfo{year}{2003}.
\newblock \bibinfo{title}{{A Sublimation Model for Martian South Polar Ice
  Features}}.
\newblock \bibinfo{journal}{Science} \bibinfo{volume}{299},
  \bibinfo{pages}{1051--1053}.
\newblock \DOIprefix\doi{10.1126/science.1080148}.
\bibitem[{{Christensen} et~al.(2009){Christensen}, {Engle}, {Anwar},
  {Dickenshied}, {Noss}, {Gorelick} and {Weiss-Malik}}]{2009AGUFMIN22A..06C}
\bibinfo{author}{{Christensen}, P.R.}, \bibinfo{author}{{Engle}, E.},
  \bibinfo{author}{{Anwar}, S.}, \bibinfo{author}{{Dickenshied}, S.},
  \bibinfo{author}{{Noss}, D.}, \bibinfo{author}{{Gorelick}, N.},
  \bibinfo{author}{{Weiss-Malik}, M.}, \bibinfo{year}{2009}.
\newblock \bibinfo{title}{{JMARS - A Planetary GIS}}.
\newblock \bibinfo{journal}{AGU Fall Meeting Abstracts} .
\bibitem[{{Clancy} et~al.(2000){Clancy}, {Sandor}, {Wolff}, {Christensen},
  {Smith}, {Pearl}, {Conrath} and {Wilson}}]{2000JGR...105.9553C}
\bibinfo{author}{{Clancy}, R.T.}, \bibinfo{author}{{Sandor}, B.J.},
  \bibinfo{author}{{Wolff}, M.J.}, \bibinfo{author}{{Christensen}, P.R.},
  \bibinfo{author}{{Smith}, M.D.}, \bibinfo{author}{{Pearl}, J.C.},
  \bibinfo{author}{{Conrath}, B.J.}, \bibinfo{author}{{Wilson}, R.J.},
  \bibinfo{year}{2000}.
\newblock \bibinfo{title}{{An intercomparison of ground-based millimeter, MGS
  TES, and Viking atmospheric temperature measurements: Seasonal and
  interannual variability of temperatures and dust loading in the global Mars
  atmosphere}}.
\newblock \bibinfo{journal}{\jgr} \bibinfo{volume}{105},
  \bibinfo{pages}{9553--9572}.
\newblock \DOIprefix\doi{10.1029/1999JE001089}.
\bibitem[{{Clifford} et~al.(2000){Clifford}, {Crisp}, {Fisher}, {Herkenhoff},
  {Smrekar}, {Thomas}, {Wynn-Williams}, {Zurek}, {Barnes}, {Bills}, {Blake},
  {Calvin}, {Cameron}, {Carr}, {Christensen}, {Clark}, {Clow}, {Cutts},
  {Dahl-Jensen}, {Durham}, {Fanale}, {Farmer}, {Forget}, {Gotto-Azuma},
  {Grard}, {Haberle}, {Harrison}, {Harvey}, {Howard}, {Ingersoll}, {James},
  {Kargel}, {Kieffer}, {Larsen}, {Lepper}, {Malin}, {McCleese}, {Murray},
  {Nye}, {Paige}, {Platt}, {Plaut}, {Reeh}, {Rice}, {Smith}, {Stoker},
  {Tanaka}, {Mosley-Thompson}, {Thorsteinsson}, {Wood}, {Zent}, {Zuber} and
  {Jay Zwally}}]{2000Icar..144..210C}
\bibinfo{author}{{Clifford}, S.M.}, \bibinfo{author}{{Crisp}, D.},
  \bibinfo{author}{{Fisher}, D.A.}, \bibinfo{author}{{Herkenhoff}, K.E.},
  \bibinfo{author}{{Smrekar}, S.E.}, \bibinfo{author}{{Thomas}, P.C.},
  \bibinfo{author}{{Wynn-Williams}, D.D.}, \bibinfo{author}{{Zurek}, R.W.},
  \bibinfo{author}{{Barnes}, J.R.}, \bibinfo{author}{{Bills}, B.G.},
  \bibinfo{author}{{Blake}, E.W.}, \bibinfo{author}{{Calvin}, W.M.},
  \bibinfo{author}{{Cameron}, J.M.}, \bibinfo{author}{{Carr}, M.H.},
  \bibinfo{author}{{Christensen}, P.R.}, \bibinfo{author}{{Clark}, B.C.},
  \bibinfo{author}{{Clow}, G.D.}, \bibinfo{author}{{Cutts}, J.A.},
  \bibinfo{author}{{Dahl-Jensen}, D.}, \bibinfo{author}{{Durham}, W.B.},
  \bibinfo{author}{{Fanale}, F.P.}, \bibinfo{author}{{Farmer}, J.D.},
  \bibinfo{author}{{Forget}, F.}, \bibinfo{author}{{Gotto-Azuma}, K.},
  \bibinfo{author}{{Grard}, R.}, \bibinfo{author}{{Haberle}, R.M.},
  \bibinfo{author}{{Harrison}, W.}, \bibinfo{author}{{Harvey}, R.},
  \bibinfo{author}{{Howard}, A.D.}, \bibinfo{author}{{Ingersoll}, A.P.},
  \bibinfo{author}{{James}, P.B.}, \bibinfo{author}{{Kargel}, J.S.},
  \bibinfo{author}{{Kieffer}, H.H.}, \bibinfo{author}{{Larsen}, J.},
  \bibinfo{author}{{Lepper}, K.}, \bibinfo{author}{{Malin}, M.C.},
  \bibinfo{author}{{McCleese}, D.J.}, \bibinfo{author}{{Murray}, B.},
  \bibinfo{author}{{Nye}, J.F.}, \bibinfo{author}{{Paige}, D.A.},
  \bibinfo{author}{{Platt}, S.R.}, \bibinfo{author}{{Plaut}, J.J.},
  \bibinfo{author}{{Reeh}, N.}, \bibinfo{author}{{Rice}, J.W.},
  \bibinfo{author}{{Smith}, D.E.}, \bibinfo{author}{{Stoker}, C.R.},
  \bibinfo{author}{{Tanaka}, K.L.}, \bibinfo{author}{{Mosley-Thompson}, E.},
  \bibinfo{author}{{Thorsteinsson}, T.}, \bibinfo{author}{{Wood}, S.E.},
  \bibinfo{author}{{Zent}, A.}, \bibinfo{author}{{Zuber}, M.T.},
  \bibinfo{author}{{Jay Zwally}, H.}, \bibinfo{year}{2000}.
\newblock \bibinfo{title}{{The State and Future of Mars Polar Science and
  Exploration}}.
\newblock \bibinfo{journal}{\icarus} \bibinfo{volume}{144},
  \bibinfo{pages}{210--242}.
\newblock \DOIprefix\doi{10.1006/icar.1999.6290}.
\bibitem[{{Cutts}(1973)}]{1973JGR....78.4231C}
\bibinfo{author}{{Cutts}, J.A.}, \bibinfo{year}{1973}.
\newblock \bibinfo{title}{{Nature and origin of layered deposits of the Martian
  polar region.}}
\newblock \bibinfo{journal}{\jgr} \bibinfo{volume}{78},
  \bibinfo{pages}{4231--4249}.
\newblock \DOIprefix\doi{10.1029/JB078i020p04231}.
\bibitem[{{de Villiers} et~al.(2012){de Villiers}, {Nermoen}, {Jamtveit},
  {Mathiesen}, {Meakin} and {Werner}}]{2012GeoRL..3913204D}
\bibinfo{author}{{de Villiers}, S.}, \bibinfo{author}{{Nermoen}, A.},
  \bibinfo{author}{{Jamtveit}, B.}, \bibinfo{author}{{Mathiesen}, J.},
  \bibinfo{author}{{Meakin}, P.}, \bibinfo{author}{{Werner}, S.C.},
  \bibinfo{year}{2012}.
\newblock \bibinfo{title}{{Formation of Martian araneiforms by gas-driven
  erosion of granular material}}.
\newblock \bibinfo{journal}{\grl} \bibinfo{volume}{39},
  \bibinfo{pages}{L13204}.
\newblock \DOIprefix\doi{10.1029/2012GL052226}.
\bibitem[{{Edwards} et~al.(2011){Edwards}, {Nowicki}, {Christensen}, {Hill},
  {Gorelick} and {Murray}}]{2011JGRE..11610008E}
\bibinfo{author}{{Edwards}, C.S.}, \bibinfo{author}{{Nowicki}, K.J.},
  \bibinfo{author}{{Christensen}, P.R.}, \bibinfo{author}{{Hill}, J.},
  \bibinfo{author}{{Gorelick}, N.}, \bibinfo{author}{{Murray}, K.},
  \bibinfo{year}{2011}.
\newblock \bibinfo{title}{{Mosaicking of global planetary image datasets: 1.
  Techniques and data processing for Thermal Emission Imaging System (THEMIS)
  multi-spectral data}}.
\newblock \bibinfo{journal}{Journal of Geophysical Research (Planets)}
  \bibinfo{volume}{116}, \bibinfo{pages}{E10008}.
\newblock \DOIprefix\doi{10.1029/2010JE003755}.
\bibitem[{{Fischer} et~al.(2012){Fischer}, {Schwamb}, {Schawinski}, {Lintott},
  {Brewer}, {Giguere}, {Lynn}, {Parrish}, {Sartori}, {Simpson}, {Smith},
  {Spronck}, {Batalha}, {Rowe}, {Jenkins}, {Bryson}, {Prsa}, {Tenenbaum},
  {Crepp}, {Morton}, {Howard}, {Beleu}, {Kaplan}, {Vannispen}, {Sharzer},
  {Defouw}, {Hajduk}, {Neal}, {Nemec}, {Schuepbach} and
  {Zimmermann}}]{2012MNRAS.419.2900F}
\bibinfo{author}{{Fischer}, D.A.}, \bibinfo{author}{{Schwamb}, M.E.},
  \bibinfo{author}{{Schawinski}, K.}, \bibinfo{author}{{Lintott}, C.},
  \bibinfo{author}{{Brewer}, J.}, \bibinfo{author}{{Giguere}, M.},
  \bibinfo{author}{{Lynn}, S.}, \bibinfo{author}{{Parrish}, M.},
  \bibinfo{author}{{Sartori}, T.}, \bibinfo{author}{{Simpson}, R.},
  \bibinfo{author}{{Smith}, A.}, \bibinfo{author}{{Spronck}, J.},
  \bibinfo{author}{{Batalha}, N.}, \bibinfo{author}{{Rowe}, J.},
  \bibinfo{author}{{Jenkins}, J.}, \bibinfo{author}{{Bryson}, S.},
  \bibinfo{author}{{Prsa}, A.}, \bibinfo{author}{{Tenenbaum}, P.},
  \bibinfo{author}{{Crepp}, J.}, \bibinfo{author}{{Morton}, T.},
  \bibinfo{author}{{Howard}, A.}, \bibinfo{author}{{Beleu}, M.},
  \bibinfo{author}{{Kaplan}, Z.}, \bibinfo{author}{{Vannispen}, N.},
  \bibinfo{author}{{Sharzer}, C.}, \bibinfo{author}{{Defouw}, J.},
  \bibinfo{author}{{Hajduk}, A.}, \bibinfo{author}{{Neal}, J.P.},
  \bibinfo{author}{{Nemec}, A.}, \bibinfo{author}{{Schuepbach}, N.},
  \bibinfo{author}{{Zimmermann}, V.}, \bibinfo{year}{2012}.
\newblock \bibinfo{title}{{Planet Hunters: the first two planet candidates
  identified by the public using the Kepler public archive data}}.
\newblock \bibinfo{journal}{\mnras} \bibinfo{volume}{419},
  \bibinfo{pages}{2900--2911}.
\newblock \DOIprefix\doi{10.1111/j.1365-2966.2011.19932.x},
  \href{http://arxiv.org/abs/1109.4621}{\tt arXiv:1109.4621}.
\bibitem[{{Fortson} et~al.(2012){Fortson}, {Masters}, {Nichol}, {Borne},
  {Edmondson}, {Lintott}, {Raddick}, {Schawinski} and
  {Wallin}}]{2012amld.book..213F}
\bibinfo{author}{{Fortson}, L.}, \bibinfo{author}{{Masters}, K.},
  \bibinfo{author}{{Nichol}, R.}, \bibinfo{author}{{Borne}, K.D.},
  \bibinfo{author}{{Edmondson}, E.M.}, \bibinfo{author}{{Lintott}, C.},
  \bibinfo{author}{{Raddick}, J.}, \bibinfo{author}{{Schawinski}, K.},
  \bibinfo{author}{{Wallin}, J.}, \bibinfo{year}{2012}.
\newblock \bibinfo{title}{{Galaxy Zoo: Morphological Classification and Citizen
  Science}}.
\newblock pp. \bibinfo{pages}{213--236}.
\bibitem[{{Genova} et~al.(2016){Genova}, {Goossens}, {Lemoine}, {Mazarico},
  {Neumann}, {Smith} and {Zuber}}]{2016Icar..272..228G}
\bibinfo{author}{{Genova}, A.}, \bibinfo{author}{{Goossens}, S.},
  \bibinfo{author}{{Lemoine}, F.G.}, \bibinfo{author}{{Mazarico}, E.},
  \bibinfo{author}{{Neumann}, G.A.}, \bibinfo{author}{{Smith}, D.E.},
  \bibinfo{author}{{Zuber}, M.T.}, \bibinfo{year}{2016}.
\newblock \bibinfo{title}{{Seasonal and static gravity field of Mars from MGS,
  Mars Odyssey and MRO radio science}}.
\newblock \bibinfo{journal}{\icarus} \bibinfo{volume}{272},
  \bibinfo{pages}{228--245}.
\newblock \DOIprefix\doi{10.1016/j.icarus.2016.02.050}.
\bibitem[{{Hansen} et~al.(2010){Hansen}, {Thomas}, {Portyankina}, {McEwen},
  {Becker}, {Byrne}, {Herkenhoff}, {Kieffer} and
  {Mellon}}]{2010Icar..205..283H}
\bibinfo{author}{{Hansen}, C.J.}, \bibinfo{author}{{Thomas}, N.},
  \bibinfo{author}{{Portyankina}, G.}, \bibinfo{author}{{McEwen}, A.},
  \bibinfo{author}{{Becker}, T.}, \bibinfo{author}{{Byrne}, S.},
  \bibinfo{author}{{Herkenhoff}, K.}, \bibinfo{author}{{Kieffer}, H.},
  \bibinfo{author}{{Mellon}, M.}, \bibinfo{year}{2010}.
\newblock \bibinfo{title}{{HiRISE observations of gas sublimation-driven
  activity in Mars'southern polar regions: I. Erosion of the surface}}.
\newblock \bibinfo{journal}{\icarus} \bibinfo{volume}{205},
  \bibinfo{pages}{283--295}.
\newblock \DOIprefix\doi{10.1016/j.icarus.2009.07.021}.
\bibitem[{{Herkenhoff} and {Plaut}(2000)}]{2000Icar..144..243H}
\bibinfo{author}{{Herkenhoff}, K.E.}, \bibinfo{author}{{Plaut}, J.J.},
  \bibinfo{year}{2000}.
\newblock \bibinfo{title}{{Surface Ages and Resurfacing Rates of the Polar
  Layered Deposits on Mars}}.
\newblock \bibinfo{journal}{\icarus} \bibinfo{volume}{144},
  \bibinfo{pages}{243--253}.
\newblock \DOIprefix\doi{10.1006/icar.1999.6287}.
\bibitem[{{Hess} et~al.(1979){Hess}, {Henry} and
  {Tillman}}]{1979JGR....84.2923H}
\bibinfo{author}{{Hess}, S.L.}, \bibinfo{author}{{Henry}, R.M.},
  \bibinfo{author}{{Tillman}, J.E.}, \bibinfo{year}{1979}.
\newblock \bibinfo{title}{{The seasonal variation of atmospheric pressure on
  Mars as affected by the south polar CAP}}.
\newblock \bibinfo{journal}{\jgr} \bibinfo{volume}{84},
  \bibinfo{pages}{2923--2927}.
\newblock \DOIprefix\doi{10.1029/JB084iB06p02923}.
\bibitem[{{Hill} et~al.(2014){Hill}, {Edwards} and
  {Christensen}}]{2014LPICo1791.1141H}
\bibinfo{author}{{Hill}, J.}, \bibinfo{author}{{Edwards}, C.S.},
  \bibinfo{author}{{Christensen}, P.R.}, \bibinfo{year}{2014}.
\newblock \bibinfo{title}{{Mapping the Martian Surface with THEMIS Global
  Infrared Mosaics}}, in: \bibinfo{booktitle}{Eighth International Conference
  on Mars}, p. \bibinfo{pages}{1141}.
\bibitem[{{James} et~al.(2001){James}, {Cantor} and
  {Davis}}]{2001JGR...10623635J}
\bibinfo{author}{{James}, P.B.}, \bibinfo{author}{{Cantor}, B.A.},
  \bibinfo{author}{{Davis}, S.}, \bibinfo{year}{2001}.
\newblock \bibinfo{title}{{Mars Orbiter Camera observations of the Martian
  south polar cap in 1999-2000}}.
\newblock \bibinfo{journal}{\jgr} \bibinfo{volume}{106},
  \bibinfo{pages}{23635--23652}.
\newblock \DOIprefix\doi{10.1029/2000JE001313}.
\bibitem[{{James} et~al.(1992){James}, {Kieffer} and
  {Paige}}]{1992mars.book..934J}
\bibinfo{author}{{James}, P.B.}, \bibinfo{author}{{Kieffer}, H.H.},
  \bibinfo{author}{{Paige}, D.A.}, \bibinfo{year}{1992}.
\newblock \bibinfo{title}{{The seasonal cycle of carbon dioxide on Mars}}.
\newblock pp. \bibinfo{pages}{934--968}.
\bibitem[{{James} et~al.(2010){James}, {Thomas} and
  {Malin}}]{2010Icar..208...82J}
\bibinfo{author}{{James}, P.B.}, \bibinfo{author}{{Thomas}, P.C.},
  \bibinfo{author}{{Malin}, M.C.}, \bibinfo{year}{2010}.
\newblock \bibinfo{title}{{Variability of the south polar cap of Mars in Mars
  years 28 and 29}}.
\newblock \bibinfo{journal}{\icarus} \bibinfo{volume}{208},
  \bibinfo{pages}{82--85}.
\newblock \DOIprefix\doi{10.1016/j.icarus.2010.02.007}.
\bibitem[{Kaufmann and Hagermann(2017)}]{Kaufmann2017118}
\bibinfo{author}{Kaufmann, E.}, \bibinfo{author}{Hagermann, A.},
  \bibinfo{year}{2017}.
\newblock \bibinfo{title}{Experimental investigation of insolation-driven dust
  ejection from mars’ \{CO2\} ice caps}.
\newblock \bibinfo{journal}{Icarus} \bibinfo{volume}{282}, \bibinfo{pages}{118
  -- 126}.
\newblock \DOIprefix\doi{http://dx.doi.org/10.1016/j.icarus.2016.09.039}.
\bibitem[{{Kelly} et~al.(2006){Kelly}, {Boynton}, {Kerry}, {Hamara}, {Janes},
  {Reedy}, {Kim} and {Haberle}}]{2006JGRE..111.3S07K}
\bibinfo{author}{{Kelly}, N.J.}, \bibinfo{author}{{Boynton}, W.V.},
  \bibinfo{author}{{Kerry}, K.}, \bibinfo{author}{{Hamara}, D.},
  \bibinfo{author}{{Janes}, D.}, \bibinfo{author}{{Reedy}, R.C.},
  \bibinfo{author}{{Kim}, K.J.}, \bibinfo{author}{{Haberle}, R.M.},
  \bibinfo{year}{2006}.
\newblock \bibinfo{title}{{Seasonal polar carbon dioxide frost on Mars:
  CO$_{2}$ mass and columnar thickness distribution}}.
\newblock \bibinfo{journal}{Journal of Geophysical Research (Planets)}
  \bibinfo{volume}{111}, \bibinfo{pages}{E03S07}.
\newblock \DOIprefix\doi{10.1029/2006JE002678}.
\bibitem[{{Kieffer}(1979)}]{1979JGR....84.8263K}
\bibinfo{author}{{Kieffer}, H.H.}, \bibinfo{year}{1979}.
\newblock \bibinfo{title}{{Mars south polar spring and summer temperatures - A
  residual CO2 frost}}.
\newblock \bibinfo{journal}{\jgr} \bibinfo{volume}{84},
  \bibinfo{pages}{8263--8288}.
\newblock \DOIprefix\doi{10.1029/JB084iB14p08263}.
\bibitem[{{Kieffer}(2000)}]{2000mpse.conf...93K}
\bibinfo{author}{{Kieffer}, H.H.}, \bibinfo{year}{2000}.
\newblock \bibinfo{title}{{Annual Punctuated CO2 Slab-Ice and Jets on Mars}},
  in: \bibinfo{booktitle}{Second International Conference on Mars Polar Science
  and Exploration}, p.~\bibinfo{pages}{93}.
\bibitem[{{Kieffer}(2007)}]{2007JGRE..112.8005K}
\bibinfo{author}{{Kieffer}, H.H.}, \bibinfo{year}{2007}.
\newblock \bibinfo{title}{{Cold jets in the Martian polar caps}}.
\newblock \bibinfo{journal}{Journal of Geophysical Research (Planets)}
  \bibinfo{volume}{112}, \bibinfo{pages}{E08005}.
\newblock \DOIprefix\doi{10.1029/2006JE002816}.
\bibitem[{{Kieffer} et~al.(2006){Kieffer}, {Christensen} and
  {Titus}}]{2006Natur.442..793K}
\bibinfo{author}{{Kieffer}, H.H.}, \bibinfo{author}{{Christensen}, P.R.},
  \bibinfo{author}{{Titus}, T.N.}, \bibinfo{year}{2006}.
\newblock \bibinfo{title}{{CO$_{2}$ jets formed by sublimation beneath
  translucent slab ice in Mars' seasonal south polar ice cap}}.
\newblock \bibinfo{journal}{\nat} \bibinfo{volume}{442},
  \bibinfo{pages}{793--796}.
\newblock \DOIprefix\doi{10.1038/nature04945}.
\bibitem[{{Kieffer} et~al.(2000){Kieffer}, {Titus}, {Mullins} and
  {Christensen}}]{2000JGR...105.9653K}
\bibinfo{author}{{Kieffer}, H.H.}, \bibinfo{author}{{Titus}, T.N.},
  \bibinfo{author}{{Mullins}, K.F.}, \bibinfo{author}{{Christensen}, P.R.},
  \bibinfo{year}{2000}.
\newblock \bibinfo{title}{{Mars south polar spring and summer behavior observed
  by TES: Seasonal cap evolution controlled by frost grain size}}.
\newblock \bibinfo{journal}{\jgr} \bibinfo{volume}{105},
  \bibinfo{pages}{9653--9700}.
\newblock \DOIprefix\doi{10.1029/1999JE001136}.
\bibitem[{{Kieffer} and {Zent}(1992)}]{1992mars.book.1180K}
\bibinfo{author}{{Kieffer}, H.H.}, \bibinfo{author}{{Zent}, A.P.},
  \bibinfo{year}{1992}.
\newblock \bibinfo{title}{{Quasi-periodic climate change on Mars}}.
\newblock pp. \bibinfo{pages}{1180--1218}.
\bibitem[{{Kraft} et~al.(1991){Kraft}, {Burrows} and
  {Nousek}}]{1991ApJ...374..344K}
\bibinfo{author}{{Kraft}, R.P.}, \bibinfo{author}{{Burrows}, D.N.},
  \bibinfo{author}{{Nousek}, J.A.}, \bibinfo{year}{1991}.
\newblock \bibinfo{title}{{Determination of confidence limits for experiments
  with low numbers of counts}}.
\newblock \bibinfo{journal}{\apj} \bibinfo{volume}{374},
  \bibinfo{pages}{344--355}.
\newblock \DOIprefix\doi{10.1086/170124}.
\bibitem[{{Kuchner} et~al.(2016){Kuchner}, {Silverberg}, {Bans},
  {Bhattacharjee}, {Kenyon}, {Debes}, {Currie}, {Garcia}, {Jung}, {Lintott},
  {McElwain}, {Padgett}, {Rebull}, {Wisniewski}, {Nesvold}, {Schawinski},
  {Thaller}, {Grady}, {Biggs}, {Bosch}, {Cernohous}, {Durantini Luca}, {Hyogo},
  {Wah}, {Piipuu} and {Pi{\~n}eiro}}]{2016arXiv160705713K}
\bibinfo{author}{{Kuchner}, M.J.}, \bibinfo{author}{{Silverberg}, S.M.},
  \bibinfo{author}{{Bans}, A.S.}, \bibinfo{author}{{Bhattacharjee}, S.},
  \bibinfo{author}{{Kenyon}, S.J.}, \bibinfo{author}{{Debes}, J.H.},
  \bibinfo{author}{{Currie}, T.}, \bibinfo{author}{{Garcia}, L.},
  \bibinfo{author}{{Jung}, D.}, \bibinfo{author}{{Lintott}, C.},
  \bibinfo{author}{{McElwain}, M.}, \bibinfo{author}{{Padgett}, D.L.},
  \bibinfo{author}{{Rebull}, L.M.}, \bibinfo{author}{{Wisniewski}, J.P.},
  \bibinfo{author}{{Nesvold}, E.}, \bibinfo{author}{{Schawinski}, K.},
  \bibinfo{author}{{Thaller}, M.L.}, \bibinfo{author}{{Grady}, C.A.},
  \bibinfo{author}{{Biggs}, J.}, \bibinfo{author}{{Bosch}, M.},
  \bibinfo{author}{{Cernohous}, T.}, \bibinfo{author}{{Durantini Luca}, H.A.},
  \bibinfo{author}{{Hyogo}, M.}, \bibinfo{author}{{Wah}, L.L.W.},
  \bibinfo{author}{{Piipuu}, A.}, \bibinfo{author}{{Pi{\~n}eiro}, F.},
  \bibinfo{year}{2016}.
\newblock \bibinfo{title}{{Disk Detective: Discovery of New Circumstellar Disk
  Candidates through Citizen Science}}.
\newblock \bibinfo{journal}{ArXiv e-prints}
  \href{http://arxiv.org/abs/1607.05713}{\tt arXiv:1607.05713}.
\bibitem[{{Landis} et~al.(2016){Landis}, {Byrne}, {Daubar}, {Herkenhoff} and
  {Dundas}}]{2016GeoRL..43.3060L}
\bibinfo{author}{{Landis}, M.E.}, \bibinfo{author}{{Byrne}, S.},
  \bibinfo{author}{{Daubar}, I.J.}, \bibinfo{author}{{Herkenhoff}, K.E.},
  \bibinfo{author}{{Dundas}, C.M.}, \bibinfo{year}{2016}.
\newblock \bibinfo{title}{{A revised surface age for the North Polar Layered
  Deposits of Mars}}.
\newblock \bibinfo{journal}{\grl} \bibinfo{volume}{43},
  \bibinfo{pages}{3060--3068}.
\newblock \DOIprefix\doi{10.1002/2016GL068434}.
\bibitem[{{Leighton} and {Murray}(1966)}]{1966Sci...153..136L}
\bibinfo{author}{{Leighton}, R.B.}, \bibinfo{author}{{Murray}, B.C.},
  \bibinfo{year}{1966}.
\newblock \bibinfo{title}{{Behavior of Carbon Dioxide and Other Volatiles on
  Mars}}.
\newblock \bibinfo{journal}{Science} \bibinfo{volume}{153},
  \bibinfo{pages}{136--144}.
\newblock \DOIprefix\doi{10.1126/science.153.3732.136}.
\bibitem[{{Lintott} et~al.(2011){Lintott}, {Schawinski}, {Bamford}, {Slosar},
  {Land}, {Thomas}, {Edmondson}, {Masters}, {Nichol}, {Raddick}, {Szalay},
  {Andreescu}, {Murray} and {Vandenberg}}]{2011MNRAS.410..166L}
\bibinfo{author}{{Lintott}, C.}, \bibinfo{author}{{Schawinski}, K.},
  \bibinfo{author}{{Bamford}, S.}, \bibinfo{author}{{Slosar}, A.},
  \bibinfo{author}{{Land}, K.}, \bibinfo{author}{{Thomas}, D.},
  \bibinfo{author}{{Edmondson}, E.}, \bibinfo{author}{{Masters}, K.},
  \bibinfo{author}{{Nichol}, R.C.}, \bibinfo{author}{{Raddick}, M.J.},
  \bibinfo{author}{{Szalay}, A.}, \bibinfo{author}{{Andreescu}, D.},
  \bibinfo{author}{{Murray}, P.}, \bibinfo{author}{{Vandenberg}, J.},
  \bibinfo{year}{2011}.
\newblock \bibinfo{title}{{Galaxy Zoo 1: data release of morphological
  classifications for nearly 900 000 galaxies}}.
\newblock \bibinfo{journal}{\mnras} \bibinfo{volume}{410},
  \bibinfo{pages}{166--178}.
\newblock \DOIprefix\doi{10.1111/j.1365-2966.2010.17432.x},
  \href{http://arxiv.org/abs/1007.3265}{\tt arXiv:1007.3265}.
\bibitem[{{Lintott} et~al.(2008){Lintott}, {Schawinski}, {Slosar}, {Land},
  {Bamford}, {Thomas}, {Raddick}, {Nichol}, {Szalay}, {Andreescu}, {Murray} and
  {Vandenberg}}]{2008MNRAS.389.1179L}
\bibinfo{author}{{Lintott}, C.J.}, \bibinfo{author}{{Schawinski}, K.},
  \bibinfo{author}{{Slosar}, A.}, \bibinfo{author}{{Land}, K.},
  \bibinfo{author}{{Bamford}, S.}, \bibinfo{author}{{Thomas}, D.},
  \bibinfo{author}{{Raddick}, M.J.}, \bibinfo{author}{{Nichol}, R.C.},
  \bibinfo{author}{{Szalay}, A.}, \bibinfo{author}{{Andreescu}, D.},
  \bibinfo{author}{{Murray}, P.}, \bibinfo{author}{{Vandenberg}, J.},
  \bibinfo{year}{2008}.
\newblock \bibinfo{title}{{Galaxy Zoo: morphologies derived from visual
  inspection of galaxies from the Sloan Digital Sky Survey}}.
\newblock \bibinfo{journal}{\mnras} \bibinfo{volume}{389},
  \bibinfo{pages}{1179--1189}.
\newblock \DOIprefix\doi{10.1111/j.1365-2966.2008.13689.x},
  \href{http://arxiv.org/abs/0804.4483}{\tt arXiv:0804.4483}.
\bibitem[{{Litvak} et~al.(2007){Litvak}, {Mitrofanov}, {Kozyrev}, {Sanin},
  {Tretyakov}, {Boynton}, {Kelly}, {Hamara} and
  {Saunders}}]{2007JGRE..112.3S13L}
\bibinfo{author}{{Litvak}, M.L.}, \bibinfo{author}{{Mitrofanov}, I.G.},
  \bibinfo{author}{{Kozyrev}, A.S.}, \bibinfo{author}{{Sanin}, A.B.},
  \bibinfo{author}{{Tretyakov}, V.I.}, \bibinfo{author}{{Boynton}, W.V.},
  \bibinfo{author}{{Kelly}, N.J.}, \bibinfo{author}{{Hamara}, D.},
  \bibinfo{author}{{Saunders}, R.S.}, \bibinfo{year}{2007}.
\newblock \bibinfo{title}{{Long-term observations of southern winters on Mars:
  Estimations of column thickness, mass, and volume density of the seasonal
  CO$_{2}$ deposit from HEND/Odyssey data}}.
\newblock \bibinfo{journal}{Journal of Geophysical Research (Planets)}
  \bibinfo{volume}{112}, \bibinfo{pages}{E03S13}.
\newblock \DOIprefix\doi{10.1029/2006JE002832}.
\bibitem[{{Malin} et~al.(2007){Malin}, {Bell}, {Cantor}, {Caplinger}, {Calvin},
  {Clancy}, {Edgett}, {Edwards}, {Haberle}, {James}, {Lee}, {Ravine}, {Thomas}
  and {Wolff}}]{2007JGRE..112.5S04M}
\bibinfo{author}{{Malin}, M.C.}, \bibinfo{author}{{Bell}, J.F.},
  \bibinfo{author}{{Cantor}, B.A.}, \bibinfo{author}{{Caplinger}, M.A.},
  \bibinfo{author}{{Calvin}, W.M.}, \bibinfo{author}{{Clancy}, R.T.},
  \bibinfo{author}{{Edgett}, K.S.}, \bibinfo{author}{{Edwards}, L.},
  \bibinfo{author}{{Haberle}, R.M.}, \bibinfo{author}{{James}, P.B.},
  \bibinfo{author}{{Lee}, S.W.}, \bibinfo{author}{{Ravine}, M.A.},
  \bibinfo{author}{{Thomas}, P.C.}, \bibinfo{author}{{Wolff}, M.J.},
  \bibinfo{year}{2007}.
\newblock \bibinfo{title}{{Context Camera Investigation on board the Mars
  Reconnaissance Orbiter}}.
\newblock \bibinfo{journal}{Journal of Geophysical Research (Planets)}
  \bibinfo{volume}{112}, \bibinfo{pages}{E05S04}.
\newblock \DOIprefix\doi{10.1029/2006JE002808}.
\bibitem[{{Malin} et~al.(2001){Malin}, {Caplinger} and
  {Davis}}]{2001Sci...294.2146M}
\bibinfo{author}{{Malin}, M.C.}, \bibinfo{author}{{Caplinger}, M.A.},
  \bibinfo{author}{{Davis}, S.D.}, \bibinfo{year}{2001}.
\newblock \bibinfo{title}{{Observational Evidence for an Active Surface
  Reservoir of Solid Carbon Dioxide on Mars}}.
\newblock \bibinfo{journal}{Science} \bibinfo{volume}{294},
  \bibinfo{pages}{2146--2148}.
\newblock \DOIprefix\doi{10.1126/science.1066416}.
\bibitem[{{Malin} et~al.(1992){Malin}, {Danielson}, {Ingersoll}, {Masursky},
  {Veverka}, {Ravine} and {Soulanille}}]{1992JGR....97.7699M}
\bibinfo{author}{{Malin}, M.C.}, \bibinfo{author}{{Danielson}, G.E.},
  \bibinfo{author}{{Ingersoll}, A.P.}, \bibinfo{author}{{Masursky}, H.},
  \bibinfo{author}{{Veverka}, J.}, \bibinfo{author}{{Ravine}, M.A.},
  \bibinfo{author}{{Soulanille}, T.A.}, \bibinfo{year}{1992}.
\newblock \bibinfo{title}{{Mars Observer camera}}.
\newblock \bibinfo{journal}{\jgr} \bibinfo{volume}{97},
  \bibinfo{pages}{7699--7718}.
\newblock \DOIprefix\doi{10.1029/92JE00340}.
\bibitem[{{Malin} and {Edgett}(2001)}]{2001JGR...10623429M}
\bibinfo{author}{{Malin}, M.C.}, \bibinfo{author}{{Edgett}, K.S.},
  \bibinfo{year}{2001}.
\newblock \bibinfo{title}{{Mars Global Surveyor Mars Orbiter Camera:
  Interplanetary cruise through primary mission}}.
\newblock \bibinfo{journal}{\jgr} \bibinfo{volume}{106},
  \bibinfo{pages}{23429--23570}.
\newblock \DOIprefix\doi{10.1029/2000JE001455}.
\bibitem[{{Malin} et~al.(2010){Malin}, {Edgett}, {Cantor}, {Caplinger},
  {Danielson}, {Jensen}, {Ravine}, {Sandoval} and
  {Supulver}}]{2010IJMSE...5....1M}
\bibinfo{author}{{Malin}, M.C.}, \bibinfo{author}{{Edgett}, K.S.},
  \bibinfo{author}{{Cantor}, B.A.}, \bibinfo{author}{{Caplinger}, M.A.},
  \bibinfo{author}{{Danielson}, G.E.}, \bibinfo{author}{{Jensen}, E.H.},
  \bibinfo{author}{{Ravine}, M.A.}, \bibinfo{author}{{Sandoval}, J.L.},
  \bibinfo{author}{{Supulver}, K.D.}, \bibinfo{year}{2010}.
\newblock \bibinfo{title}{{An overview of the 1985-2006 Mars Orbiter Camera
  science investigation}}.
\newblock \bibinfo{journal}{International Journal of Mars Science and
  Exploration} \bibinfo{volume}{5}, \bibinfo{pages}{1--60}.
\newblock \DOIprefix\doi{10.1555/mars.2010.0001}.
\bibitem[{{Mangold}(2005)}]{2005Icar..174..336M}
\bibinfo{author}{{Mangold}, N.}, \bibinfo{year}{2005}.
\newblock \bibinfo{title}{{High latitude patterned grounds on Mars:
  Classification, distribution and climatic control}}.
\newblock \bibinfo{journal}{\icarus} \bibinfo{volume}{174},
  \bibinfo{pages}{336--359}.
\newblock \DOIprefix\doi{10.1016/j.icarus.2004.07.030}.
\bibitem[{{Marshall} et~al.(2015){Marshall}, {Lintott} and
  {Fletcher}}]{2015ARA&A..53..247M}
\bibinfo{author}{{Marshall}, P.J.}, \bibinfo{author}{{Lintott}, C.J.},
  \bibinfo{author}{{Fletcher}, L.N.}, \bibinfo{year}{2015}.
\newblock \bibinfo{title}{{Ideas for Citizen Science in Astronomy}}.
\newblock \bibinfo{journal}{\araa} \bibinfo{volume}{53},
  \bibinfo{pages}{247--278}.
\newblock \DOIprefix\doi{10.1146/annurev-astro-081913-035959},
  \href{http://arxiv.org/abs/1409.4291}{\tt arXiv:1409.4291}.
\bibitem[{{Matsuo} and {Heki}(2009)}]{2009Icar..202...90M}
\bibinfo{author}{{Matsuo}, K.}, \bibinfo{author}{{Heki}, K.},
  \bibinfo{year}{2009}.
\newblock \bibinfo{title}{{Seasonal and inter-annual changes of volume density
  of martian CO $_{2}$ snow from time-variable elevation and gravity}}.
\newblock \bibinfo{journal}{\icarus} \bibinfo{volume}{202},
  \bibinfo{pages}{90--94}.
\newblock \DOIprefix\doi{10.1016/j.icarus.2009.02.023}.
\bibitem[{{McEwen} et~al.(2007){McEwen}, {Eliason}, {Bergstrom}, {Bridges},
  {Hansen}, {Delamere}, {Grant}, {Gulick}, {Herkenhoff}, {Keszthelyi}, {Kirk},
  {Mellon}, {Squyres}, {Thomas} and {Weitz}}]{2007JGRE..112.5S02M}
\bibinfo{author}{{McEwen}, A.S.}, \bibinfo{author}{{Eliason}, E.M.},
  \bibinfo{author}{{Bergstrom}, J.W.}, \bibinfo{author}{{Bridges}, N.T.},
  \bibinfo{author}{{Hansen}, C.J.}, \bibinfo{author}{{Delamere}, W.A.},
  \bibinfo{author}{{Grant}, J.A.}, \bibinfo{author}{{Gulick}, V.C.},
  \bibinfo{author}{{Herkenhoff}, K.E.}, \bibinfo{author}{{Keszthelyi}, L.},
  \bibinfo{author}{{Kirk}, R.L.}, \bibinfo{author}{{Mellon}, M.T.},
  \bibinfo{author}{{Squyres}, S.W.}, \bibinfo{author}{{Thomas}, N.},
  \bibinfo{author}{{Weitz}, C.M.}, \bibinfo{year}{2007}.
\newblock \bibinfo{title}{{Mars Reconnaissance Orbiter's High Resolution
  Imaging Science Experiment (HiRISE)}}.
\newblock \bibinfo{journal}{Journal of Geophysical Research (Planets)}
  \bibinfo{volume}{112}, \bibinfo{pages}{E05S02}.
\newblock \DOIprefix\doi{10.1029/2005JE002605}.
\bibitem[{{McEwen} et~al.(2011){McEwen}, {Ojha}, {Dundas}, {Mattson}, {Byrne},
  {Wray}, {Cull}, {Murchie}, {Thomas} and {Gulick}}]{2011Sci...333..740M}
\bibinfo{author}{{McEwen}, A.S.}, \bibinfo{author}{{Ojha}, L.},
  \bibinfo{author}{{Dundas}, C.M.}, \bibinfo{author}{{Mattson}, S.S.},
  \bibinfo{author}{{Byrne}, S.}, \bibinfo{author}{{Wray}, J.J.},
  \bibinfo{author}{{Cull}, S.C.}, \bibinfo{author}{{Murchie}, S.L.},
  \bibinfo{author}{{Thomas}, N.}, \bibinfo{author}{{Gulick}, V.C.},
  \bibinfo{year}{2011}.
\newblock \bibinfo{title}{{Seasonal Flows on Warm Martian Slopes}}.
\newblock \bibinfo{journal}{Science} \bibinfo{volume}{333},
  \bibinfo{pages}{740}.
\newblock \DOIprefix\doi{10.1126/science.1204816}.
\bibitem[{{Ojha} et~al.(2014){Ojha}, {McEwen}, {Dundas}, {Byrne}, {Mattson},
  {Wray}, {Masse} and {Schaefer}}]{2014Icar..231..365O}
\bibinfo{author}{{Ojha}, L.}, \bibinfo{author}{{McEwen}, A.},
  \bibinfo{author}{{Dundas}, C.}, \bibinfo{author}{{Byrne}, S.},
  \bibinfo{author}{{Mattson}, S.}, \bibinfo{author}{{Wray}, J.},
  \bibinfo{author}{{Masse}, M.}, \bibinfo{author}{{Schaefer}, E.},
  \bibinfo{year}{2014}.
\newblock \bibinfo{title}{{HiRISE observations of Recurring Slope Lineae (RSL)
  during southern summer on Mars}}.
\newblock \bibinfo{journal}{\icarus} \bibinfo{volume}{231},
  \bibinfo{pages}{365--376}.
\newblock \DOIprefix\doi{10.1016/j.icarus.2013.12.021}.
\bibitem[{{Phillips} et~al.(2011){Phillips}, {Davis}, {Tanaka}, {Byrne},
  {Mellon}, {Putzig}, {Haberle}, {Kahre}, {Campbell}, {Carter}, {Smith},
  {Holt}, {Smrekar}, {Nunes}, {Plaut}, {Egan}, {Titus} and
  {Seu}}]{2011Sci...332..838P}
\bibinfo{author}{{Phillips}, R.J.}, \bibinfo{author}{{Davis}, B.J.},
  \bibinfo{author}{{Tanaka}, K.L.}, \bibinfo{author}{{Byrne}, S.},
  \bibinfo{author}{{Mellon}, M.T.}, \bibinfo{author}{{Putzig}, N.E.},
  \bibinfo{author}{{Haberle}, R.M.}, \bibinfo{author}{{Kahre}, M.A.},
  \bibinfo{author}{{Campbell}, B.A.}, \bibinfo{author}{{Carter}, L.M.},
  \bibinfo{author}{{Smith}, I.B.}, \bibinfo{author}{{Holt}, J.W.},
  \bibinfo{author}{{Smrekar}, S.E.}, \bibinfo{author}{{Nunes}, D.C.},
  \bibinfo{author}{{Plaut}, J.J.}, \bibinfo{author}{{Egan}, A.F.},
  \bibinfo{author}{{Titus}, T.N.}, \bibinfo{author}{{Seu}, R.},
  \bibinfo{year}{2011}.
\newblock \bibinfo{title}{{Massive CO$_{2}$ Ice Deposits Sequestered in the
  South Polar Layered Deposits of Mars}}.
\newblock \bibinfo{journal}{Science} \bibinfo{volume}{332},
  \bibinfo{pages}{838}.
\newblock \DOIprefix\doi{10.1126/science.1203091}.
\bibitem[{{Pilorget} et~al.(2013){Pilorget}, {Edwards}, {Ehlmann}, {Forget} and
  {Millour}}]{2013JGRE..118.2520P}
\bibinfo{author}{{Pilorget}, C.}, \bibinfo{author}{{Edwards}, C.S.},
  \bibinfo{author}{{Ehlmann}, B.L.}, \bibinfo{author}{{Forget}, F.},
  \bibinfo{author}{{Millour}, E.}, \bibinfo{year}{2013}.
\newblock \bibinfo{title}{{Material ejection by the cold jets and temperature
  evolution of the south seasonal polar cap of Mars from THEMIS/CRISM
  observations and implications for surface properties}}.
\newblock \bibinfo{journal}{Journal of Geophysical Research (Planets)}
  \bibinfo{volume}{118}, \bibinfo{pages}{2520--2536}.
\newblock \DOIprefix\doi{10.1002/2013JE004513}.
\bibitem[{{Pilorget} et~al.(2011){Pilorget}, {Forget}, {Millour}, {Vincendon}
  and {Madeleine}}]{2011Icar..213..131P}
\bibinfo{author}{{Pilorget}, C.}, \bibinfo{author}{{Forget}, F.},
  \bibinfo{author}{{Millour}, E.}, \bibinfo{author}{{Vincendon}, M.},
  \bibinfo{author}{{Madeleine}, J.B.}, \bibinfo{year}{2011}.
\newblock \bibinfo{title}{{Dark spots and cold jets in the polar regions of
  Mars: New clues from a thermal model of surface CO $_{2}$ ice}}.
\newblock \bibinfo{journal}{\icarus} \bibinfo{volume}{213},
  \bibinfo{pages}{131--149}.
\newblock \DOIprefix\doi{10.1016/j.icarus.2011.01.031}.
\bibitem[{{Piqueux} et~al.(2015a){Piqueux}, {Byrne}, {Kieffer}, {Titus} and
  {Hansen}}]{2015Icar..251..332P}
\bibinfo{author}{{Piqueux}, S.}, \bibinfo{author}{{Byrne}, S.},
  \bibinfo{author}{{Kieffer}, H.H.}, \bibinfo{author}{{Titus}, T.N.},
  \bibinfo{author}{{Hansen}, C.J.}, \bibinfo{year}{2015}a.
\newblock \bibinfo{title}{{Enumeration of Mars years and seasons since the
  beginning of telescopic exploration}}.
\newblock \bibinfo{journal}{\icarus} \bibinfo{volume}{251},
  \bibinfo{pages}{332--338}.
\newblock \DOIprefix\doi{10.1016/j.icarus.2014.12.014}.
\bibitem[{{Piqueux} et~al.(2003){Piqueux}, {Byrne} and
  {Richardson}}]{2003JGRE..108.5084P}
\bibinfo{author}{{Piqueux}, S.}, \bibinfo{author}{{Byrne}, S.},
  \bibinfo{author}{{Richardson}, M.I.}, \bibinfo{year}{2003}.
\newblock \bibinfo{title}{{Sublimation of Mars's southern seasonal CO$_{2}$ ice
  cap and the formation of spiders}}.
\newblock \bibinfo{journal}{Journal of Geophysical Research (Planets)}
  \bibinfo{volume}{108}, \bibinfo{pages}{3--1}.
\newblock \DOIprefix\doi{10.1029/2002JE002007}.
\bibitem[{{Piqueux} and {Christensen}(2008)}]{2008JGRE..113.6005P}
\bibinfo{author}{{Piqueux}, S.}, \bibinfo{author}{{Christensen}, P.R.},
  \bibinfo{year}{2008}.
\newblock \bibinfo{title}{{North and south subice gas flow and venting of the
  seasonal caps of Mars: A major geomorphological agent}}.
\newblock \bibinfo{journal}{Journal of Geophysical Research (Planets)}
  \bibinfo{volume}{113}, \bibinfo{pages}{E06005}.
\newblock \DOIprefix\doi{10.1029/2007JE003009}.
\bibitem[{{Piqueux} et~al.(2015b){Piqueux}, {Kleinb{\"o}hl}, {Hayne}, {Kass},
  {Schofield} and {McCleese}}]{2015Icar..251..164P}
\bibinfo{author}{{Piqueux}, S.}, \bibinfo{author}{{Kleinb{\"o}hl}, A.},
  \bibinfo{author}{{Hayne}, P.O.}, \bibinfo{author}{{Kass}, D.M.},
  \bibinfo{author}{{Schofield}, J.T.}, \bibinfo{author}{{McCleese}, D.J.},
  \bibinfo{year}{2015}b.
\newblock \bibinfo{title}{{Variability of the martian seasonal CO$_{2}$ cap
  extent over eight Mars Years}}.
\newblock \bibinfo{journal}{\icarus} \bibinfo{volume}{251},
  \bibinfo{pages}{164--180}.
\newblock \DOIprefix\doi{10.1016/j.icarus.2014.10.045}.
\bibitem[{{Pommerol} et~al.(2011){Pommerol}, {Portyankina}, {Thomas}, {Aye},
  {Hansen}, {Vincendon} and {Langevin}}]{2011JGRE..116.8007P}
\bibinfo{author}{{Pommerol}, A.}, \bibinfo{author}{{Portyankina}, G.},
  \bibinfo{author}{{Thomas}, N.}, \bibinfo{author}{{Aye}, K.M.},
  \bibinfo{author}{{Hansen}, C.J.}, \bibinfo{author}{{Vincendon}, M.},
  \bibinfo{author}{{Langevin}, Y.}, \bibinfo{year}{2011}.
\newblock \bibinfo{title}{{Evolution of south seasonal cap during Martian
  spring: Insights from high-resolution observations by HiRISE and CRISM on
  Mars Reconnaissance Orbiter}}.
\newblock \bibinfo{journal}{Journal of Geophysical Research (Planets)}
  \bibinfo{volume}{116}, \bibinfo{pages}{E08007}.
\newblock \DOIprefix\doi{10.1029/2010JE003790}.
\bibitem[{{Portyankina}(2005)}]{Portyankina2005}
\bibinfo{author}{{Portyankina}, G.}, \bibinfo{year}{2005}.
\newblock \bibinfo{title}{{Atmosphere-surface vapor exchange and ices in the
  Martian polar regions}}.
\newblock \bibinfo{journal}{PhD Thesis,ISBN 3-936586-47-0} .
\bibitem[{Portyankina et~al.(2017)Portyankina, Hansen and
  Aye}]{Portyankina201793}
\bibinfo{author}{Portyankina, G.}, \bibinfo{author}{Hansen, C.J.},
  \bibinfo{author}{Aye, K.M.}, \bibinfo{year}{2017}.
\newblock \bibinfo{title}{Present-day erosion of martian polar terrain by the
  seasonal \{CO2\} jets}.
\newblock \bibinfo{journal}{Icarus} \bibinfo{volume}{282}, \bibinfo{pages}{93
  -- 103}.
\newblock \DOIprefix\doi{http://dx.doi.org/10.1016/j.icarus.2016.09.007}.
\bibitem[{{Portyankina} et~al.(2010){Portyankina}, {Markiewicz}, {Thomas},
  {Hansen} and {Milazzo}}]{2010Icar..205..311P}
\bibinfo{author}{{Portyankina}, G.}, \bibinfo{author}{{Markiewicz}, W.J.},
  \bibinfo{author}{{Thomas}, N.}, \bibinfo{author}{{Hansen}, C.J.},
  \bibinfo{author}{{Milazzo}, M.}, \bibinfo{year}{2010}.
\newblock \bibinfo{title}{{HiRISE observations of gas sublimation-driven
  activity in Mars' southern polar regions: III. Models of processes involving
  translucent ice}}.
\newblock \bibinfo{journal}{\icarus} \bibinfo{volume}{205},
  \bibinfo{pages}{311--320}.
\newblock \DOIprefix\doi{10.1016/j.icarus.2009.08.029}.
\bibitem[{{Portyankina} et~al.(2012){Portyankina}, {Pommerol}, {Aye}, {Hansen}
  and {Thomas}}]{2012JGRE..117.2006P}
\bibinfo{author}{{Portyankina}, G.}, \bibinfo{author}{{Pommerol}, A.},
  \bibinfo{author}{{Aye}, K.M.}, \bibinfo{author}{{Hansen}, C.J.},
  \bibinfo{author}{{Thomas}, N.}, \bibinfo{year}{2012}.
\newblock \bibinfo{title}{{Polygonal cracks in the seasonal semi-translucent
  CO$_{2}$ ice layer in Martian polar areas}}.
\newblock \bibinfo{journal}{Journal of Geophysical Research (Planets)}
  \bibinfo{volume}{117}, \bibinfo{pages}{E02006}.
\newblock \DOIprefix\doi{10.1029/2011JE003917}.
\bibitem[{{Prettyman} et~al.(2009){Prettyman}, {Feldman} and
  {Titus}}]{2009JGRE..114.8005P}
\bibinfo{author}{{Prettyman}, T.H.}, \bibinfo{author}{{Feldman}, W.C.},
  \bibinfo{author}{{Titus}, T.N.}, \bibinfo{year}{2009}.
\newblock \bibinfo{title}{{Characterization of Mars' seasonal caps using
  neutron spectroscopy}}.
\newblock \bibinfo{journal}{Journal of Geophysical Research (Planets)}
  \bibinfo{volume}{114}, \bibinfo{pages}{E08005}.
\newblock \DOIprefix\doi{10.1029/2008JE003275}.
\bibitem[{{Robbins} et~al.(2014){Robbins}, {Antonenko}, {Kirchoff}, {Chapman},
  {Fassett}, {Herrick}, {Singer}, {Zanetti}, {Lehan}, {Huang} and
  {Gay}}]{2014Icar..234..109R}
\bibinfo{author}{{Robbins}, S.J.}, \bibinfo{author}{{Antonenko}, I.},
  \bibinfo{author}{{Kirchoff}, M.R.}, \bibinfo{author}{{Chapman}, C.R.},
  \bibinfo{author}{{Fassett}, C.I.}, \bibinfo{author}{{Herrick}, R.R.},
  \bibinfo{author}{{Singer}, K.}, \bibinfo{author}{{Zanetti}, M.},
  \bibinfo{author}{{Lehan}, C.}, \bibinfo{author}{{Huang}, D.},
  \bibinfo{author}{{Gay}, P.L.}, \bibinfo{year}{2014}.
\newblock \bibinfo{title}{{The variability of crater identification among
  expert and community crater analysts}}.
\newblock \bibinfo{journal}{\icarus} \bibinfo{volume}{234},
  \bibinfo{pages}{109--131}.
\newblock \DOIprefix\doi{10.1016/j.icarus.2014.02.022},
  \href{http://arxiv.org/abs/1404.1334}{\tt arXiv:1404.1334}.
\bibitem[{{Robbins} and {Hynek}(2012)}]{2012JGRE..117.5004R}
\bibinfo{author}{{Robbins}, S.J.}, \bibinfo{author}{{Hynek}, B.M.},
  \bibinfo{year}{2012}.
\newblock \bibinfo{title}{{A new global database of Mars impact craters {$\ge$}1  km: 1. Database creation, properties, and parameters}}.
\newblock \bibinfo{journal}{Journal of Geophysical Research (Planets)}
  \bibinfo{volume}{117}, \bibinfo{pages}{E05004}.
\newblock \DOIprefix\doi{10.1029/2011JE003966}.
\bibitem[{{Schorghofer} and {Edgett}(2006)}]{2006Icar..180..321S}
\bibinfo{author}{{Schorghofer}, N.}, \bibinfo{author}{{Edgett}, K.S.},
  \bibinfo{year}{2006}.
\newblock \bibinfo{title}{{Seasonal surface frost at low latitudes on Mars}}.
\newblock \bibinfo{journal}{\icarus} \bibinfo{volume}{180},
  \bibinfo{pages}{321--334}.
\newblock \DOIprefix\doi{10.1016/j.icarus.2005.08.022}.
\bibitem[{{Schwamb} et~al.(2012){Schwamb}, {Lintott}, {Fischer}, {Giguere},
  {Lynn}, {Smith}, {Brewer}, {Parrish}, {Schawinski} and
  {Simpson}}]{2012ApJ...754..129S}
\bibinfo{author}{{Schwamb}, M.E.}, \bibinfo{author}{{Lintott}, C.J.},
  \bibinfo{author}{{Fischer}, D.A.}, \bibinfo{author}{{Giguere}, M.J.},
  \bibinfo{author}{{Lynn}, S.}, \bibinfo{author}{{Smith}, A.M.},
  \bibinfo{author}{{Brewer}, J.M.}, \bibinfo{author}{{Parrish}, M.},
  \bibinfo{author}{{Schawinski}, K.}, \bibinfo{author}{{Simpson}, R.J.},
  \bibinfo{year}{2012}.
\newblock \bibinfo{title}{{Planet Hunters: Assessing the Kepler Inventory of
  Short-period Planets}}.
\newblock \bibinfo{journal}{\apj} \bibinfo{volume}{754}, \bibinfo{pages}{129}.
\newblock \DOIprefix\doi{10.1088/0004-637X/754/2/129},
  \href{http://arxiv.org/abs/1205.6769}{\tt arXiv:1205.6769}.
\bibitem[{{Smith} et~al.(2001a){Smith}, {Zuber}, {Frey}, {Garvin}, {Head},
  {Muhleman}, {Pettengill}, {Phillips}, {Solomon}, {Zwally}, {Banerdt},
  {Duxbury}, {Golombek}, {Lemoine}, {Neumann}, {Rowlands}, {Aharonson}, {Ford},
  {Ivanov}, {Johnson}, {McGovern}, {Abshire}, {Afzal} and
  {Sun}}]{2001JGR...10623689S}
\bibinfo{author}{{Smith}, D.E.}, \bibinfo{author}{{Zuber}, M.T.},
  \bibinfo{author}{{Frey}, H.V.}, \bibinfo{author}{{Garvin}, J.B.},
  \bibinfo{author}{{Head}, J.W.}, \bibinfo{author}{{Muhleman}, D.O.},
  \bibinfo{author}{{Pettengill}, G.H.}, \bibinfo{author}{{Phillips}, R.J.},
  \bibinfo{author}{{Solomon}, S.C.}, \bibinfo{author}{{Zwally}, H.J.},
  \bibinfo{author}{{Banerdt}, W.B.}, \bibinfo{author}{{Duxbury}, T.C.},
  \bibinfo{author}{{Golombek}, M.P.}, \bibinfo{author}{{Lemoine}, F.G.},
  \bibinfo{author}{{Neumann}, G.A.}, \bibinfo{author}{{Rowlands}, D.D.},
  \bibinfo{author}{{Aharonson}, O.}, \bibinfo{author}{{Ford}, P.G.},
  \bibinfo{author}{{Ivanov}, A.B.}, \bibinfo{author}{{Johnson}, C.L.},
  \bibinfo{author}{{McGovern}, P.J.}, \bibinfo{author}{{Abshire}, J.B.},
  \bibinfo{author}{{Afzal}, R.S.}, \bibinfo{author}{{Sun}, X.},
  \bibinfo{year}{2001}a.
\newblock \bibinfo{title}{{Mars Orbiter Laser Altimeter: Experiment summary
  after the first year of global mapping of Mars}}.
\newblock \bibinfo{journal}{\jgr} \bibinfo{volume}{106},
  \bibinfo{pages}{23689--23722}.
\newblock \DOIprefix\doi{10.1029/2000JE001364}.
\bibitem[{{Smith} et~al.(2001b){Smith}, {Zuber} and
  {Neumann}}]{2001Sci...294.2141S}
\bibinfo{author}{{Smith}, D.E.}, \bibinfo{author}{{Zuber}, M.T.},
  \bibinfo{author}{{Neumann}, G.A.}, \bibinfo{year}{2001}b.
\newblock \bibinfo{title}{{Seasonal Variations of Snow Depth on Mars}}.
\newblock \bibinfo{journal}{Science} \bibinfo{volume}{294},
  \bibinfo{pages}{2141--2146}.
\newblock \DOIprefix\doi{10.1126/science.1066556}.
\bibitem[{{Tanaka} et~al.(2014){Tanaka}, {Skinner}, {Dohm}, {Irwin},
  III~{Kolb}, {Fortezzo}, T., {Michael} and {Hare}}]{Tanaka:2014wd}
\bibinfo{author}{{Tanaka}, K.L.}, \bibinfo{author}{{Skinner}, J.A.},
  \bibinfo{author}{{Dohm}, J.M.}, \bibinfo{author}{{Irwin}, R.P.},
  \bibinfo{author}{III~{Kolb}, E.J.}, \bibinfo{author}{{Fortezzo}, C.M.},
  \bibinfo{author}{T., P.}, \bibinfo{author}{{Michael}, G.G.},
  \bibinfo{author}{{Hare}, T.M.}, \bibinfo{year}{2014}.
\newblock \bibinfo{title}{{Geologic map of Mars: U.S. Geological Survey
  Scientific Investigations Map 3292, scale 1:20,000,000, pamphle}}.
\newblock \bibinfo{type}{Technical Report}. \bibinfo{address}{Reston, VA C6 -
  ET -}.
\bibitem[{{Thomas} et~al.(2010){Thomas}, {Hansen}, {Portyankina} and
  {Russell}}]{2010Icar..205..296T}
\bibinfo{author}{{Thomas}, N.}, \bibinfo{author}{{Hansen}, C.J.},
  \bibinfo{author}{{Portyankina}, G.}, \bibinfo{author}{{Russell}, P.S.},
  \bibinfo{year}{2010}.
\newblock \bibinfo{title}{{HiRISE observations of gas sublimation-driven
  activity in Mars southern polar regions: II. Surficial deposits and their
  origins}}.
\newblock \bibinfo{journal}{\icarus} \bibinfo{volume}{205},
  \bibinfo{pages}{296--310}.
\newblock \DOIprefix\doi{10.1016/j.icarus.2009.05.030}.
\bibitem[{{Thomas} et~al.(2011){Thomas}, {Portyankina}, {Hansen} and
  {Pommerol}}]{2011GeoRL..38.8203T}
\bibinfo{author}{{Thomas}, N.}, \bibinfo{author}{{Portyankina}, G.},
  \bibinfo{author}{{Hansen}, C.J.}, \bibinfo{author}{{Pommerol}, A.},
  \bibinfo{year}{2011}.
\newblock \bibinfo{title}{{Sub-surface CO$_{2}$ gas flow in Mars' polar
  regions: Gas transport under constant production rate conditions}}.
\newblock \bibinfo{journal}{\grl} \bibinfo{volume}{38},
  \bibinfo{pages}{L08203}.
\newblock \DOIprefix\doi{10.1029/2011GL046797}.
\bibitem[{{Thomas} et~al.(2016){Thomas}, {Calvin}, {Cantor}, {Haberle}, {James}
  and {Lee}}]{2016Icar..268..118T}
\bibinfo{author}{{Thomas}, P.C.}, \bibinfo{author}{{Calvin}, W.},
  \bibinfo{author}{{Cantor}, B.}, \bibinfo{author}{{Haberle}, R.},
  \bibinfo{author}{{James}, P.B.}, \bibinfo{author}{{Lee}, S.W.},
  \bibinfo{year}{2016}.
\newblock \bibinfo{title}{{Mass balance of Mars' residual south polar cap from
  CTX images and other data}}.
\newblock \bibinfo{journal}{\icarus} \bibinfo{volume}{268},
  \bibinfo{pages}{118--130}.
\newblock \DOIprefix\doi{10.1016/j.icarus.2015.12.038}.
\bibitem[{{Thomas} et~al.(2013){Thomas}, {Calvin}, {Gierasch}, {Haberle},
  {James} and {Sholes}}]{2013Icar..225..923T}
\bibinfo{author}{{Thomas}, P.C.}, \bibinfo{author}{{Calvin}, W.M.},
  \bibinfo{author}{{Gierasch}, P.}, \bibinfo{author}{{Haberle}, R.},
  \bibinfo{author}{{James}, P.B.}, \bibinfo{author}{{Sholes}, S.},
  \bibinfo{year}{2013}.
\newblock \bibinfo{title}{{Time scales of erosion and deposition recorded in
  the residual south polar cap of Mars}}.
\newblock \bibinfo{journal}{\icarus} \bibinfo{volume}{225},
  \bibinfo{pages}{923--932}.
\newblock \DOIprefix\doi{10.1016/j.icarus.2012.08.038}.
\bibitem[{{Thomas} et~al.(2009){Thomas}, {James}, {Calvin}, {Haberle} and
  {Malin}}]{2009Icar..203..352T}
\bibinfo{author}{{Thomas}, P.C.}, \bibinfo{author}{{James}, P.B.},
  \bibinfo{author}{{Calvin}, W.M.}, \bibinfo{author}{{Haberle}, R.},
  \bibinfo{author}{{Malin}, M.C.}, \bibinfo{year}{2009}.
\newblock \bibinfo{title}{{Residual south polar cap of Mars: Stratigraphy,
  history, and implications of recent changes}}.
\newblock \bibinfo{journal}{\icarus} \bibinfo{volume}{203},
  \bibinfo{pages}{352--375}.
\newblock \DOIprefix\doi{10.1016/j.icarus.2009.05.014}.
\bibitem[{{Thomas} et~al.(2000){Thomas}, {Malin}, {Edgett}, {Carr}, {Hartmann},
  {Ingersoll}, {James}, {Soderblom}, {Veverka} and
  {Sullivan}}]{2000Natur.404..161T}
\bibinfo{author}{{Thomas}, P.C.}, \bibinfo{author}{{Malin}, M.C.},
  \bibinfo{author}{{Edgett}, K.S.}, \bibinfo{author}{{Carr}, M.H.},
  \bibinfo{author}{{Hartmann}, W.K.}, \bibinfo{author}{{Ingersoll}, A.P.},
  \bibinfo{author}{{James}, P.B.}, \bibinfo{author}{{Soderblom}, L.A.},
  \bibinfo{author}{{Veverka}, J.}, \bibinfo{author}{{Sullivan}, R.},
  \bibinfo{year}{2000}.
\newblock \bibinfo{title}{{North-south geological differences between the
  residual polar caps on Mars}}.
\newblock \bibinfo{journal}{\nat} \bibinfo{volume}{404},
  \bibinfo{pages}{161--164}.
\bibitem[{{Thomas} et~al.(2005){Thomas}, {Malin}, {James}, {Cantor}, {Williams}
  and {Gierasch}}]{2005Icar..174..535T}
\bibinfo{author}{{Thomas}, P.C.}, \bibinfo{author}{{Malin}, M.C.},
  \bibinfo{author}{{James}, P.B.}, \bibinfo{author}{{Cantor}, B.A.},
  \bibinfo{author}{{Williams}, R.M.E.}, \bibinfo{author}{{Gierasch}, P.},
  \bibinfo{year}{2005}.
\newblock \bibinfo{title}{{South polar residual cap of Mars: Features,
  stratigraphy, and changes}}.
\newblock \bibinfo{journal}{\icarus} \bibinfo{volume}{174},
  \bibinfo{pages}{535--559}.
\newblock \DOIprefix\doi{10.1016/j.icarus.2004.07.028}.
\bibitem[{{Titus} et~al.(2003){Titus}, {Kieffer} and
  {Christensen}}]{2003Sci...299.1048T}
\bibinfo{author}{{Titus}, T.N.}, \bibinfo{author}{{Kieffer}, H.H.},
  \bibinfo{author}{{Christensen}, P.R.}, \bibinfo{year}{2003}.
\newblock \bibinfo{title}{{Exposed Water Ice Discovered near the South Pole of
  Mars}}.
\newblock \bibinfo{journal}{Science} \bibinfo{volume}{299},
  \bibinfo{pages}{1048--1051}.
\newblock \DOIprefix\doi{10.1126/science.299.5609.1048}.
\bibitem[{{Vasavada} et~al.(2000){Vasavada}, {Williams}, {Paige}, {Herkenhoff},
  {Bridges}, {Greeley}, {Murray}, {Bass} and {McBride}}]{2000JGR...105.6961V}
\bibinfo{author}{{Vasavada}, A.R.}, \bibinfo{author}{{Williams}, J.P.},
  \bibinfo{author}{{Paige}, D.A.}, \bibinfo{author}{{Herkenhoff}, K.E.},
  \bibinfo{author}{{Bridges}, N.T.}, \bibinfo{author}{{Greeley}, R.},
  \bibinfo{author}{{Murray}, B.C.}, \bibinfo{author}{{Bass}, D.S.},
  \bibinfo{author}{{McBride}, K.S.}, \bibinfo{year}{2000}.
\newblock \bibinfo{title}{{Surface properties of Mars' polar layered deposits
  and polar landing sites}}.
\newblock \bibinfo{journal}{\jgr} \bibinfo{volume}{105},
  \bibinfo{pages}{6961--6970}.
\newblock \DOIprefix\doi{10.1029/1999JE001108}.
\bibitem[{{Willett} et~al.(2013){Willett}, {Lintott}, {Bamford}, {Masters},
  {Simmons}, {Casteels}, {Edmondson}, {Fortson}, {Kaviraj}, {Keel}, {Melvin},
  {Nichol}, {Raddick}, {Schawinski}, {Simpson}, {Skibba}, {Smith} and
  {Thomas}}]{2013MNRAS.435.2835W}
\bibinfo{author}{{Willett}, K.W.}, \bibinfo{author}{{Lintott}, C.J.},
  \bibinfo{author}{{Bamford}, S.P.}, \bibinfo{author}{{Masters}, K.L.},
  \bibinfo{author}{{Simmons}, B.D.}, \bibinfo{author}{{Casteels}, K.R.V.},
  \bibinfo{author}{{Edmondson}, E.M.}, \bibinfo{author}{{Fortson}, L.F.},
  \bibinfo{author}{{Kaviraj}, S.}, \bibinfo{author}{{Keel}, W.C.},
  \bibinfo{author}{{Melvin}, T.}, \bibinfo{author}{{Nichol}, R.C.},
  \bibinfo{author}{{Raddick}, M.J.}, \bibinfo{author}{{Schawinski}, K.},
  \bibinfo{author}{{Simpson}, R.J.}, \bibinfo{author}{{Skibba}, R.A.},
  \bibinfo{author}{{Smith}, A.M.}, \bibinfo{author}{{Thomas}, D.},
  \bibinfo{year}{2013}.
\newblock \bibinfo{title}{{Galaxy Zoo 2: detailed morphological classifications
  for 304 122 galaxies from the Sloan Digital Sky Survey}}.
\newblock \bibinfo{journal}{\mnras} \bibinfo{volume}{435},
  \bibinfo{pages}{2835--2860}.
\newblock \DOIprefix\doi{10.1093/mnras/stt1458},
  \href{http://arxiv.org/abs/1308.3496}{\tt arXiv:1308.3496}.
\bibitem[{{Wood} and {Paige}(1992)}]{1992Icar...99....1W}
\bibinfo{author}{{Wood}, S.E.}, \bibinfo{author}{{Paige}, D.A.},
  \bibinfo{year}{1992}.
\newblock \bibinfo{title}{{Modeling the Martian seasonal CO2 cycle. I - Fitting
  the Viking Lander pressure curves. II - Interannual variability}}.
\newblock \bibinfo{journal}{\icarus} \bibinfo{volume}{99},
  \bibinfo{pages}{1--27}.
\newblock \DOIprefix\doi{10.1016/0019-1035(92)90166-5}.
\bibitem[{{Zuber} et~al.(1992){Zuber}, {Smith}, {Solomon}, {Muhleman}, {Head},
  {Garvin}, {Abshire} and {Bufton}}]{1992JGR....97.7781Z}
\bibinfo{author}{{Zuber}, M.T.}, \bibinfo{author}{{Smith}, D.E.},
  \bibinfo{author}{{Solomon}, S.C.}, \bibinfo{author}{{Muhleman}, D.O.},
  \bibinfo{author}{{Head}, J.W.}, \bibinfo{author}{{Garvin}, J.B.},
  \bibinfo{author}{{Abshire}, J.B.}, \bibinfo{author}{{Bufton}, J.L.},
  \bibinfo{year}{1992}.
\newblock \bibinfo{title}{{The Mars Observer laser altimeter investigation}}.
\newblock \bibinfo{journal}{\jgr} \bibinfo{volume}{97},
  \bibinfo{pages}{7781--7797}.
\newblock \DOIprefix\doi{10.1029/92JE00341}.

\end{thebibliography}

\end{document}